\documentclass[seqeqn]{elsart}
\usepackage{cite}
\usepackage{color}
\usepackage{slashed}
\usepackage{amsfonts}
\usepackage{amsmath}
\usepackage[colorlinks=true, pdfstartview=FitV, linkcolor=blue,  citecolor=blue, urlcolor=blue]{hyperref}
\usepackage{tocloft}
\usepackage{ulem}

\definecolor{lightgrey}{gray}{0.9}
\def\btab#1\etab{\begin{tabular}{p{45mm}p{65mm}}#1\end{tabular}}
\def\btabx#1\etabx{\begin{tabular}{p{60mm}p{50mm}}#1\end{tabular}}
\def\btaby#1\etaby{\begin{tabular}{p{15mm}p{95mm}}#1\end{tabular}}
\def\bcen{\begin{center}}
\def\ecen{\end{center}}
\def\bgfb#1\egfb{\bcen\fcolorbox{black}{lightgrey}{\parbox{118mm}{\btab#1\etab}}\ecen}
\def\bgfbx#1\egfbx{\bcen\fcolorbox{black}{lightgrey}{\parbox{118mm}{\btabx#1\etabx}}\ecen}
\def\bgfbalign#1\egfbalign{\bcen\fcolorbox{black}{lightgrey}{\parbox{118mm}{\btaby#1\etaby}}\ecen}
\newcommand{\comment}[1]{}
\def\beq{\begin{equation}}
\def\eeq{\end{equation}}
\def\bsp#1\esp{\begin{split}#1\end{split}}

\definecolor{lightgrey}{gray}{0.9}

\def\be{\begin{equation}}
\def\ee{\end{equation}}
\def\bea{\begin{eqnarray}}
\def\eea{\end{eqnarray}}
\def\bsp#1\esp{\begin{split}#1\end{split}}
\def\bpm{\begin{pmatrix}} 
\def\epm{\end{pmatrix}} 
\def\bcen{\begin{center}}
\def\ecen{\end{center}}
\def\nn{\nonumber}
\def\bv{\begin{verbatim}}
\def\ev{\end{verbatim}}

\newcommand{\etc}{{\it etc.}}
\newcommand{\ie}{{\it i.e.}}
\newcommand{\eg}{{\it e.g.}}

\newcommand{\del}{\partial}

\newcommand{\Qbar}{{\bar Q}}
\newcommand{\Dbar}{{\bar D}}
\newcommand{\Wbar}{{\overline W}}
\newcommand{\sibar}{{\bar\sigma}}

\newcommand{\lambar}{{\bar\lambda}} 
\newcommand{\thetabar}{{\bar\theta}}

\newcommand{\alphadot}{{\dot\alpha}} 
\newcommand{\betadot}{{\dot\beta}}

\newcommand{\feynrules}{{\sc FeynRules}}
\newcommand{\FRversion}{2.0}
\newcommand{\asperge}{{\sc ASperGe}}
\newcommand{\gsl}{{\sc Gsl}}
\newcommand{\feynarts}{{\sc FeynArts}}
\newcommand{\mathematica}{{\sc Mathematica}}
\newcommand{\calchep}{{\sc CalcHep}}
\newcommand{\comphep}{{\sc CompHep}}

\newcommand{\sherpa}{{\sc Sherpa}}
\newcommand{\whizard}{{\sc Whizard}}
\newcommand{\formcalc}{{\sc Form\-Calc}}
\newcommand{\gosam}{{\sc GoSam}}
\newcommand{\python}{{\sc Python}}
\newcommand{\madgraph}{{\sc MadGraph}}

\newcommand{\madanalysis}{{\sc MadAnalysis}}
\newcommand{\aloha}{{\sc Aloha}}
\newcommand{\herwig}{{\sc Herwig}}
\newcommand{\lanhep}{{\sc LanHep}}
\newcommand{\ohmega}{{\sc Omega}}

\newcommand{\cpp}{{\sc C++}}
\newcommand{\pythia}{{\sc Pythia}}
\newcommand{\ufo}{{\sc UFO}}
\newcommand{\hc}{{\rm h.c.}}

\newcommand{\heshe}{he/she}
\newcommand{\hisher}{his/her}

\definecolor{darkgreen}{rgb}{0.0, 0.45, 0.0}

\begin{document}

\begin{frontmatter}
\begin{flushright}\tiny
CERN-PH-TH/2013-239,
MCNET-13-14,
IPPP/13/71, DCPT/13/142,
PITT-PACC-1308 
\end{flushright}

\title{\feynrules\ \FRversion  - A complete toolbox for tree-level phenomenology}

\author[a]{Adam Alloul},
\author[b]{Neil D.\ Christensen},
\author[c,d]{C\'eline Degrande},
\author[d]{Claude Duhr},
\author[e,f]{Benjamin Fuks}

\address[a]{Groupe de Recherche de Physique des Hautes \'Energies (GRPHE), Universit\'e de Haute-Alsace, IUT Colmar, 34 rue du Grillenbreit BP 50568, 68008 Colmar Cedex, France, E-mail: adam.alloul@iphc.cnrs.fr}
\address[b]{PITTsburgh Particle physics, Astrophysics and Cosmology Center (PITT PACC),\\
     University of Pittsburgh, Pittsburgh, PA 15260 USA,
     E-mail : neilc@pitt.edu}
\address[c]{Department of Physics, University of Illinois at Urbana-Champaign,\\
    1110 West green Street, Urbana, IL 61801, United States,\\ 
    E-mail : cdegrand@illinois.edu}
\address[d]{Institute for Particle Physics Phenomenology, University of Durham,\\ 
    Durham, DH1 3LE, United Kingdom, E-mail: duhrc@itp.phys.ethz.ch}
  \address[e]{Theory Division, Physics Department, CERN, CH-1211 Geneva 23, 
    Switzerland}
\address[f]{Institut Pluridisciplinaire Hubert Curien/D\'epartement Recherches Subatomiques, Universit\'e de Strasbourg/CNRS-IN2P3, 23 Rue du Loess, F-67037 Strasbourg, France, E-mail : benjamin.fuks@iphc.cnrs.fr}

\begin{abstract}
\feynrules\ is a \mathematica-based package which addresses the implementation
of particle physics models, which are given in the form of a list of fields,
parameters and a Lagrangian, into high-energy physics tools.
It calculates the underlying Feynman rules and outputs them to a form appropriate for
various programs such as
\calchep, \feynarts, \madgraph, \sherpa\ and \whizard. 
Since the original version, many new features have been added: support for
two-component fermions, spin-3/2 and spin-2 fields,
superspace notation and calculations, automatic mass diagonalization,
completely general \feynarts\ output, a new universal \feynrules\ output interface, a new \whizard\ interface, automatic $1\to2$ decay width calculation, improved speed and efficiency, new guidelines for validation and a new web-based validation package.  With this feature set, \feynrules\ enables models to go from theory to simulation and comparison with experiment quickly, efficiently and accurately.  

\begin{keyword}
Model building 
\sep Feynman rules 
\sep Monte Carlo programs.
\end{keyword}

\end{abstract}

\end{frontmatter}


\noindent {\bf PROGRAM SUMMARY}                                               \\
  \begin{small}
  {\bf Manuscript Title:} \feynrules\ \FRversion  - A complete toolbox for tree-level phenomenology          \\
  {\bf Authors:} Adam Alloul, Neil.~D.~Christensen, C\'eline Degrande, 
     Claude Duhr, Benjamin Fuks.                                              \\
  {\bf Program Title:} \feynrules\ \FRversion                                          \\
  {\bf Journal Reference:}                                                    \\
  {\bf Catalogue identifier:}                                                 \\
  {\bf Licensing provisions:} None.                                           \\
  {\bf Programming language:} \mathematica.                                   \\
  {\bf Computer:} Platforms on which Mathematica is available.                \\
  {\bf Operating system:} Operating systems on which Mathematica is available.\\
  {\bf Keywords:} Model building, Feynman rules, Monte Carlo simulations.     \\
  {\bf Classification:} 11.1 General, High Energy Physics and Computing.      \\
  \phantom{{\bf Classification:}} 11.6 Phenomenological and Empirical Models
                                 and Theories.                                \\
  {\bf External routines/libraries:} None.                                    \\
  {\bf Nature of problem:} The program computes the Feynman rules of any 
    quantum field theory, expressed in four-dimensional spacetime, directly 
    from the Lagrangian of the model. Various interfaces to Feynman diagram 
    calculators are included that allow to export the interaction vertices in 
    a format readable by different Monte Carlo event generators or symbolic
    calculation tools.                                                        \\
  {\bf Solution method:}  \feynrules\ works in three steps:                   \\
    \begin{enumerate}\vspace{-1cm}
      \item If necessary, the model Lagrangian is written in terms of 
        four-component fermions and the usual fields of particle physics,
        instead of Weyl fermions or superfields.
      \item Derivation of the Feynman rules directly form the Lagrangian 
        using canonical commutation relations among fields and creation 
        operators.
      \item Implementation of the new physics model into \feynarts\ as well 
        as into various Monte Carlo programs via dedicated interfaces.
    \end{enumerate}\vspace{-.5cm} 
  {\bf Restrictions:} \mathematica\ version 7.0 or higher. The Lagrangian 
    must fulfill basic quantum field theory requirements, such as locality 
    and Lorentz and gauge invariance. Fields with spin 0, 1/2, 1, 3/2 
    and 2 are supported.                                                    \\
  {\bf Unusual features:} Translation interfaces to various Feynman 
    diagram generators exist. Superfields are also supported and can be 
    expanded in terms of their component fields, which allows to perform 
    various sets of superspace computations.                                  \\
  {\bf Running time:} The computation of the Feynman rules from a Lagrangian
    varies with the complexity of the model, and runs from a few seconds to 
    several minutes. See Section \ref{sec:bench} of the present manuscript for 
    more information.\\
\end{small}

\newpage



\section{Introduction} \label{sec:intro}

The current era of theoretical particle physics is on the cusp of making several important discoveries.  Although the Standard Model (SM) has been amazingly successful at describing and predicting the outcome of most experiments, there are some significant experiments and observations that can not be explained by the SM.  As a result, new as-yet unknown physics beyond the SM is required and is expected to be discovered in the coming decades at current and future experiments.  Among these puzzles is the instability of the Higgs boson generated vacuum to quantum corrections, the nature of dark matter, neutrino mass, the baryon/antibaryon asymmetry and the hierarchy of fermion masses.


The common feature among these puzzles is that each requires new theoretical models to be built and implemented into simulation software to be compared with experiment and observation.  There are several powerful packages that simulate the effects of new models at particle colliders.  Each of these has its strengths and may be appropriate for particular kinds of calculations.  Among the multi-purpose matrix element generators are \calchep~\cite{Pukhov:1999gg,Boos:2004kh,Pukhov:2004ca,Belyaev:2012qa}, \feynarts/\formcalc~\cite{Hahn:1998yk,Hahn:2000kx,Hahn:2006zy,Hahn:2009bf,Agrawal:2011tm}, {\sc Helac}~\cite{Kanaki:2000ey,Cafarella:2007pc}, \madgraph~\cite{Stelzer:1994ta,Maltoni:2002qb,Alwall:2007st,Alwall:2008pm,Alwall:2011uj}, \sherpa \cite{Gleisberg:2003xi,Gleisberg:2008ta} and \whizard~\cite{Moretti:2001zz,Kilian:2007gr}.  The parton-level collision is often followed by the important step of radiation and hadronization by packages such as \herwig~\cite{Corcella:2000bw,Corcella:2002jc,Bahr:2008pv,Arnold:2012fq}, \pythia~\cite{Sjostrand:2000wi,Sjostrand:2006za,Sjostrand:2007gs} and \sherpa.  However, the step of implementing a new theoretical model into the simulation software has been difficult in the past.  The chief problems have been that: 
\begin{enumerate}
\item[1.] Each event generator uses its own syntax for a new model.  Learning one syntax did not transfer to another generator; 
\item[2.] Implementing a new model often required the modification of the event generator code itself.  These implementations did not transfer well between theorists or to experimentalists; 
\item[3.] These implementations required the hand coding of each vertex one-by-one.  Doing this for the hundreds or thousands of vertices in a model was tedious and very error prone.
\end{enumerate}

\lanhep~\cite{Semenov:1996es,Semenov:1998eb,Semenov:2002jw,Semenov:2008jy,Semenov:2010qt}, \feynrules~\cite{Christensen:2008py} and {\sc Sarah}~\cite{Staub:2008uz,Staub:2012pb} were each created to overcome these challenges. In this paper we present version 2.0 of the \feynrules\ package.  Even at the stage of the first version, \feynrules\ had a significant feature set, allowing theorists to quickly implement their new models into simulation packages.  It allowed the theorist to implement their model once in a single unified syntax, which was Mathematica based and closely related to the syntax of \feynarts.  It also allowed the user to enter the Lagrangian for their model rather than the individual vertices.  It then did the work of calculating the vertices from the Lagrangian, thus saving theorists time and errors.  It came with dedicated export interfaces which wrote the model to file in a form appropriate for the event generators \calchep, \feynarts, \madgraph\ and \sherpa.  The user was not expected to know the syntax of any of these event generators.  Furthermore, for any supported vertex, the final product could be used in the standard version of these event generators and did not require modifications of their code.  Thus, these implementations could be passed between theorists and experimentalists with great success.
This early version was thoroughly tested both between different simulation tools, between different gauges and with independent implementations of a standard set of models \cite{Christensen:2009jx}.
A complete chain was established that went from theoretical model to parton-level simulation package to hadronization, at which point comparison with experiment could be done \cite{Ask:2012sm}.  Since then, many new important features have been added.  We will summarize the major improvements now.

The core was updated to support two-component Weyl fermion notation~\cite{Butterworth:2010ym}. The Lagrangian can now be written in terms of two-component fermions, both left- and right-handed, which may greatly facilitate the implementation of complicated models. Matrix element generators, however, in general work at the level of four-component fermions, and \feynrules\ has the capability to convert the Lagrangian to four-component fermions in an automated way. See Section~\ref{sec:parts} and~\ref{sec:auto susy L} for further details.

Support for spin-$3/2$ was added to the core as well as to the \calchep\ and \ufo\ interfaces \cite{Christensen:2013aua,Alloul:2013jea}.  The resulting output was tested in the context of three models.  The first contained a spin-$3/2$ Majorana gravitino, the second contained a spin-$3/2$ Dirac top-quark excitation, and the third contained a general effective operator approach for a spin-$3/2$ Dirac quark excitation.  The total cross sections, high energy growth cancellations and angular distributions were all tested with theory and agreement was found.  See Section~\ref{sec:parts} for further details.

A superspace module was added \cite{Duhr:2011se} allowing the user to implement a supersymmetric (SUSY) model using the superfields directly. The user has the possibility to implement the superspace action, which can then be automatically expanded into component fields. In addition, the equations of motion for the auxiliary fields can be solved by \feynrules. The superspace module was tested on various non-trivial supersymmetric models, including the Minimal Supersymmetric Standard Model (MSSM) and the left-right symmetric supersymmetric model \cite{Alloul:2013fra}.
See Sections~\ref{sec:superparts}, \ref{sec:auto susy L}, \ref{sec:manipsusy} and \ref{sec:susy computations} for further details.

An automatic mass diagonalization module was created \cite{Alloul:2013fw,Alloul:2013jea}  allowing {\feynrules} to extract automatically from any Lagrangian the tree-level mass matrices and to export them through the {\asperge} interface for a numerical extraction of the spectrum. In practice, one re-organizes slightly the {\feynrules} model file and then uses the new built-in functions to extract the mass matrices and generate the {\asperge} source code. The latter is an ensemble of {\sc C++} routines able to diagonalize the mass matrices and store the results in a SUSY Les-Houches Accord (SLHA)-like file.  See Sections~\ref{sec:mixdecl}, \ref{sec:asperge} and \ref{sec:aspergeinter} for further details.

The \feynarts\ interface was greatly extended to allow the implementation of operators that give rise to Lorentz structures that are not SM or MSSM-like, as is generally the case for higher-dimensional operators. In particular, the new version of the interface creates, on the fly, the generic model file required by \feynarts\ to be able to use vertices with non-standard Lorentz structures. The new interface was tested on dimension-six operators for top-quark production and decay and for anomalous electroweak gauge boson couplings.  See Section~\ref{sec:feynarts} for further details.

The universal \feynrules\ output (\ufo) interface was created~\cite{Degrande:2011ua}.  The idea of this interface is to create a model format that is independent of any individual Feynman diagram calculator and which contains the full information contained in the \feynrules\ model.  It is currently supported by \madgraph~5, \madanalysis~\cite{Conte:2012fm} and \gosam~\cite{Cullen:2011ac}, and will be used in the future by \herwig++~\cite{Bahr:2008pv}.  See Section~\ref{sec:ufo} for further details.

A new export interface to \whizard\ was written \cite{Christensen:2010wz} that allows the output of models in the native format of \whizard.  One highlight of this interface is that it not only supports unitary and Feynman gauge, but will automatically generate $R_\xi$ gauge model files from a model implemented in Feynman gauge.  
The resulting output was tested between gauges and with \calchep\ and \madgraph\ for the SM, MSSM and 3-Site model.
In conjunction with these validations, the first phenomenological study with this interface was performed \cite{Christensen:2012wk}.  See Section~\ref{sec:whizard} for further details.

A module for automatically computing the $1\to2$ widths was added \cite{decaypaper,Alloul:2013jea}. 
\feynrules\ computes the squared matrix elements for all possible $1\to2$ decays present in the model and multiplies by the appropriate phase-space factor. In addition, the analytic formulas can be included into the \ufo\ output and used by the matrix element generators to obtain the tree-level two-body widths and branching ratios for all the particles defined in the model.
See Section~\ref{sec:decays} for further details.

The speed and efficiency of \feynrules\ was greatly improved.  Parts of the core were rewritten to make better use of more efficient Mathematica functions.  Parallelization was also added to some of the most labor-intensive parts of the code.  See Section~\ref{sec:bench} for further details.

New guidelines were established to improve the quality of \feynrules\ based model implementations \cite{Butterworth:2010ym}.  Additionally, a new web-based validation platform was created \cite{Brooijmans:2012yi} to help accomplish this task.  This validation platform allows any user to upload their model implementation.  It then generates a list of $2\to2$ processes which it compares between matrix element generators and between gauges.  It also has the ability to compare with existing, independently implemented versions of the model, if available.  The user is not required to know the details of the different matrix element generators.  Processes with large discrepancies between gauges or between matrix element generators are flagged for the user to look into.  See Section~\ref{sec: web validation} for further details.


The purpose of this manual is to give an up-to-date, integrated and complete instruction set for using \feynrules\ and all its new features.  The organization is as follows.  Sections \ref{sec:modelfile} and \ref{sec:lag} describe the implementation of the model and the Lagrangian.  Section \ref{sec:running} describes how to run \feynrules\ in a Mathematica session.  Section \ref{sec:simpleexample} gives a simple example of a model implementation.  Section \ref{sec:interfaces} describes the use of the export interfaces to generate code that can be run by one of the Feynman diagram calculators.  Section \ref{sec:bench} discusses the speed and efficiency improvements.  Section \ref{sec: web validation} discusses the web based validation package and its use to improve the accuracy of the models built with \feynrules.  In Section \ref{sec:conclusion}, we conclude.


\section{The Model Description}
\label{sec:modelfile}

To use \feynrules, the user must start by entering the details of the new model in a form that can be parsed by \feynrules.  First of all, this means that, since \feynrules{} is a \mathematica\
package, the model must be written in a valid \mathematica\ syntax.  Secondly, \feynrules{} specifies a set of special variables and macros for the user to enter each aspect of a new model.  In this section, we will describe each of these in turn.  These definitions can be placed in a pure text file or in a Mathematica notebook.  The latter can be helpful during the development of the model details, however, we suggest storing the final version of any model in a pure text file.

\subsection{Model Information}
The user has the possibility to include general information about the model implementation
into the model file via the variables \texttt{M\$Mo\-del\-Na\-me} and \texttt{M\$In\-for\-ma\-ti\-on}.
While the first of these variables is a string giving the name of the model, the 
variable \verb+M$Information+ acts as an electronic signature of the model file 
that allows to store, \eg, the names and addresses of the authors of the model 
file, references to the papers used for the implementation, the version number of the 
model file, \etc\ This information is stored as a \mathematica\ replacement list as shown in the following example
\begin{verbatim}
M$ModelName = "my_new_model";

M$Information = {
          Authors      -> {"Mr. X", "Ms. Y"},
          Institutions -> {"UC Louvain"},
          Emails       -> {"X@uclouvain.be", "Y@uclouvain.be},
          Date         -> "01.03.2013",
          References   -> {"reference 1", "reference 2"},
          URLs         -> {"http://feynrules.irmp.ucl.ac.be"},
          Version      -> "1.0"
                };
\end{verbatim}
A summary and complete set of options available for \verb+M$Information+ can be found in Table~\ref{fig:Model Information}.

The model information will be printed on the screen whenever the model is loaded 
into \mathematica. In addition, the contents of \verb+M$Information+ can be 
retrieved by issuing the command \verb+ModelInformation[]+ in a \mathematica\ 
session, after the model has been loaded.

\begin{table}
\bgfb
\multicolumn{2}{c}{\textbf{Table~\ref{fig:Model Information}: Model Information}}\\
\\
\tt{M\$ModelName} &  A string, specifying the name of the model. The default value is the name of the model file. \\
\tt{M\$Information}  & A replacement list, acting as an electronic signature of the model implementation.\\
\tt{ModelInformation[]} & Prints the contents of {\tt M\$Information}.\\
\\
\multicolumn{2}{l}{Options for \texttt{M\$Information}}\\ 
\\
\tt{Authors} & A list of strings, specifying the authors of the model implementation.\\
\tt{Institutions} & A list of strings, specifying the institutions of the authors.\\
\tt{Emails} & A list of strings, specifying the email addresses of the authors. The order of the email addresses is the same as the order in which the names of the authors have been specified.\\
\tt{Date} & A string specifying the date.\\
\tt{References} & A list of strings, containing the references that the authors 
  would like to be cited whenever the model implementation is used.\\
\tt{URLs} & A list of strings, containing Internet addresses where more
  information about the model can be found.\\
\tt{Version} & A version number for the model.  Each time the model is improved, this version number should be increased and a note written in the model file describing the change.
\egfb
\textcolor{white}{\caption{\label{fig:Model Information}}}
\end{table}

 \subsection{Index Definitions}\label{sec:indices}
In general the Lagrangian describing a model is a polynomial in the fields (and their derivatives) 
as well as in the parameters of the model. Very often, these quantities carry 
indices specifying their members and/or how the different quantities transform 
under symmetry operations. For example, the gauge field $G_\mu^a$ of an unbroken 
gauge group $SU(N)$ carries two different types of indices:
 \begin{itemize}
\item[-] a Lorentz index $\mu$ ranging from $0$ to $3$;
\item[-] an adjoint gauge index $a$ ranging from $1$ to $N^2-1$.
\end{itemize}
It is therefore crucial to define at the beginning of each model file the types of indices that appear in the model, together with the range of values each type of index may take. 

A field $\psi_{i_1i_2\dots}(x)$ carrying indices $i_1$, $i_2$, \ldots  is represented inside \feynrules\ by an expression of the form {\tt psi[}$index_1$, $index_2$, \ldots{\tt]}. Each $index_i$ denotes an object of the form \verb+Index[+name, i\verb+]+, and represents an index of type name taking the value i. In this expression name is a symbol and value can be both a symbol or an integer. In general the name can be chosen freely by the user, but we emphasize that there are predefined names for the index types describing four-vectors ({\tt Lorentz}), four-component spinors ({\tt Spin}) and two-component left and right-handed Weyl spinors ({\tt Spin1} and {\tt Spin2}).

For \feynrules\ to run properly, the different types of indices that appear in the model have to be declared at the beginning of the model file, together with the range of values they can take. 
This is achieved like in the following examples
\begin{verbatim}
IndexRange[ Index[Colour] ] = Range[3];
IndexRange[ Index[SU2W] ]  = Unfold[ Range[3] ];
IndexRange[ Index[Gluon] ]  = NoUnfold[ Range[8] ];
\end{verbatim}
These commands declare three types of indices named \verb+Colour+, \verb+SU2W+ and \verb+Gluon+ ranging form $1$ to $3$ and $1$ to $8$ respectively. The function \verb+Range+ is an internal \mathematica\ command taking an integer n as input and returning the range $\{1,\ldots,n\}$. 
Moreover, the indices of type {\tt Lorentz}, {\tt Spin}, {\tt Spin1} and {\tt Spin2} are defined internally and do not need to be defined by the user.

At this stage we have to comment on the functions \verb+Unfold+ and \verb+NoUnfold+ used in the declaration of the indices of type \verb+SU2W+ and \verb+Gluon+:
\begin{enumerate}
\item The \verb+Unfold+ command instructs \feynrules\ that if an index of this type appears contracted inside a monomial, then it should be expanded, \ie, the monomial with the contracted pair of indices should be replaced by the explicit sum over the indices.
Any index that expands in terms of non-physical states must be wrapped in \verb+Unfold+.
For instance, the $SU(2)_L$ indices in the Standard Model or in the Minimal Supersymmetric 
Standard Model must always be expanded
in order to get the Feynman rules in terms of the physical states of the theory. Otherwise, wrong results could be obtained when employing matrix element generators.
We refer to Section~\ref{sec:running} for more details.
\item The \verb+NoUnfold+ is ignored by \feynrules. It however plays a role in \feynarts, and we refer to Section~\ref{sec:feynarts} or to the \feynarts\ manual 
\cite{Hahn:2000kx} for more details. 
\end{enumerate} 

While indices are represented internally inside \feynrules\ by expressions of the form \verb+Index[+name, i\verb+]+, the user does not need to enter indices in this form. Since it is always possible to reconstruct the type of an index from its position inside the expression {\tt psi[}$index_1$, $index_2$, \ldots{\tt]}. For example, the gluon field {\tt G[mu, a]} has been declared as carrying 
two indices, the first one being of type {\tt Lorentz} and the second one of type {\tt Gluon}
(see Section \ref{sec:parts}). \feynrules\ 
can then employ particle class properties to restore the correct notation internally, as in
\vskip0.25cm\noindent
{\tt G[mu, a]} $\longrightarrow$ {\tt G[Index[Lorentz, mu], Index[Gluon, a]]}\,.
\vskip0.25cm
In addition, it is possible to specify how the different types of indices should be printed on the screen.
This is done via the \verb+IndexStyle+ command, \eg,
\begin{verbatim}
IndexStyle[ Colour, i ];
IndexStyle[ Gluon, a ];
\end{verbatim}
Issuing these commands at the beginning of a model file instructs \feynrules\ to 
print indices of type \verb+Colour+ and \verb+Gluon+ with symbols starting with
the letters \verb+i+ and  \verb+a+, respectively, followed by an integer number.

A summary of information for the index can be found in Table~\ref{fig:Index Information}.

\begin{table}
\bgfb
\multicolumn{2}{c}{\textbf{Table~\ref{fig:Index Information}: Index Information}}\\
\\
{\tt Index[}name, value{\tt]} & Represents an index of type 
  name with a value value.\\
{\tt IndexRange[Index[}name{\tt]]} & Declares an index of type name along with its range.\\
{\tt Range[}n{\tt]} & Internal \mathematica\ command, returning the range 
  of integers $\{1,\ldots,n\}$. Allows to declare the range of an index to be 
  from 1 to n.\\
{\tt IndexStyle[}name, label{\tt]} & Fixes the {\tt StandardForm} 
  and {\tt TraditionalForm} of an index of type name to be printed as a 
  symbol starting with label.\\
{\tt NoUnfold} & \feynarts\ command. This is only used by the \feynarts\ 
  interface.\\
{\tt Unfold} & Indices of this type will always be expanded out in the interfaces. See Section~\ref{sec:running}.\\
\\
\multicolumn{2}{l}{Predefined types of indices}\\ 
\\
{\tt Lorentz} & Name for Lorentz indices, ranging from $1$ to $4$, and printed as
   $\mu$.\\
{\tt Spin} & Name for Dirac indices, ranging from $1$ to $4$, and printed as 
  $s$.\\
{\tt Spin1} & Name for left-handed Weyl indices, ranging from $1$ to $2$, and 
  printed as $\alpha$.\\
{\tt Spin2} & Name for right-handed Weyl indices, ranging from $1$ to $2$, and 
  printed as $\dot\alpha$.
\egfb
\textcolor{white}{\caption{\label{fig:Index Information}}}
\end{table}

\subsection{The model parameters}
\label{sec:parameters}
All the model parameters  (coupling constants, mixing angles and matrices,
masses, \etc) are implemented as elements of the list \texttt{M\$Parameters},

\begin{verbatim}
 M$Parameters = { 
   param1 == { options1 },  
   param2 == { options2 },  
           ...  
 };
\end{verbatim}
Each component of this list consists of an equality whose left-hand side is
a label and the right-hand side is a list of \mathematica\ replacement rules. The
labels (\texttt{param1} and \texttt{param2} in the example) are
user-defined names to be used when building the
Lagrangian. The sets of replacement rules (\texttt{options1}
and \texttt{options2} in the example) contain optional information allowing to 
define each parameter together with its properties.
The model parameters are split into two categories according to whether they
carry indices or not. We start by reviewing 
in Section \ref{sec:scpar}
the implementation of scalar
parameters, \ie, parameters that do not carry any index. 
Tensorial parameters, \ie, parameters carrying one or
several indices, are then discussed in Section \ref{sec:tepar}.

\subsubsection{Scalar parameters}\label{sec:scpar}
To illustrate the implementation of scalar parameters, we focus on the
example of the strong coupling constant. 
The declaration of any other parameter is similar.
Although the strong coupling constant $g_s$ usually appears in
the Lagrangian, it is in
general more convenient to use the quantity $\alpha_s= g_s^2/4 \pi$ as an
input parameter, since its numerical value (\eg, at the electroweak scale) 
has been precisely determined from experiments.
It is therefore desirable to have both parameters in the \feynrules\ model
file. This motivates us to choose, in our example, 
$\alpha_s$ as a free parameter of the model, \ie,
as an \textit{external} parameter (in \feynrules\ parlance)
or equivalently as an independent parameter.
In contrast, $g_s$ is an \textit{internal} parameter, or in other words, a
parameter depending on one or several of the other internal and/or external
parameters of the model. 
As a result,  $\alpha_s$ ({\tt aS}) and $g_s$ ({\tt gs})
could be implemented as in the example 
\begin{verbatim}
 aS == {                               
   TeX              -> Subscript[\[Alpha],s], 
   ParameterType    -> External,       
   InteractionOrder -> {QCD, 2},       
   Value            -> 0.1184,         
   BlockName        -> SMINPUTS,       
   OrderBlock       -> 3,
   Description      -> "Strong coupling constant at the Z pole"  
 };

 gs == {
   TeX              -> Subscript[g,s],
   ParameterType    -> Internal,
   ComplexParameter -> False,
   InteractionOrder -> {QCD, 1},
   Value            -> Sqrt[4 Pi aS],
   ParameterName    -> G,
   Description      -> "Strong coupling constant at the Z pole"
  } 
\end{verbatim}
The external or internal nature of a parameter can be specified by setting
the attribute \texttt{ParameterType} to the value
\texttt{External} or \texttt{Internal}.
Another difference between external and internal parameters lies 
in the value taken by the attribute \texttt{Value} of the parameter class. For 
external parameters, it refers
to a \textit{real} number\footnote{Complex external parameters have to be split 
into their real and imaginary parts and declared
individually.} while for internal parameters, it contains a formula. This
formula is given in standard \mathematica\ syntax and provides the way an
internal parameter is connected to the other model parameters. It is important
to note that only previously declared parameters can be employed when
implementing the formula.
Internal parameters can be either real or complex variables, which
is specified by setting the attribute \texttt{ComplexParameter} to the value
\texttt{True} or \texttt{False} (default). 

Instead of providing the value of a parameter (a real number for an external
parameter or a formula for an internal parameter) through the attribute
\texttt{Value}, the user has the option to use the attribute
\texttt{Definitions}. This option of the parameter class takes as argument a
\mathematica\ replacement 
rule. 
Returning to our example above, we could replace the {\tt Value} attribute of
{\tt gs} by
\begin{verbatim}
Definitions -> { gs -> Sqrt[4 Pi aS] }
\end{verbatim}
The difference between these two
choices appears at a later stage and concerns the derivation of the interaction
vertices by
\feynrules. A parameter provided with a definition is removed from the
vertices and replaced by its definition, whilst otherwise, 
the associated symbol is kept.

\begin{table}
\bgfb
\multicolumn{2}{c}{\textbf{Table~\ref{tab:ParameterOptions1}: Attributes of the
 parameter class common for}}\\
\multicolumn{2}{c}{\textbf{scalar and tensorial parameters.}}\\
 {\tt ParameterType} & Specifies the nature of a parameter. The
   allowed choices are {\tt External} or {\tt
   Internal}. By default, scalar parameters are considered as external and
   tensorial parameters as internal.\\
 {\tt Value} & Refers to a real number (for external
   parameters) or to an analytical formula defining the parameter  
   that can be expressed in terms of other parameters (for internal parameters). 
   By default, the \texttt{Value} attribute takes the value 1.\\
 {\tt Definitions} & Refers to a list of \mathematica\
   replacement rules that are applied by \feynrules\ before computing the
   interaction vertices of a model.\\
 {\tt ComplexParameter} & Defines whether a parameter is a real ({\tt
   False}) or complex ({\tt True}) quantity. By default, scalar parameters are
   real while tensorial parameters as complex. External parameters must be real.\\
 {\tt BlockName} & This provides information about the name of the Les Houches 
   block containing an external parameter (see Section \ref{sec:paramin}). By
   default, the block name is taken as {\tt FRBlock}.\\
 {\tt OrderBlock} & Provides information about the position of an external
   parameter within a given Les Houches block. 
   By default, this number starts at one as is
   incremented after the declaration of each parameter. It must be left
   unspecified for tensorial parameters (see Section \ref{sec:paramin}).\\
 {\tt TeX} & Teaches \feynrules\ how to write the \TeX{} form of
   a parameter. By default, it refers to (the string of) the associated \mathematica\ symbol.\\
 {\tt Description} & Refers to a string containing a
   description of the physical meaning of the parameter.
\egfb
\textcolor{white}{\caption{\label{tab:ParameterOptions1}}}
\end{table}

The list of all the attributes of the parameter class 
for scalar quantities is given in 
Table \ref{tab:ParameterOptions1} and Table \ref{tab:ParameterOptions2}. 
With the exception of the {\tt TeX} attribute allowing for the 
\TeX-form of a parameter and the {\tt Description} attribute which gives, as 
a string, the physical meaning of a parameter,
the options non-described so far  
are related to the interfaces to Feynman diagram generators and discussed
in greater detail in Section~\ref{sec:interfaces}.

\begin{table}
\bgfb
\multicolumn{2}{c}{\textbf{Table~\ref{tab:ParameterOptions2}: Additional
attributes of the parameter class}}\\
\multicolumn{2}{c}{\textbf{related to Feynman diagram calculators}}\\
 {\tt ParameterName} & Specifies what to replace the symbol by
   before writing out the Feynman diagram calculator model files. By default,
   it is taken equal to the symbol representing the parameter. See Section
   \ref{sec:namerestr} for more details.\\
 {\tt InteractionOrder} & Specifies the order of the parameter according
   to a specific interaction. It refers to a pair with the interaction name,
   followed by the order, or a list of such pairs. This option has no default 
   value. See Section \ref{sec:intorder} for more details.
\egfb
\textcolor{white}{\caption{\label{tab:ParameterOptions2}}}
\end{table}

\subsubsection{Tensorial parameters}\label{sec:tepar}
The second category of parameters are tensorial parameters, \ie, parameters 
carrying indices.
The index structure can be specified through the attribute \texttt{Indices}
of the parameter class as in the following example,
\begin{verbatim}
Indices -> {Index[Scalar], Index[Generation]}
\end{verbatim}
The parameter under consideration carries two indices, one
index of type {\tt Scalar} and one index of type {\tt Generation}. 
In many cases tensorial parameters correspond to unitary, Hermitian or
orthogonal matrices. These properties can be implemented in the
\feynrules\ model description by turning the values of the attributes 
\texttt{Unitary}, \texttt{Hermitian} or \texttt{Orthogonal} to \texttt{True}, 
the default value being \texttt{False}.

In contrast to scalar parameters, the attribute \texttt{Value}
(\texttt{Definitions}) is now a list of values (replacement rules) 
for all possible numerical value of the indices. Moreover,   
tensorial parameters are by default complex quantities, \ie, the
default value for the attribute \texttt{ComplexParameter} is \texttt{True}.

For example, the Cabibbo-Kobayashi-Maskawa (CKM) matrix
could be defined as follows,
\begin{verbatim}
 CKM == {
   ParameterType    -> Internal,
   Indices          -> {Index[Generation], Index[Generation]},
   Unitary          -> True,
   ComplexParameter -> True,
   Definitions      -> {
     CKM[i_,3] :> 0 /; i!=3, 
     CKM[3,i_] :> 0 /; i!=3,
     CKM[3,3]  -> 1 }, 
   Value            -> {
     CKM[1,1] ->  Cos[cabi],
     CKM[1,2] ->  Sin[cabi], 
     CKM[2,1] -> -Sin[cabi], 
     CKM[2,2] -> Cos[cabi] },
   Description      -> "CKM-Matrix"
 } 
\end{verbatim}
In these declarations, \texttt{cabi} stands for the Cabibbo angle, assumed to 
be declared previously in the model file. 
We have also simultaneously employed the attributes
\texttt{Value} and \texttt{Definitions} so that vanishing elements of the
CKM matrix are removed at the time of the extraction of the interaction vertices
by \feynrules. In this way, zero vertices are removed from the output.

\feynrules\ assumes that
all indices appearing inside a Lagrangian are contracted, \ie, all indices must come in pairs. 
However, it may sometimes be 
convenient to be able
to break this rule. For example, in the case of a diagonal Yukawa matrix $y$,
the associated Lagrangian term 
\beq
 {\cal L}_{\rm yuk} = H\ \bar{\psi}_f\ y_{ff'}\ \psi_{f'}\ , 
\eeq 
where $\psi$ denotes a generic fermionic field and $H$ a generic scalar field, 
can be written in a more compact form as
\beq
 {\cal L}_{\rm yuk} =   \tilde y_f\ H\ \bar{\psi}_f\ \psi_f \ .
\eeq
In the second expression, the Yukawa matrix has been replaced by its diagonal 
form 
$y_{ff'} = \tilde y_f \delta_{ff'}$, rendering the summation
convention explicitly violated since the index $f$ appears three times.
In order to allow for such violations in \feynrules, the attribute 
\verb+AllowSummation+ must be set to the value {\tt True} when implementing the
Yukawa coupling $\tilde y$. Consequently, this allows \feynrules\ to sum the single index
carried by the vectorial parameter $y$ along with the two other indices 
fulfilling the 
summation conventions (in our example, the indices of the fermions). We stress
that this option is only available for parameters carrying one single index. 

In the case of contracted Weyl spin indices, the upper and lower position of the
indices also plays an important role, since 
\beq
  \chi^\alpha \chi'_\alpha = -
  \chi_\alpha \chi'{}^\alpha\ ,
\eeq
where $\chi$ and $\chi'$ are two generic left-handed Weyl fermions. This issue
is described in details in Section \ref{sec:manipsusy}.  

The complete list of specific attributes related to tensorial parameters can be found
in Table \ref{tab:ParameterOptions3}, while all the attributes of Table
\ref{tab:ParameterOptions1} and Table \ref{tab:ParameterOptions2} can also
be employed. 

\begin{table}
\bgfb
\multicolumn{2}{c}{\textbf{Table~\ref{tab:ParameterOptions3}: Attributes of the
parameter class associated}}\\
\multicolumn{2}{c}{\textbf{to tensorial parameters.}}\\
 {\tt Indices} & Mandatory. Provides, as a list, the
   indices carried by a parameter.\\
 {\tt Unitary} & Defines a matrix as unitary ({\tt True}) or not
   ({\tt False}). The default value is {\tt False}.\\
 {\tt Hermitian} & Defines a matrix as Hermitian ({\tt True}) or not
   ({\tt False}). The default value is {\tt False}.\\
 {\tt Orthogonal} & Defines a matrix as orthogonal ({\tt True}) or not
   ({\tt False}). The default value is {\tt False}.\\
 {\tt AllowSummation} & See the description in the text. By default this
   option is set to {\tt False}. Moreover, it is only available for parameters
   with one single index.
\egfb
\textcolor{white}{\caption{\label{tab:ParameterOptions3}}}
\end{table}


\subsection{Particle Classes}
\label{sec:parts}
Particle classes can be instantiated in a similar way to 
parameter classes. The main difference is that, following the original \feynarts\ 
syntax \cite{Hahn:2000kx}, the particle classes are labeled according to the 
spins of the particles rather than to their symbol,
\begin{verbatim}
M$ClassesDescription = {
   spin1[1] == { options1 },
   spin1[2] == { options2 },
   spin2[1] == { options3 },
   ...}
\end{verbatim}
The symbols {\tt spin1}, {\tt spin2},  \etc, refer each to one of 
the field type supported by \feynrules\footnote{The classes {\tt R}, {\tt W} and 
{\tt RW} are specific to \feynrules\ and not supported by \feynarts.}:
\begin{itemize}
\item[-] \verb+S+: scalar fields;
\item[-] \verb+F+: Dirac and Majorana spinor fields;
\item[-] \verb+W+: Weyl fermions (both left- and right-handed);
\item[-] \verb+V+: vector fields;
\item[-] \verb+R+: four-component Rarita-Schwinger fields (spin-3/2 fields);
\item[-] \verb+RW+: two-component Rarita-Schwinger fields (both left- and right-handed spin-3/2 fields);
\item[-] \verb+T+: spin-2 fields;
\item[-] \verb+U+: ghost fields (only complex ghosts are supported).
\end{itemize}
Similar to the declaration of the parameter classes, the quantities {\tt options1}, {\tt options2}, {\tt options3},
\etc, are sets of replacement rules defining field properties.
Following the spirit of the original \feynarts\ model file format,
each particle class should be thought of as a `multiplet' consisting of particles that carry the same quantum numbers but might differ in mass. 
This implies that all fields belonging to the same class necessarily carry the same 
indices. The main advantage of collecting particles with the same indices into 
classes is that it allows the user to write compact expressions for Lagrangians. 
This is illustrated in the example Lagrangian
\begin{equation}\label{eq:LQCD}
  \mathcal{L}=\bar q_f i\slashed{\partial} q_f + g_s \bar q_f\gamma^{\mu}T_a q_f 
    G_\mu^a\ ,
\end{equation}
where $q_f$ denotes the ``quark class'', $g_s$ the strong coupling constant, $T_a$
the fundamental representation matrices of $SU(3)$ and $G_\mu$ stands for the
gluon field. The notation of Eq.~\eqref{eq:LQCD} avoids having to write out
explicitly a Lagrangian term for each quark flavor.

Just like for the parameter classes, each particle class can be given a number of options which specify the properties of the field. In particular, there are two mandatory options that have to be defined for every field:
\begin{enumerate}
\item Each particle class must be given a name, specified by the \verb+ClassName+ option, which is at the same time the symbol by which the particle class is denoted in the Lagrangian.
\item The option \verb+SelfConjugate+ specifies whether the particle has an antiparticle or not.  The possible values for this option are \verb+True+ or \verb+False+.
\end{enumerate}
In addition, the particle class comes with various other options, as shown in 
Table~\ref{fig:Particle Class Options}. The set of all options can be divided into two groups:
\begin{enumerate}
\item[-] options which are directly used by \feynrules\ in the \mathematica\ session when computing the Feynman rules. Examples include the indices carried by a field, its mass, \etc;
\item[-] options which are mostly ignored by \feynrules, but describe information required by some Feynman diagram calculators. These options will be passed on to the Feynman diagram calculators via the corresponding \feynrules\ interfaces (see Section~\ref{sec:interfaces}).
\end{enumerate}
In the following we only describe the options directly used by \feynrules\ itself.
The interface specific options are discussed in Section~\ref{sec:interfaces}.

Besides the name of the particle class, the user may also provide names for the individual members of the class. For example, if {\tt uq} denotes the class of all up-type quarks with members {\tt \{u,c,t\}}, then this class and its members can be defined in the model file as
\begin{verbatim}
ClassName    -> uq,
ClassMembers -> {u, c, t}
\end{verbatim}
If a field is not self-conjugate, an instance of the particle class associated 
with the antiparticle is created automatically by appending `{\tt bar}' to the name of the particle. For example, after the model file has been loaded into \feynrules, the symbols {\tt uqbar}, as well as {\tt \{ubar, cbar, tbar\}}, are available and can be used to construct a Lagrangian. 
For bosonic fields, the field representing the antiparticle corresponds to the 
Hermitian conjugate of the field, while for fermionic fields, the symbol 
{\tt psibar} ({\tt psi} representing any fermionic field $\psi$) corresponds to 
the quantity $\bar{\psi} = \psi^\dagger\gamma^0$.

Fields transform in general as tensors under some symmetry groups, and
they usually carry indices indicating the transformation laws. Just like for
parameters, the indices carried by a field can be specified via the {\tt Indices}
option, \eg,
\begin{verbatim}
Indices -> {Index[ Colour ]}
Indices -> {Index[ Colour ], Index[ SU2D ]} 
\end{verbatim}
Indices labeling the transformation laws of the fields under the Poincar\'e 
symmetry (spin and/or Lorentz indices) are automatically inferred by \feynrules\ 
and do not need to be specified.
In addition to indices labeling how a field transforms under symmetries, each 
field may have additional indices such as flavor indices. One of these can be 
distinguished as the {\it flavor index} of the class and labels its members. It is
declared in the model file via the {\tt FlavorIndex} option. For example, the 
up-type quark class {\tt uq} previously introduced is usually defined carrying two
indices supplementing the spin index (automatically handled by \feynrules), one of
type {\tt Colour} ranging from 1 to 3 and specifying the color of the quark, and 
another index of type {\tt Flavour} ranging from 1 to 3. The latter is specified 
as the flavor index of the class (via the {\tt FlavourIndex} option) so that it 
labels the members of the class,
\begin{verbatim}
Indices -> { Index[ Colour ], Index[ Flavour ] },
FlavorIndex -> Flavour
\end{verbatim}

Quantum fields are not always only characterized by the tensor indices they carry,
but also by their charges under the discrete and / or abelian groups of the model.
\feynrules\ allows the user to define an arbitrary number of $U(1)$ charges 
carried by a field, as, \eg, in
\begin{verbatim}
QuantumNumbers -> {Q -> -1, LeptonNumber -> 1}
QuantumNumbers -> {Q -> 2/3}  
\end{verbatim}

Next, the user can specify the symbol and the numerical value for the masses and the decay widths of the different members of a particle class using the {\tt Mass} and {\tt Width} options\footnote{In the following we only discuss the masses of the particles. Widths however work in exactly the same way.}.
The argument of \verb+Mass+ is a list with masses for each of the class members, 
as in 
\begin{verbatim}
Mass -> {MW}
Mass -> {MU, MC, MT}
Mass -> {Mu, MU, MC, MT}
\end{verbatim}
where in the last example, the symbol \verb+Mu+ is given for the entire class, while the symbols \verb+MU+, \verb+MC+ and \verb+MT+ are given to the members.  
The symbol for the generic mass ({\tt Mu} in this case) is by default a tensorial 
parameter carrying a single index corresponding to the {\tt FlavorIndex} of the 
class. In addition, the {\tt AllowSummation} property is internally set to 
{\tt True}. The user can not only specify the symbols used for the masses but also their numerical value as in
\begin{verbatim}
Mass -> {MW, Internal} 
Mass -> {MZ, 91.188}
Mass -> {{MU,0}, {MC,0}, {MT, 174.3}}
Mass -> {Mu, {MU, 0}, {MC, 0}, {MT, 174.3}}
\end{verbatim}
If no numerical value is given, the default value 1 is assumed. 
In the first example, \verb+MW+ is given the value \verb+Internal+, which instructs \feynrules\ that the mass is defined as an internal parameter in \verb+M$Parameters+.  This is the only case in which a user needs to define a mass in the parameter list.  All other masses given in the definition of the particles should
not be included in this list.

While Lagrangians can often be written in a compact form in the gauge eigenbasis, the user usually wants to obtain Feynman rules in the mass eigenbasis. More generally, if the mass matrix appearing in the Lagrangian is not diagonal, one has to perform a field rotation that diagonalizes the mass matrix, 
\ie, the user has to replace certain fields by a linear combination of new fields such that in this new basis the mass matrix is diagonal. For fields in the gauge 
eigenbasis, the {\tt Mass} and {\tt Width} options do not have to be set, but
these options are replaced by the option
\begin{verbatim}
Unphysical -> True
\end{verbatim}
The diagonalization of the gauge eigenbasis can be performed automatically by 
means of the \asperge\ package which is interfaced to \feynrules. We refer to 
Section \ref{sec:asperge} and Section \ref{sec:aspergeinter} for more details. 
The values of these rotation matrices can also be expressed analytically (if available) or as numerical values,
relying on other tools to perform the diagonalization. The rotation to the 
mass eigenbasis in \feynrules\ are, in this case, implemented by using the 
{\tt Definitions} option of the 
particle classes, which consists of a set of replacement rules that are applied to
the Lagrangian before the computation of the Feynman rules. 
As an example, the relation between the hypercharge gauge boson ({\tt B}) and the
photon ({\tt A}) and $Z$-boson ({\tt Z}) could be included in a model description
as
\begin{verbatim}
Definitions -> {B[mu_] -> -sw Z[mu] + cw A[mu]}
\end{verbatim}
where the symbols {\tt sw} and {\tt cw} stand for the sine and cosine of the weak
mixing angle and are declared alongside the other model parameters. 
The replacement is purely symbolic and can be performed at the \mathematica\ level, even if the numerical values of the mixing parameters are unknown.

We conclude this section by giving some options specific to certain kinds of fields.
For ghost particles, there is an option \verb+Ghost+ which tells \feynrules\ the name of the gauge 
boson the ghost field is connected to. There is a similar option \verb+Goldstone+ 
in the case of scalar fields.

Majorana fermions are by definition eigenstates of the charge conjugation 
operator, and the associated eigenvalue must be a phase, since charge conjugation 
is unitary. If $\lambda$ denotes a Majorana fermion, then
\beq
\lambda^c = C\,\bar{\lambda}^T = e^{i\phi}\,\lambda\,.
\eeq
The information on the phase $\phi$ can be provided by the option 
{\tt MajoranaPhase} of the particle class, 
\begin{verbatim}
MajoranaPhase -> Phi
\end{verbatim}
where {\tt Phi} stands for the phase of the Majorana fermion $\lambda$. 
This phase can be further retrieved in the \mathematica\ session by typing
\begin{verbatim}
MajoranaPhase[lambda]
\end{verbatim}
where the symbol {\tt lambda} represents the Majorana fermion $\lambda$. 
The default value for the Majorana phase is 0.
 
Weyl fermion classes have an additional attribute that allows to specify their 
chirality. In the following we only discuss the case of spin-1/2 two-component fermions ({\tt W}), keeping in mind that the same attributes are also available for spin-3/2 two-component fermions ({\tt RW}). For example, the sets of rules
\begin{verbatim}
W[1] == {
   ClassName     -> chi,
   Chirality     -> Left,
   SelfConjugate -> False
   },
W[2] == {
   ClassName     -> xibar,
   Chirality     -> Right,
   SelfConjugate -> False
   }
\end{verbatim}
define two Weyl fermions $\chi$ ({\tt chi}) and $ \bar{\xi}$ ({\tt xibar}), the 
first one being left-handed and the second one right-handed. While Weyl fermions are in general not supported by any Feynman diagram calculator, it is often simpler to write down the Lagrangian in terms of two-component fermions and to transform them into four-component spinors in a second step. For this reason, it is possible to add the option {\tt WeylComponents} to an instance of the particle class 
{\tt F} associated with four-component Dirac and Majorana fermions. This option 
instructs \feynrules\ how to perform the mapping of the the two-component to the four-component spinors. For example, the Dirac fermion $\psi = (\chi, \bar{\xi})^T$ could be defined in \feynrules\ as
\begin{verbatim}
F[1] == {
   ClassName      -> psi,
   SelfConjugate  -> False,
   WeylComponents -> {chi, xibar}
   }
\end{verbatim}
where the Weyl fermions {\tt chi} and {\tt xibar} are defined above.
In the case of a four-component Majorana fermion only the left handed component 
needs to be specified. In Section~\ref{sec:running}, we will see how we can 
instruct \feynrules\ to transform a Lagrangian written in terms of two-component 
spinors to its counterpart expressed in terms of four-component fermions.

\begin{table}
\bgfb
\multicolumn{2}{c}{\textbf{Table~\ref{fig:Particle Class Options}: Particle Class Options}}\\
{\tt S}, {\tt F}, {\tt W}, {\tt V}, {\tt R}, {\tt RW}, {\tt T}, {\tt U} & Particle classes. \\
{\tt ClassName} & Mandatory. This option gives the symbol by which the class is represented.\\
{\tt SelfConjugate} & Mandatory. Takes the values {\tt True} or {\tt False}.\\
{\tt Indices} & The list of indices carried by the field. Note that Lorentz indices ({\tt Lorentz}, {\tt Spin}, {\tt Spin1}, {\tt Spin2}) are implicit and not included in this list.\\
{\tt FlavorIndex} & The name of the index making the link between the generic class symbol and the class members.\\
{\tt QuantumNumbers} & A replacement rule list, containing the $U(1)$ quantum numbers carried by the class.\\
{\tt ClassMembers} & A list with all the members of a class. If the class contains only one member, this is by default the {\tt ClassName}.\\
{\tt Mass} & A list with the masses for the class members. A mass can be entered 
  either as the symbol which is representing it, or by a two-component list, the 
  first element being the associated symbol and the second one its numerical 
  value. A generic symbol with a default numerical value equal to $1$ is generated
  if omitted.\\
{\tt Width} & A list with the decay rates for the class members. Similar to {\tt Mass}.\\
{\tt Ghost} & Option for ghost fields specifying the {\tt ClassName} of the gauge boson the ghost is associated to.\\
{\tt Goldstone} & For a scalar field, tagging it as the 
  Goldstone boson related to the gauge boson which this option is referring 
  to.\\
{\tt MajoranaPhase} & The Majorana phase of a Majorana fermion. The default is 0.\\
{\tt Chirality} & Option for Weyl fermions, specifying their chirality, 
  {\tt Left} or {\tt Right}. The default is {\tt Left}.
\egfb
\textcolor{white}{\caption{\label{fig:Particle Class Options}}}
\end{table}
\begin{table}
\bgfb
\multicolumn{2}{c}{\textbf{Particle Class Options (continued)}}\\
{\tt WeylComponents} & Option for the Dirac and Majorana field classes, mapping 
  them to their left- and right-handed Weyl components. For Majorana particles, 
  only the left-handed piece needs to be specified.\\
{\tt Definitions} & A list of replacement rules that should be applied by \feynrules\ before calculating the interaction vertices.\\
{\tt ParticleName} & A list of strings, corresponding to the particle names as they should appear in the output files for the Feynman diagram calculation programs. By default, the value is the same as the one of {\tt ClassMembers}. This name must satisfy the constraints of whatever Feynman diagram calculation package the user wishes to use it with. See Section \ref{sec:namerestr}.\\
{\tt AntiParticleName} & Similar to {\tt ParticleName}.\\
{\tt TeXParticleName} & A list of strings with the \TeX-form of each class member
  name. The default is the same as {\tt ParticleName}. See Section 
  \ref{sec:namerestr}.\\
{\tt TeXAntiParticleName} & Similar to {\tt TeXParticleName}.\\
{\tt Unphysical} & If {\tt True}, declares that the field does not have to be 
  included in the particle list written for another code by a \feynrules\ 
  interface but replaced instead by the value of the {\tt Definitions} attribute. 
  The default is {\tt False}.\\
{\tt PDG} & A list of the PDG codes of the particles. An automatically generated PDG code is assigned if this option is omitted. See Section \ref{sec:PDG}.\\
{\tt PropagatorLabel} & A list of strings propagators should be labeled with when drawing Feynman diagrams. The default value is the same as the {\tt ParticleName}. See Section~\ref{sec:propagators}.
\egfb
\textcolor{white}{\caption{\label{fig:Particle Class Options continued}}}
\end{table}
\begin{table}
\bgfb
\multicolumn{2}{c}{\textbf{Particle Class Options (continued)}}\\
{\tt PropagatorArrow} & Whether to put an arrow on the propagator ({\tt True}) or not ({\tt False}).  {\tt False} by default. See Section \ref{sec:propagators}.\\
{\tt PropagatorType} & This specifies how to draw the propagator line for this field. The default
  value is inferred from the class. The allowed choices are {\tt ScalarDash} (straight dashed line),
  {\tt Sine} (sinusoidal line), {\tt Straight} (straight solid line), {\tt GhostDash}
  (dashed line), and {\tt Curly} (curly \textit{gluonic} line).
  See Section~\ref{sec:propagators}.\\
{\tt FullName} & A string, specifying the full name of the particle, or a list containing the names for each class member. By default {\tt FullName} is the same as {\tt ParticleName}.\\
{\tt MixingPartners} & \feynarts\ option. See Ref.~\cite{Hahn:2000kx}.\\
{\tt InsertOnly} & \feynarts\ option. See Ref.~\cite{Hahn:2000kx}.\\
{\tt MatrixTraceFactor} & \feynarts\ option. See Ref.~\cite{Hahn:2000kx}.
\egfb
\textcolor{white}{\caption{\label{fig:Particle Class Options continued 2}}}
\end{table}


\subsection{\label{sec:superparts}Implementing superfields in \feynrules}
Supersymmetric theories can usually be written in a very compact form when using 
superfields, 
which are the natural objects to describe fields in supersymmetry. Feynman diagram calculators 
usually require a model to be implemented in terms of the component fields, \ie,
the standard fields of particle physics. 
\feynrules\ allows the user to implement the model in terms of superfields, and the code then takes care of deriving internally the component field Lagrangian~\cite{Duhr:2011se}.

Most of the phenomenologically relevant supersymmetric theories can be entirely
built in terms of chiral and vector superfields\footnote{Vector superfields must
be constructed in the Wess-Zumino gauge.}. Superfields are declared in \feynrules\ in a way
similar to ordinary fields,
\begin{verbatim}
 M$Superfields = { 
   superfield1[1] == { options1 }, 
   superfield2[2] == { options2 }, 
              ...  
 }
\end{verbatim}
The elements in the left hand side specify the type of superfield (similar to the way
ordinary particle classes are defined by their spin). Currently, \feynrules\ supports two
types of superfields:
\begin{itemize}
\item[-] {\tt CSF}: chiral superfields;
\item[-] {\tt VSF}: vector superfields in the Wess-Zumino gauge.
\end{itemize}
To illustrate the main features of the superfield declaration, 
we discuss the implementation of a left-handed chiral
superfield $\Phi$, a right-handed 
chiral superfield $\Xi$ and a non-abelian vector superfield $V_{\rm w.z.}$ in the 
Wess-Zumino gauge, 
\begin{verbatim}
 CSF[1] == {          
   ClassName -> PHI, 
   Chirality -> Left, 
   Weyl      -> psi,  
   Scalar    -> z,    
   Auxiliary -> FF    
 }                    
 CSF[2] == {           
   ClassName -> XI,  
   Chirality -> Right, 
   Weyl      -> psibar,
   Scalar    -> zbar   
 }                       
 VSF[1] == { 
   ClassName  -> VWZ,
   GaugeBoson -> V,
   Gaugino    -> lambda,
   Indices    -> {Index[SU2W]}
   }
\end{verbatim}
This declares two chiral superfields labeled by the
tags \texttt{CSF[1]} and \texttt{CSF[2]} and one
vector superfield labeled by the tag \texttt{VSF[1]}. For all three superfields,
the symbols
used when constructing a Lagrangian or performing computations in superspace
are specified through the attribute \texttt{ClassName}. 
Moreover, as for Weyl fermions, the chirality of the fermionic component of a chiral superfield can be assigned
by setting the attribute \texttt{Chirality} to \texttt{Left} or \texttt{Right}.

It is mandatory to match each superfield to its
component fields. The fermionic and scalar components of a chiral superfield are
hence referred to as the values of the attributes \texttt{Weyl} and
\texttt{Scalar}, respectively, 
whilst the vector and gaugino components of a vector superfield
are stored in the attributes 
{\tt GaugeBoson} and {\tt Gaugino}. Each of these options must point to the
symbol associated to the relevant fields. In contrast, the declaration of
the auxiliary ($F$- and $D$-) fields is optional. If not specified by the user, 
\feynrules\ takes care of it internally (as for the \texttt{XI} and
\texttt{VWZ} superfields above). If the user however desires to
match the auxiliary component of a superfield to an existing unphysical scalar 
field, he/she has to provide a value for the attribute \texttt{Auxiliary}
(as for the \texttt{PHI} superfield above). All the components (\texttt{psi},
\texttt{z}, \texttt{FF}, \texttt{V} and \texttt{lambda} in the examples above)
are assumed to be properly declared at the time of the field declaration
as described in Section~\ref{sec:parts}.

In addition, the attributes \texttt{Indices} and \texttt{QuantumNumbers} are
also available for the superfield class. We refer to Section \ref{sec:parts}
for more details. The complete list of the attributes of the superfield class
is indicated in Table \ref{tab:SF}.

\begin{table}
\bgfb
\multicolumn{2}{c}{\textbf{Table~\ref{tab:SF}: Superfield class options}}\\
 {\tt ClassName} & Refers to the symbol associated to a
    superfield.\\
 {\tt Chirality} & Refers to the chirality of a chiral
   superfield. It has to be set to the value {\tt Left} or {\tt Right}. \\
 {\tt GaugeBoson} & Refers to the symbol associated to the vector
   component of a vector superfield.\\
 {\tt Gaugino} & Refers to the symbol associated to the gaugino 
   component of a vector superfield.\\
 {\tt Weyl} & Refers to the symbol associated to the fermionic
   component of a chiral superfield.\\
 {\tt Scalar} & Refers to the symbol associated to the scalar
   component of a chiral superfield.\\
\multicolumn{2}{l}{\textbf{Optional attributes}}\\
 {\tt Auxiliary} & Refers to the symbol associated to the auxiliary
   component of a superfield. If not specified, \feynrules\ handles it
   internally. \\
 {\tt Indices} & Refers to the list of indices carried by a
   superfield. \\
 {\tt QuantumNumbers} & A list with the $U(1)$ quantum numbers
   carried by a superfield.\\
\egfb
\textcolor{white}{\caption{\label{tab:SF}}}
\end{table}


\subsection{Gauge Groups}
\label{sec:gaugegroups}

\subsubsection{Gauge group declaration}
The structure of the interactions of a model is in general dictated by gauge
symmetries. Fields are chosen to transform in specific representations of the gauge
group, and the invariance of the action under gauge transformation then governs the form of the interactions. In this section,
we focus on the declaration of the gauge groups in \feynrules. Doing so allows the user to 
use several functions dedicated to an efficient implementation of the
Lagrangian, such as, \eg, covariant derivatives or (super)field strength
tensors.

The gauge group of a model is either simple or semi-simple, and the different simple factors are defined 
independently in the list \texttt{M\$GaugeGroups},
\begin{verbatim}
  M$GaugeGroups = { 
    Group1 == { options1 },  
    Group2 == { options2 }, 
          ...  
  };
\end{verbatim}
Each element of this list consists of an equality defining
one specific direct
factor of the full symmetry group. The left-hand sides of these equalities
are labels associated to the different subgroups (\texttt{Group1},
\texttt{Group2}, \etc) while the right-hand sides contain
sets of replacement rules defining the properties of the subgroups
(\texttt{options1}, \texttt{options2}, \etc). For example, let us
consider the implementation of the SM gauge group in \feynrules,
\begin{verbatim}
  M$GaugeGroups = { 
    U1Y  == {                         
      Abelian          -> True,      
      CouplingConstant -> gp,        
      GaugeBoson       -> B,       
      Charge           -> Y          
    },
    SU2L == { 
      Abelian           -> False, 
      CouplingConstant  -> gw,
      GaugeBoson        -> Wi,
      StructureConstant -> ep,
      Representations   -> {{Ta,SU2D}}, 
      Definitions       -> {Ta[a__]->PauliSigma[a]/2, ep->Eps}
    },
    SU3C ==  {                   
      Abelian           -> False,
      CouplingConstant  -> gs,  
      GaugeBoson        -> G, 
      StructureConstant -> f,    
      Representations   -> {{T,Colour}},
      SymmetricTensor   -> dSUN  
    }
  }                       
\end{verbatim}
Abelian and non-abelian groups are distinguished by the option {\tt Abelian},
which may take the values {\tt True} or {\tt False}.

In the case of abelian groups, the user has the possibility to define the
associated $U(1)$
charge by setting the attribute \texttt{Charge} of the gauge group class
to a symbol related to this operator. This allows \feynrules\ to check for 
charge conservation at the
Lagrangian or at the Feynman rules level, assuming that the charges of the
different (super)fields have been properly declared via the option 
{\tt QuantumNumbers} of the particle class.

Non-abelian groups have specific attributes to define
the generators and structure constants. 
For example, the symbols of the symmetric and 
antisymmetric structure constants (whose indices are those of the adjoint representation) of the group are defined as the values of
the attributes \texttt{SymmetricTensor} and \texttt{StructureConstant}. 
Furthermore, the representations of the group necessary to define all the fields
of the model are listed
through the attribute \texttt{Representations}, which takes as a value a list of
pairs. The first component of each pair is a symbol defining  
the symbol for the generator, while the second one consists of the type of  
indices it acts on. For example, in the declaration of the $SU(3)$ gauge
group above, \feynrules\ internally creates 
a matrix of $SU(3)$ denoted by 
\texttt{T[Gluon,Colour,Colour]}, where we have explicitly indicated the index
types. The first index, represented here by the symbol \texttt{Gluon}, stands 
for the adjoint gauge
index. Even if this information is not explicitly provided at the time of the
gauge group declaration, \feynrules\ infers it from the value of the
\texttt{GaugeBoson} (or \texttt{Superfield}) attribute (see below).
Finally, the attribute \texttt{Definitions} allows the user to
provide analytical formulas expressing representation matrices and structure
constants in terms of the model parameters and/or standard \mathematica\
variables (see the $SU(2)_L$ declaration in the example above). As
long as the user consistently provides definitions for each set of
representation matrices that he/she wants to use, there is no limitation on
those that can be used within \feynrules. However, care must be taken when
exporting the model to a specific Monte Carlo generator as it in general has
specific restrictions on the supported gauge groups and representations.
Typically, these can be satisfied by fully expanding the vertices in terms of
the component fields of the new gauge group. In addition, the \feynrules\ package
comes with internal definitions for the most common $SU(3)$ representations for
which we refer to Section~\ref{sec:conventionsSMgauge} for more information.

At this stage we have to make a comment about how \feynrules\ deals with 
group representations. As should be apparent from the previous definition,
\feynrules\ does not make any distinction between a representation and its complex conjugate.
Indeed, if $T_R$ denote the generators of some irreducible representation
and $T_{\overline{R}}$ are the generator of the complex conjugate representation,
then it is always possible to eliminate all the generators $T_{\overline{R}}$ from the Lagrangian 
using the formula
\beq
\left(T_{\overline{R}}\right)_{ij} = -\left(T_R\right)_{ij}^\ast = -\left(T_R\right)_{ji}\,.
\eeq

Furthermore, if a field transforms in the representation $\overline{R}$, then the antifield
transforms in the representation $R$. We can thus always choose the particles 
as the ones transforming in the representation $R$. If the user wants to introduce an $SU(3)$ antitriplet, for example, \heshe{} can add a field \texttt{phi} transforming as a triplet. Consequently, \texttt{phibar} transforms as an antitriplet and can be used as such when writing the Lagrangian.

\begin{table}
\bgfb
\multicolumn{2}{c}{\textbf{Table~\ref{tab:GaugeGroupOptions}: Gauge group options}}\\
\\
  {\tt Abelian} & Mandatory, specifying whether the gauge group is
    abelian ({\tt True}) or not ({\tt False}).\\
  {\tt GaugeBoson} & Refers to the name of the gauge boson
    associated with the gauge group. This attribute must be specified, except if a
    superfield is related to the gauge group through the attribute
    \texttt{Superfield}. In addition, the gauge boson must be properly
    declared in the particle list (see Section~\ref{sec:parts}). \\
  {\tt Superfield} & Refers to the name of the vector superfield
    associated with the gauge group. This attribute must be specified, except if a
    vector boson is related to the gauge group through the attribute
    \texttt{GaugeBoson}. The superfield must be declared in the superfield list (see Section~\ref{sec:superparts}).\\
  {\tt CouplingConstant} & Refers to the parameter standing for the
    coupling constant associated with the gauge group.\\
  {\tt Charge} & Mandatory for abelian groups, referring to
    the symbol associated with the $U(1)$ charge connected with this gauge group.\\
  {\tt StructureConstant} & Mandatory for non-abelian groups,
    referring to the symbol associated with the structure constants of the gauge 
    group.\\
  {\tt SymmetricTensor} & Refers to the symbol associated with
    the totally symmetric tensor of a non-abelian group.\\
\egfb
\textcolor{white}{\caption{\label{tab:GaugeGroupOptions}}}
\end{table}
\begin{table}
\bgfb
\multicolumn{2}{c}{\textbf{Table~\ref{tab:GaugeGroupOptions}: Gauge group
options (continued)}}\\
\\
  {\tt Representations} & Refers to a list of two-component lists
    containing all the representations defined for this gauge group. The first
    component of these lists consists of the
    symbol by which the generators of the representation are denoted, while the
    second component is the name of the index it acts on.\\
  {\tt Definitions} & Contains a  list of replacement rules that
    should be applied by \feynrules\ before calculating vertices, expressing
    representation matrices and/or structure constants in terms of the model
    parameters and \mathematica\ standard objects.\\
\egfb
\textcolor{white}{\caption{}}
\end{table}

Information on the gauge boson responsible for the mediation of the gauge
interactions is provided through the
\texttt{GaugeBoson} attribute of the gauge group class. It refers to the symbol 
of the vector field associated to the group, defined separately in 
\verb+M$ClassesDescriptions+. For non-abelian symmetries, the
(non-Lorentz) index carried by this field consists of the adjoint index. As
briefly mentioned above, this is the way used by \feynrules\ to define adjoint
gauge indices. For supersymmetric models, it is also possible to link
a vector superfield instead of a gauge boson through the attribute
\texttt{Superfield}. If both the
\texttt{Superfield} and \texttt{GaugeBoson} attributes are specified by the
user, they must be consistent, \ie, the gauge boson must be the vector component 
of the vector superfield.

The last attribute appearing in the examples above consists of the model parameter
to be used as the gauge coupling constant, which is specified through the option
\texttt{CouplingConstant}.

The list of all the options described above is summarized in Table
\ref{tab:GaugeGroupOptions}.

\subsubsection{\feynrules\ functions related to gauge groups}
The declaration of a gauge group enables \feynrules\ to
automatically construct field strength tensors, superfield strength tensors and 
covariant derivatives associated with this group, so that they can be further 
used when building Lagrangians. In the case of abelian gauge groups, the
field strength tensor is invoked by issuing 
\begin{verbatim}
FS[A, mu, nu]
\end{verbatim}
where \verb+A+ is the corresponding gauge boson and \verb+mu+ and \verb+nu+
denote Lorentz indices. Its supersymmetric counterparts can be called by the command  
\begin{verbatim}
SuperfieldStrengthL[ V, sp    ]
SuperfieldStrengthR[ V, spdot ]
\end{verbatim}
respectively. In these commands, {\tt V} stands for the vector superfield associated
with the gauge group and
\texttt{sp} and \texttt{spdot} are left-handed and right-handed spin indices.
These three functions can be easily generalized to the non-abelian case, 
\begin{verbatim}
FS[ A, mu, nu, a ]
SuperfieldStrengthL[ V, sp   , a ]
SuperfieldStrengthR[ V, spdot, a ]
\end{verbatim}
where \texttt{a} stands for an adjoint gauge index. Following the \feynrules\
conventions, these quantities are defined as
\be\bsp\label{eq:field_strength}
  F_{\mu\nu}^a =&\ \del_\mu A_\nu^a - \del_\nu A_\mu^a + g f^a{}_{bc} 
    A_\mu^b A_\nu^c \ ,\\
  W_\alpha =&\ -\frac14 \Dbar\!\cdot\!\Dbar e^{2gV} D_\alpha e^{-2gV}
   \ ,\\
  \Wbar_\alphadot =&\ -\frac14 D\!\cdot\!D e^{-2gV} \Dbar_\alphadot e^{2gV} \ ,
\esp\ee
where $g$ and $f$ denote the coupling constant and the structure constants of
the gauge group and $D$ and $\Dbar$ are the superderivatives defined below, 
in Section
\ref{sec:manipsusy}. The abelian limit is trivially derived from these
expressions. We emphasize that the
spinorial superfields $W_\alpha$ and $\Wbar_\alphadot$ are not hard-coded in \feynrules\ and are recalculated
each time. However, they are evaluated  only when an 
expansion in terms of the component fields of the vector superfield \texttt{V} 
is performed.

\begin{table}
\bgfb
\multicolumn{2}{c}{\textbf{Table~\ref{tab:FieldStrengthTensors}: Field strength tensor and covariant derivative}}\\
\\
  {\tt FS[A, mu, nu]} & Constructs the field strength tensor
    $F_{\mu\nu}$ connected with the $U(1)$ gauge boson {\tt A}.\\
  {\tt FS[A, mu, nu, a]} & Constructs the field strength tensor
    $F_{\mu\nu}^a$ connected with the non abelian gauge boson {\tt A}.\\
\multicolumn{2}{l}{{\tt SuperfieldStrengthL[V, sp]}}\\
& Calculates the left-handed superfield strength tensor associated
with the abelian vector superfield {\tt V}. \\
\multicolumn{2}{l}{{\tt SuperfieldStrengthL[V, sp, a]}}\\
& Calculates the left-handed superfield strength tensor associated
with the non-abelian vector superfield {\tt V}. \\
\multicolumn{2}{l}{{\tt SuperfieldStrengthR[V, spdot]}}\\
& Calculates the right-handed superfield strength tensor associated
with the abelian vector superfield {\tt V}. \\
\multicolumn{2}{l}{{\tt SuperfieldStrengthR[V, spdot, a]}}\\
& Calculates the right-handed superfield strength tensor associated
with the non-abelian vector superfield {\tt V}. \\
{\tt DC[phi, mu]} & The covariant derivative acting on the field
\texttt{phi} with a Lorentz index \texttt{mu}.
\egfb
\textcolor{white}{\caption{\label{tab:FieldStrengthTensors}}}
\end{table}

From the information provided at the time of the declaration of the gauge group,
\feynrules\ can also define, in an automated way, 
gauge covariant derivatives. These can be
accessed through the symbol \verb+DC[phi, mu]+, where \verb+phi+ is the field
that it acts on and \verb+mu+ the Lorentz index. In our conventions, the
covariant derivative reads 
\be\label{eq:covder}
  D_\mu\phi = \del_\mu\phi - i g A_\mu^a T_a \phi
\ee
where $T_a$ stands for the representation matrices associated to the
representation of the gauge group in which the field $\phi$ lies. The sign
convention in Eq.~\eqref{eq:covder} is consistent with the sign convention
in Eq.~\eqref{eq:field_strength}.

All the functions presented in this section are summarized
in Table \ref{tab:FieldStrengthTensors}.



 \subsection{Model restrictions}
\label{sec:restrictions}
In phenomenological studies, it can sometimes be useful to consider restricted models which are obtained from a parent model by putting some of the external parameters to zero. As an example, one might be interested in the Standard Model 
with a diagonal CKM matrix. While it is of course always possible to make the CKM matrix numerically diagonal, it is desirable to remove the interaction terms proportional to the off-diagonal terms altogether in order to avoid a proliferation of vanishing diagrams in Feynman diagram calculations.
This can be achieved by the use of the so-called restriction files in \feynrules. Restriction files are text files (with the extension \texttt{.rst}) that contain a list, named \verb+M$Restrictions+, of replacement rules to be applied to the Lagrangian before the evaluation of the Feynman rules. As an example, a possible restriction file to restrict the CKM matrix to a diagonal matrix reads
\begin{verbatim}
M$Restrictions = {
            CKM[i_,i_] -> 1,
            CKM[i_?NumericQ, j_?NumericQ] :> 0 /; (i =!= j)
}
\end{verbatim}
Applying these rules to the Lagrangian (or to the vertices) obviously removes all the flavor-changing interactions among quark fields. 

Restriction files can be loaded at run time after the model file has been 
loaded, \ie, after the {\tt LoadModel[ ]} command has been issued. The corresponding command is
\begin{verbatim}
LoadRestriction[ file1.rst, file2.rst, ... ]
\end{verbatim}
After this function has been called, the parameter definitions inside \feynrules\ are updated and the parameters appearing in the left-hand side of the replacement rules are removed from the model. In addition, the rules are kept in memory and are applied automatically to the Lagrangian when computing the Feynman rules. Note that this process is irreversible, and the restrictions cannot be undone after the {\tt LoadRestriction[]} command has been issued (unless the kernel is restarted).

For complicated models, one often chooses the benchmark point in a special way such that many of the new parameters in the model are zero, and only a few new interactions are considered. \feynrules\ provides a command that allows to identify the parameters that are numerically zero and to create a restriction file that removes all the vanishing parameters from the model. More precisely, the command 
\begin{verbatim}
WriteRestrictionFile[]
\end{verbatim}
creates a restriction file called {\tt ZeroValues.rst} which can be loaded just like any other restriction file.
We emphasize that for complicated model the use of this restriction file may considerably speed up calculation performed with Feynman diagram calculators.

\subsection{Mixing declaration}\label{sec:mixdecl}
\feynrules\ comes with a module allowing for the derivation of the mass spectrum included in any Lagrangian (see
Section \ref{sec:asperge}). The mass matrices
calculated in this way can be further passed to the \asperge\ program~\cite{Alloul:2013fw} for a numerical evaluation of the necessary rotations
yielding a diagonal mass basis (see Section \ref{sec:aspergeinter}). In
principle, kinetic mixing among gauge bosons related to different $U(1)$ factors
of the gauge group is also possible. This is not supported by the new module,
so that such a mixing has to be implemented manually by the user. We refer to
Ref.~\cite{Christensen:2009jx} for a detailed example.
In order to employ the functionalities related to the \asperge\ package, the user has
to implement particle mixings in a specific way, using instances of the mixing class collected into a list
dubbed \texttt{M\$MixingsDescription},
\begin{verbatim}
  M$MixingsDescription = {
     Mix["l1"] == { options1 },
     Mix["l2"] == { options2 },
            ...
  }
\end{verbatim}
This maps a label given as a string (\texttt{"l1"}, \texttt{"l2"}, \etc) to a set of \mathematica\ replacement
rules defining mixing properties (\texttt{options1}, \texttt{options2},
\etc). To illustrate the available options,  we take the example of the $W_1$ and $W_2$ gauge fields of $SU(2)_L$ that mix,
in the Standard Model, to the $W^\pm$-bosons
\be
  W_\mu^+ = \frac{W_\mu^1 - i W^2_\mu}{\sqrt{2}}   \quad\text{and}\quad
  W_\mu^- = \frac{W_\mu^1 + i W^2_\mu}{\sqrt{2}} \ .
\ee
As this mixing relation does not depend on any model parameter and is fully numerical, it can be declared in a very compact
form, 
\begin{verbatim}
  Mix["l1"] == {
    MassBasis  -> {W, Wbar}, 
    GaugeBasis -> {Wi[1], Wi[2]},
    Value      -> {{1/Sqrt[2],-I/Sqrt[2]},{1/Sqrt[2],I/Sqrt[2]}}
  }
\end{verbatim}
The previous mixing declaration should be understood as the matrix product
\begin{verbatim}
  MassBasis = Value . GaugeBasis
\end{verbatim}
Information on
the gauge and mass bases is provided through the attributes \texttt{GaugeBasis} and \texttt{MassBasis}
and the mixing matrix is given in a numerical form as the argument of the attribute \texttt{Value}.
In the case the user already knows analytical formulas for the element of the mixing matrix,
the syntax above cannot be employed. This matrix
has instead to be declared as a standard model parameter (see Section~\ref{sec:parameters}) and referred
to via the \texttt{MixingMatrix} attribute of the mixing class (see below).
As the gauge basis is unphysical, it has to  only contain fields declared as such, while the mass basis can in contrast
contain either physical fields, unphysical
fields or both. When unphysical fields are present, particle mixings are effectively implemented in
several steps. We refer to Ref.~\cite{Alloul:2013fw} for examples.
In order to implement the bases in a more compact form,
spin and Lorentz indices can be omitted. Moreover, if some indices
are irrelevant, \ie, if they are identical
for all the involved fields, underscores can be employed, as in the example 
\begin{verbatim}
  Mix["l2"] == { 
    MassBasis  -> {sdL[1,_], sdL[2,_], sdL[3,_]}, 
    GaugeBasis -> {QLs[2,1,_], QLs[2,2,_], QLs[2,3,_]}, 
                ... 
  }
\end{verbatim}
Underscores reflect that the last index of each field has the same type and is not affected by the mixing (such as
color indices for instance).

Most of the time, the mixing matrix is not known numerically before implementing the model, which leads to a slightly
different syntax at the level of the implementation. The declaration
\begin{verbatim}
  Mix["l3"] == {
    MassBasis    -> {A, Z}, 
    GaugeBasis   -> {B, Wi[3]}, 
    MixingMatrix -> UW, 
    BlockName    -> WEAKMIX
  } 
\end{verbatim}
describes the rotation of the third weak boson $W_3$ and the hypercharge
gauge boson $B$ to the photon and $Z$-boson states, 
\be
  \bpm  A_\mu \\ Z_\mu \epm = U_w \bpm B_\mu \\ W_\mu^3 \epm \ ,
\ee
where we have introduced the mixing matrix $U_w$, the associated symbol being \texttt{UW}. The latter is
referred to by means of the attribute \texttt{MixingMatrix} of the mixing class.
The matrix \texttt{UW} does not have to be declared separately as
this task is internally handled by \feynrules.
Note that it is assumed that
{\tt UW} is a complex matrix with two indices, and 
according to our conventions we declare separately its real and imaginary parts.
In other words, \feynrules\ creates two real external tensorial parameters, 
the real and imaginary parts of the matrix, and 
one internal complex tensorial parameter
for the matrix {\tt UW} itself.
The user must specify the name of a Les Houches block  \cite{Skands:2003cj,Allanach:2008qq}
which will contain the numerical values associated with the elements of the matrix (see Section \ref{sec:paramin}).
In the example above,
we impose the real part of the elements of $U_w$ to be stored in
a block named \texttt{WEAKMIX} and their imaginary part in 
a block \texttt{IMWEAKMIX}.

It must be noted that mixing matrices declared through a mixing declaration 
cannot be employed explicitly in Lagrangians (such as for the CKM matrix in the Minimal Supersymmetric
Standard Model). If the user wants to use the mixing matrices explicitly in the Lagrangian,  \heshe{} must declare the matrices following the standard syntax
(see Section \ref{sec:parameters}), numerical values being  provided from the beginning.

Finally, it is sometimes more practical to implement a mixing relation under the form
\begin{verbatim}
  GaugeBasis = MixingMatrix . MassBasis
\end{verbatim}
as for the CKM rotation in the Standard Model, 
\be
  d_L^0 = V_{\rm CKM} \cdot d_L \ ,
\label{eq:dia2}\ee
where $d_L^0$ and $d_L$ are respectively gauge- and mass-eigenstates.
In order to avoid a cumbersome use of inverse matrices in mixing declarations, the user can make use of the
optional attribute \texttt{Inverse} of the mixing class, setting it to \texttt{True}.

All the attributes of the mixing class are summarized in Table \ref{tab:mix} while more involved cases are now detailed below.

\begin{table}
\bgfb
\multicolumn{2}{c}{\textbf{Table~\ref{tab:mix}: Attributes of the
 mixing class.}}\\
 {\tt MassBasis} & Specifies the field basis in which the mass matrix is diagonal. Takes a list of bases as argument
    when multiple bases are involved.\\
 {\tt GaugeBasis} & Specifies the field basis that must be diagonalized. Takes a list of bases as
    argument when several bases are involved.\\
 {\tt Value} & Refers to the numerical value of a mixing matrix, when it is known at the time of the model implementation.
    Relates several mixing matrices to their values (given as a list) when mass diagonalization requires several rotations.\\
 {\tt MixingMatrix} & Refers to the symbol associated with a mixing matrix.
    Relates several mixing matrices to their symbols (given as a list) when mass diagonalization requires several rotations.\\
 {\tt BlockName} & Provides information about the name of the Les Houches 
   block containing the mixing matrices (see Section \ref{sec:paramin}). Relates several mixing matrices to their Les
   Houches blocks (given as a list) when mass diagonalization requires several rotations.\\
 {\tt Inverse} & When set to {\tt True}, allows to use an inverse matrix for diagonalizing a basis.
    Takes a list of Boolean numbers as argument when mass diagonalization requires several rotations.\\
\egfb
\textcolor{white}{\caption{\label{tab:mix}}}
\end{table}

When complex scalar fields are split
into their real and imaginary degrees of freedom, it is necessary to declare one scalar
and one pseudoscalar mass basis.  In this case, the \texttt{MassBasis} attribute is given a list of the
two bases.  The first basis gives a list of the real scalar fields in the mass basis, while the second gives a list of the pseudoscalar fields.
Consistently, the arguments of the
attributes \texttt{Value}, \texttt{BlockName}, \texttt{MixingMatrix} and
\texttt{Inverse} are now lists as well, the first
element of these lists always referring to scalar fields and the second one
to pseudoscalar fields. When some of the elements
of these lists are irrelevant, as for instance when the scalar mixing matrix is
unknown (\texttt{MixingMatrix} and \texttt{BlockName} are used) and the
pseudoscalar mixing matrix is known (\texttt{Value} is used),
underscores have to be used to indicate the unnecessary entries of the lists, such as in
\begin{verbatim}
  Mix["scalar"] == { 
    MassBasis    -> { {h1, h2}, {a1, a2} }, 
    GaugeBasis   -> { phi1, phi2 }, 
    BlockName    -> { SMIX, _ }, 
    MixingMatrix -> { US, _ },
    Value        -> { _, ... }
  } 
\end{verbatim}

In the case of four-component fermion mixing, two choices are possible.
We recall that the mass terms (for Dirac-type fermions) can be generically written as
\be
  m \bar\psi \psi = m (\bar\psi_R \psi_L + \bar\psi_L \psi_R)\ ,
\ee
all indices being suppressed for clarity.
As a first option, the user can use a single gauge basis and \feynrules\
internally takes care of the chirality projectors appearing in the mass
terms. In this case, the attribute \texttt{GaugeBasis} contains a list with the $\psi$
fields above. As a second option,
two bases with different particle classes standing for the left-handed and
right-handed pieces of the fields can be employed. The user here sets the value of the
\texttt{GaugeBasis}
attribute to a two-component list. The first (second) element of this list is another list
containing the names of the left-handed (right-handed) components of the fields, \ie, the instance
of the particle class describing the $\psi_L$ ($\psi_R$) fields above.
Additionally,
the arguments of the \texttt{Value}, \texttt{BlockName}, \texttt{MixingMatrix} and
\texttt{Inverse} attributes are consistently upgraded to lists too, the first component being
associated with left-handed fermions and the second one with
right-handed fermions. As for neutral scalar mixing, underscores can be used for
irrelevant list elements. This second option is more commonly employed in the existing models
although it may seem more complicated. We therefore give an explicit example:
\begin{verbatim}
  Mix["dq"] == {
    MassBasis    -> {dq[1, _], dq[2, _], dq[3, _]},
    GaugeBasis   -> { 
                      {QL[2, 1, _], QL[2, 2, _], QL[2, 3, _]},
                      {dR[1, _], dR[2, _], dR[3, _]}
                    },
    MixingMatrix -> {CKM, _},
    Value        -> {_, {{1,0,0}, {0,1,0}, {0,0,1}} },
    Inverse      -> {True, _}
  }
\end{verbatim}
In this example, we describe the implementation of the mixing relations
associated with the down-type quarks in the Standard Model. The mass eigenstates
are denoted by the three fields \texttt{dq} standing
for the three generations of down-type quarks.
The color indices being irrelevant, they are replaced by underscores everywhere.
The left-handed  gauge eigenstates are the
second piece of the $SU(2)_L$ doublets of quarks (the \texttt{QL} fields) whereas
the right-handed gauge eigenstates are the $SU(2)_R$ singlets
(the \texttt{dR} quarks). Finally, the last three arguments of the instance
of the mixing class above indicate that the left-handed fields mix trough Eq.~\eqref{eq:dia2}
(the first elements of the attributes \texttt{MixingMatrix} and \texttt{Inverse} are set
accordingly)
and that the right-handed fields do not have to be rotated (the argument \texttt{Value} contains
an identity matrix as its second element).

Lagrangian mass terms for charged Weyl fermions are generically
written as
\be
   \Big(\psi_1^-, \ldots, \psi_n^-\Big)\ M\ \bpm \chi_1^+\\ \vdots \\ \chi_n^+
    \epm \ ,
\ee 
where $M$ stands for the mass matrix and $\psi_i^-$ and $\chi_i^+$ are Weyl
fermions with opposite electric charges.
The diagonalization of the matrix $M$ proceeds through two unitary
rotations $U$ and $V$,
\be
   \bpm \tilde\psi^-_1\\ \vdots \\ \tilde\psi^-_n \epm  = U\ 
    \bpm \psi^-_1\\ \vdots \\ \psi^-_n\epm \quad\text{and}\quad
   \bpm \tilde\chi^+_1\\ \vdots \\ \tilde\chi^+_n \epm  = V\ 
    \bpm \chi^+_1\\ \vdots \\ \chi^+_n\epm\ , 
\ee
which introduces two mass bases. Therefore, all the attributes
\texttt{MassBasis}, \texttt{GaugeBasis}, \texttt{Value},
\texttt{Mi\-xing\-Ma\-trix}, and \texttt{Block\-Na\-me} now take lists as
arguments (with underscores included where relevant). It is mandatory to consistently order these lists.

Finally, in realistic models, the ground state of the theory is non-trivial
and fields must be shifted by their vacuum expectation value (vev). The case of
electrically  neutral scalar fields can be treated in a specific way and included easily in particle
mixing handling. Information on the vevs is included in a list of two-component elements denoted by \texttt{M\$vevs}.
The first of these elements refers to an unphysical field while the second one to the symbol associated with its vev,
as in
\begin{verbatim}
  M$vevs = { { phi1, vev1 }, { phi2, vev2 } }
\end{verbatim}
the vacuum expectation values \texttt{vev1} and \texttt{vev2} being declared
as any other model parameter (see Section \ref{sec:parameters}). In this way, the information on the
vevs of the different fields is consistently taken into account when \feynrules\ internally rewrites
the neutral scalar fields in terms of their real degrees of freedom. For the sake of the example,
we associate with the vev declaration above the following definition
of the mixing of the $\phi_1$ and $\phi_2$ gauge eigenstates to the real scalar (pseudoscalar)
mass eigenstates $h_1$ and $h_2$ ($a_1$ and $a_2$),
\begin{verbatim}
  Mix["phi"] == {
    MassBasis    -> { {h1, h2}, {a1, a2} },
    GaugeBasis   -> { phi1, phi2 },
    MixingMatrix -> { US, UP }
  }
\end{verbatim}
This teaches \feynrules\ to internally understand the mixing pattern of the $\phi_i$ fields as
\be
  \bpm \phi_1\\ \phi_2\epm  = \frac{1}{\sqrt{2}} \Bigg[ \bpm v_1 \\ v_2\epm + U_s^\dag \bpm h_1\\ h_2\epm
  + i U_p^\dag \bpm a_1\\ a_2\epm \Bigg] \ ,
\ee
the scalar and pseudoscalar mixing matrices $U_s$ and $U_p$ being represented by the symbols \texttt{US}
and \texttt{UP}, respectively.

\section{The Lagrangian}\label{sec:lag}

An essential ingredient of a model implementation is the Lagrangian of the model. The Lagrangian can be built out of the fields of the models,  augmented by some internal \feynrules\ and \mathematica\ functions. All the fields can be accessed by their {\tt ClassName}, for example \verb+psi[a,b,...]+ where \verb+a,b,...+ are the indices of the field starting with its Lorentz indices for vector and tensor bosons, or its spin index for fermions followed by a Lorentz index for a spin-$3/2$ fermion\footnote{Spin-3/2 fields always have their spin index before their 
Lorentz index.}.
The remaining indices are the ones defined in the particle definition through the {\tt Indices} option, given in the same order. The different flavors can also be accessed using the names given in {\tt ClassMembers}. They have the same indices as the full flavor multiplet, with the flavor index omitted. We recall that, if a field is not self-conjugate, \feynrules\ automatically creates the symbol for the 
conjugate field by adding `\verb+bar+' at the end of the particle name, \ie, the antiparticle associated to {\tt psi} is denoted by {\tt psibar}. For a fermion $\psi$, the conjugate field is  $\bar\psi\equiv \psi^\dagger \gamma^0$. Alternatively, the conjugate field can be obtained by issuing \verb+anti[psi]+.

Fields (and their derivatives) can be combined into polynomials. By convention, all the indices appearing inside a monomial in \feynrules\ must be contracted, \ie, all indices must appear pairwise\footnote{With the exception of single-index parameters for which the {\tt AllowSummation} option is set to {\tt True} (see Section~\ref{sec:parameters}).}. Furthermore, all indices must be spelled out explicitly. For anticommuting fields (fermions and ghosts), the \mathematica\ {\tt Dot} function has 
to be used, in order to keep the relative order among them fixed.
For example, the interaction between the gluon and all the down quarks can be written as
 \begin{verbatim}
gs Ga[mu, s, r] T[a, i, j] dqbar[s, f, i].dq[r, f, j] G[mu, a]
\end{verbatim}
There is however one case where indices do not need to be spelled out completely 
but can be omitted. 
If in a fermion bilinear, all the indices of the rightmost fermion are connected to all the indices of the leftmost fermion (perhaps with intermediate matrices), then these indices can be suppressed and \feynrules\ takes care of restoring them internally, such as in

{\tt dqbar.Ga[mu].T[a].dq} \\
$\phantom{aaaaaaaaaaaa}\rightarrow$ {\tt Ga[mu,s,r] T[a,i,j] dqbar[s,f,i].dq[r,f,j]}\,.

In case of doubt, the user should always spell out all indices explicitly.

The  \verb+Dot+  product is mandatory for anticommuting fields or parameters. It should be noted that \mathematica\ does not keep the \verb+Dot+ product between the components of vectors or matrices after computing their product explicitly
\begin{verbatim}
{ubar, dbar}.{u, d}  =  u ubar + d dbar
\end{verbatim}
The appropriate treatment requires, therefore, use of the {\tt Inner} function for each {\tt Dot}, {\textit e.g.}
\begin{verbatim}
Inner[Dot, {ubar, dbar}, {u, d}]  =  ubar.u + dbar.d
\end{verbatim}
or for more than one multiplication,
\begin{verbatim}
Inner[Dot, Inner[Dot, {ubar, dbar} , 
    {{a11, a12} , {a21, a22}}] , {u, d}].
\end{verbatim}

As already discussed in Section~\ref{sec:gaugegroups}, gauge invariant derivatives
can be conveniently defined via the functions  \verb+DC[phi,mu]+ and 
\verb+FS[G, mu, nu, a]+. The first argument of both functions  is the relevant
field, \verb+mu+ and \verb+nu+ are Lorentz indices and \verb+a+ represents an
index of the adjoint representation of the associated gauge group. The gauge 
fields and generators that appear in covariant derivatives of a particular field 
are fixed by its indices and by the definition of the gauge group. 
For example, the QCD Lagrangian for massless down quarks,
\begin{equation}
\mathcal{L^{QCD}}\equiv -\frac{1}{4}G^{\mu\nu}_a G_{\mu\nu}^a + i \bar{d}
  \slashed{D}d, \label{eq:QCD Lagrangian}
\end{equation}
is written as
\begin{verbatim}
L = -1/4 FS[G, mu, nu, a] FS[G, mu, nu, a]  
           +  I dqbar.Ga[mu].DC[dq, mu]  
\end{verbatim}  
All the predefined \feynrules\ functions useful for the building of the Lagrangian are given in Table~\ref{fig:Special Symbols for the Lagrangian}.

Finally, it is often convenient to write a Lagrangian in terms of two-component fermions and to let \feynrules\ perform the transformations to four-component fermions. We note that this operation is mandatory for most Feynman diagram calculators, which in general only work with four-component spinors. More precisely, if $\chi$ and $\bar{\xi}$ are left and right-handed Weyl spinors, and $\psi=(\chi,\bar{\xi})^T$ is a Dirac fermion, we can easily switch to four-component fermions by using the replacements
\beq\bsp
\chi\rightarrow\frac{1-\gamma^5}{2}\psi\,,&
\qquad
\xi\rightarrow\frac{1-\gamma^5}{2}\psi^c\,,\\
\bar\chi\rightarrow\frac{1+\gamma^5}{2}\psi^c\,,&
\qquad
\bar\xi\rightarrow\frac{1+\gamma^5}{2}\psi\,.
\esp\eeq
These transformation rules are implemented in \feynrules\ via the
{\tt Weyl\-To\-Di\-rac} function, which takes as an argument a Lagrangian written in terms of two-component fermions, and returns the same Lagrangian in terms of four-component fermions.

\begin{table}
\bgfb
\multicolumn{2}{c}{\textbf{Table~\ref{fig:Special Symbols for the Lagrangian}: Special symbols for Lagrangians}}\\
{\tt del[}$\phi$, $\mu${\tt ]} & Partial derivative of $\phi$ with respect to the space-time coordinate $x^\mu$.\\
{\tt DC[}$\phi$, $\mu${\tt ]} & Gauge covariant derivative of $\phi$ with respect to the space-time coordinate $x^\mu$.\\
{\tt ME[}$\mu$, $\nu${\tt ]} & Minkowski metric tensor $\eta_{\mu\nu}$.\\
{\tt IndexDelta[}$i$,$j${\tt ]} & Kronecker delta $\delta_{ij}$.\\
{\tt Eps[}a,\ldots, b{\tt ]} & Totally antisymmetric Levi-Civita tensor with respect to the indices {\tt a,...,b}. In the case of four indices, the conventions on the $\epsilon$-tensors
are important when employing interfaces to matrix element generators. They are based on
$\epsilon_{0123} = +1$.\\
{\tt HC[}expr {\tt ]} & Hermitian conjugate of expr.\\
{\tt CC[}$\psi${\tt]} & The charge conjugate $\psi^c$ of a Dirac field $\psi$.\\
{\tt FS} & Field strength tensor. \\
{\tt Ga[}$\mu${\tt ]}, {\tt Ga[}$\mu$, $i$, $j${\tt ]} & Dirac Matrix $\gamma^\mu$, $\gamma^\mu_{ij}$.\\
{\tt ProjP}, {\tt ProjP[}$i$, $j${\tt ]} & Chiral projection operator $\frac{1+\gamma^5}{2}$, $\left(\frac{1+\gamma^5}{2}\right)_{ij}$.\\
{\tt ProjM}, {\tt ProjM[}$i$, $j${\tt ]} & Chiral projection operator $\frac{1-\gamma^5}{2}$, $\left(\frac{1-\gamma^5}{2}\right)_{ij}$.\\
{\tt ProjP[}$\mu${\tt ]}, {\tt ProjP[}$\mu$, $i$, $j${\tt ]} & $\gamma^\mu\frac{1+\gamma^5}{2}$, $\left(\gamma^\mu\frac{1+\gamma^5}{2}\right)_{ij}$.\\
{\tt ProjM[}$\mu${\tt ]}, {\tt ProjM[}$\mu$, $i$, $j${\tt ]} & $\gamma^\mu\frac{1-\gamma^5}{2}$, $\left(\gamma^\mu\frac{1-\gamma^5}{2}\right)_{ij}$.\\
{\tt right[}$\psi${\tt ]} & The right-handed part of the four-component fermion field $\psi$, $\frac{1+\gamma^5}{2}\psi$.\\
{\tt left[}$\psi${\tt ]} & The left-handed part of the four-component fermion field $\psi$, $\frac{1-\gamma^5}{2}\psi$.\\
{\tt si[}$\mu${\tt ]}, {\tt si[}$\mu$, $\alpha$, $\dot\alpha${\tt ]} & Pauli matrix $\sigma^\mu$, $\sigma^\mu{}_{\alpha\dot\alpha}$.\\
{\tt sibar[}$\mu${\tt ]}, {\tt sibar[}$\mu$, $\dot\alpha$, $\alpha${\tt ]} & Pauli matrix $\bar\sigma^\mu$, $\bar\sigma^{\mu\dot\alpha\alpha}$.
\egfb
\textcolor{white}{\caption{\label{fig:Special Symbols for the Lagrangian}}}
\end{table}

\subsection{Tools for Lagrangians}

\feynrules\ provides functions, collected in Table 
\ref{fig:Manipulating a Lagrangian}, that can be used while constructing 
Lagrangians. For example, the function \verb+ExpandIndices[]+ returns the Lagrangian with all the indices
written explicitly. 
Each of the other functions return a different part of the Lagrangian as described in the table.
\begin{table}
\bgfbx
\multicolumn{2}{c}{\textbf{Table~\ref{fig:Manipulating a Lagrangian}: Manipulating a Lagrangian}}\\
\multicolumn{2}{l}{All the functions below share the same options as
{\tt FeynmanRules},}\\
\multicolumn{2}{l}{which can be found in Table~\ref{fig:FR and V Options}.}\\
{\tt ExpandIndices[}$\mathcal{L}$, options\tt{]} & Restores all the suppressed indices in the Lagrangian $\mathcal{L}$.\\
{\tt GetKineticTerms[}$\mathcal{L}$, options{\tt]} & Returns the kinetic terms in the Lagrangian $\mathcal{L}$.\\
{\tt GetMassTerms[}$\mathcal{L}$, options{\tt]} & Returns the mass terms in the Lagrangian $\mathcal{L}$.\\
{\tt GetQuadraticTerms[}$\mathcal{L}$, options{\tt]} & Returns the quadratic terms in the Lagrangian $\mathcal{L}$.\\
{\tt GetInteractionTerms[}$\mathcal{L}$, options{\tt]} & Returns the interaction terms in the Lagrangian $\mathcal{L}$.\\
{\tt SelectFieldContent[}$\mathcal{L}$, list\tt{]} & Returns the part of the Lagrangian $\mathcal{L}$ corresponding to the field content specified in list.
\egfbx
\textcolor{white}{\caption{\label{fig:Manipulating a Lagrangian}}}
\end{table}

Once the Lagrangian is implemented, several sanity checks can be performed by means
of the functions presented in Table \ref{fig:Checks Lagrangian}. First, the 
function

{\tt CheckHermiticity[ }$\mathcal{L}$ {\tt ];}

checks if the Lagrangian $\mathcal{L}$ is Hermitian. Next, three functions are 
available to check if the kinetic terms and the mass terms are diagonal, 
{\tt Check\-Di\-a\-go\-nal\-Ki\-ne\-tic\-Terms}, {\tt CheckDiagonalMassTerms} and 
{\tt CheckDiagonalQuadraticTerms}. Finally, two functions, 
{\tt CheckKineticTermNormalisation} and {\tt Check\-Mass\-Spec\-trum}, allow to 
check 
the normalization of the kinetic terms and compare the masses computed from the 
Lagrangian to those of the model description.
The \feynrules\ conventions on the normalization of the kinetic and mass terms for
the scalar, spin 1/2 and vector fields are
\begin{enumerate}
\item Scalar fields:
\begin{itemize}
\item[-] Real: 
\begin{equation}
\frac{1}{2}\partial_\mu \phi\partial^\mu\phi -\frac{1}{2}m^2\phi^2,\nonumber
\end{equation}
\item[-] Complex (including ghost fields): 
\begin{equation}
\partial_\mu \phi^\dagger\partial^\mu\phi -m^2\phi^\dagger\phi,\nonumber
\end{equation}
\end{itemize}
\item Spin-1/2 fermions:
\begin{itemize}
\item[-] Majorana: 
\begin{equation}
\frac{1}{2}\bar\lambda i\slashed{\partial}\lambda -\frac{1}{2}m\bar\lambda\lambda,\nonumber
\end{equation}
\item[-] Dirac: 
\begin{equation}
\bar\psi i\slashed{\partial}\psi -m\bar\psi\psi,\nonumber
\end{equation}
\end{itemize}
\item Vectors:
\begin{itemize}
\item[-] Real: 
\begin{equation}
-\frac{1}{4}F_{\mu\nu}F^{\mu\nu}-\frac{1}{2}m^2A_{\mu}A^{\mu},\nonumber
\end{equation}
\item[-] Complex: 
\begin{equation}
-\frac{1}{2}F_{\mu\nu}^\dagger F^{\mu\nu}-m^2A_{\mu}^\dagger A^{\mu}.\nonumber
\end{equation}
\end{itemize}
\end{enumerate}
\feynrules\ does not use the quadratic pieces of a Lagrangian. However, the 
propagators hard-coded either in \feynrules\ or in the event generators assume 
that the quadratic piece of the Lagrangian follow the above-mentioned 
conventions. Furthermore, since the kinetic and mass terms for spin-3/2 and 
spin-2 fields are model dependent, they are therefore not implemented. 
Finally, checks on Weyl fermion kinetic and mass terms are also not supported since there
exist several ways to write them down.

\begin{table}
\bgfbalign
\multicolumn{2}{c}{\textbf{Table~\ref{fig:Checks Lagrangian}: Checking the consistency of a Lagrangian}}\\
\multicolumn{2}{l}{All the functions below share the same options as
{\tt FeynmanRules},}\\
\multicolumn{2}{l}{which can be found in Table~\ref{fig:FR and V Options}.}\\
\multicolumn{2}{l}{{\tt CheckHermiticity[} $\mathcal{L}$, options {\tt]}}\\
 & Checks if the Lagrangian $\mathcal{L}$ is Hermitian.\\
\multicolumn{2}{l}{{\tt CheckDiagonalKineticTerms[} $\mathcal{L}$, options {\tt]}}\\
 & Checks if all the kinetic terms in the Lagrangian $\mathcal{L}$ are diagonal.\\
\multicolumn{2}{l}{{\tt CheckDiagonalMassTerms[} $\mathcal{L}$, options {\tt]}}\\
 & Checks if all the mass terms in the Lagrangian $\mathcal{L}$ are diagonal.\\
\multicolumn{2}{l}{{\tt CheckDiagonalQuadraticTerms[} $\mathcal{L}$, options {\tt]}}\\
 & Checks if all the quadratic terms in the Lagrangian $\mathcal{L}$ are diagonal.\\
\multicolumn{2}{l}{{\tt CheckKineticTermNormalisation[} $\mathcal{L}$, options {\tt]}}\\
 & Checks if all the kinetic terms in the Lagrangian $\mathcal{L}$ are correctly diagonalized and normalized.\\
\multicolumn{2}{l}{{\tt CheckMassSpectrum[} $\mathcal{L}$, options {\tt]}}\\
 & Checks if all the mass terms in the Lagrangian $\mathcal{L}$ are correctly diagonalized and if their value corresponds to the numerical value given in the model file.\\
\egfbalign
\textcolor{white}{\caption{\label{fig:Checks Lagrangian}}}
\end{table}

\subsection{\label{sec:auto susy L}Automatic generation of supersymmetric Lagrangians}
The implementation of supersymmetric Lagrangians in \feynrules\ can be highly
facilitated by
means of a series of dedicated built-in functions. 
The Lagrangian describing the kinetic terms and the gauge interactions of 
the chiral content of any supersymmetric theory
is given by 
\beq\bsp
  {\cal L}_{\rm chiral} &\ =
     \bigg[ \Phi_i^\dag \, e^{-2 g_j V^j} \,
      \Phi^i \bigg]_{\theta\cdot\theta \bar\theta\cdot\bar\theta} \\
 &\ =
      D_\mu \phi_i^\dag D^\mu\phi^i + F_i^\dag F^i
    - \frac{i}{2} \big(D_\mu \bar \psi_i \bar \sigma^\mu \psi^i - 
        \bar \psi_i\bar \sigma^\mu  D_\mu \psi^i \big) \\
  &\  + 
      i\sqrt{2}g_j  \bar \lambda^{ja} \cdot \bar \psi_i T_a \phi^i  
    - i\sqrt{2}g_j \phi_i^\dag T_a  \psi^i \cdot  \lambda^{ja} - g_j D^{ja}
      \phi_i^\dag T_a \phi^i \  ,
\esp \eeq
where the superfield content of the theory is represented by a set of
chiral superfields $\{\Phi^i=(\phi^i, \psi^i, F^i)\}$ and vector
 superfields $\{V^j = (v^j, \lambda^j,
D^j)\}$. In the first line of the equation above, the
$[\ .\ ]_{\theta\cdot\theta \bar\theta\cdot\bar\theta}$ indicates that one has to
extract the highest-order coefficient in $\theta$ and $\thetabar$ from the
expansion of the
superfield expression included in the brackets. We recall that the covariant 
derivatives are defined in Eq.\ \eqref{eq:covder} and that the matrices 
$T_a$ stand for
representation matrices of the gauge group relevant to the fields
they act on. This Lagrangian can be computed in \feynrules\ by issuing
\begin{verbatim} 
Theta2Thetabar2Component[ CSFKineticTerms[ ] ]
\end{verbatim}
The function \texttt{CSFKineticTerms} returns the kinetic and gauge
interaction Lagrangian terms for all the chiral supermultiplets of the model. As a
result, the Lagrangian is computed in terms
of superfields. The method \texttt{Theta2\-Theta\-bar2\-Com\-po\-nent} is then
needed to ensure that only the terms
in $\theta\!\cdot\!\theta \bar\theta\!\cdot\!\bar\theta$ are kept, after the
superfields have been expanded in terms of their component fields. 
If the user is interested in one specific superfield, represented by,
\eg, the symbol
\texttt{Phi}, the corresponding Lagrangian can be obtained by specifying this superfield as the
argument of the function \texttt{CSFKineticTerms},
\begin{verbatim} 
Theta2Thetabar2Component[ CSFKineticTerms[ Phi ] ]
\end{verbatim}

Non-gauge interactions among the chiral superfields are driven by the superpotential, a
holomorphic function $W(\Phi)$ of the chiral superfields $\Phi$ of the theory.
The associated interaction Lagrangian is given by
\beq\bsp
  {\cal L}_W = \big[W(\Phi)\big]_{\theta\cdot\theta} + \hc =  F^i \frac{\del
    W(\phi)}{\del \phi^i} + \frac12 \frac{\del^2 W(\phi)}{\del \phi^i \del \phi^j}  
    \psi^i\!\cdot\!\psi^j  +\hc \ ,
\esp\eeq
where in our notations, $[\ .\ ]_{\theta\cdot\theta}$ indicates that one has to
select the component in $\theta\!\cdot\!\theta$ from the expansion of the
superpotential in terms of the Grassmann variables $\theta$ and $\thetabar$. 
Assuming that $W(\Phi)$ is represented in the \feynrules\ model file by 
a variable \texttt{SuperW}, the Lagrangian can be
calculated by issuing
\begin{verbatim}
Theta2Component[SuperW] + Thetabar2Component[HC[SuperW]] 
\end{verbatim}
where the two functions \texttt{Theta2Component} and
\texttt{Theta\-bar2\-Component} allow to extract the correct coefficients of the
expansion of the superpotential in $\theta$ and $\thetabar$. 

We now turn to the Lagrangian associated with the gauge sector. Kinetic and gauge
interaction terms are built from squaring the superfield strength tensors and
read, in the non-abelian case\footnote{The abelian limit
can be easily derived. In the case where
one has several abelian groups, kinetic mixing is often present. Such a feature
is not covered in the automatic extraction of a supersymmetric Lagrangian by
\feynrules\ and we leave to the user to implement it manually if he/she finds it
necessary.},
\beq\bsp
  {\cal L}_V =&\ \frac{1}{16 g^2} \big[W^{\alpha a} W_{\alpha a}
    \big]_{\theta\cdot\theta} + \hc \\ = &\ 
    -\frac14 F_a^{\mu\nu} F^a_{\mu\nu}  
     + \frac{i}{2}(\lambda_a \sigma^\mu D_\mu\lambar^a - D_\mu
       \lambda_a\sigma^\mu \lambar^a) + \frac12 D^a D_a  \ .
\esp\eeq
Implementing this Lagrangian into \feynrules\ can be achieved by issuing
\begin{verbatim}
LV = Theta2Component[ VSFKineticTerms[ ] ] + 
    Thetabar2Component[ VSFKineticTerms[ ] ]
\end{verbatim}
where an implicit sum over the whole vector superfield content of the theory is performed.
The function \texttt{VSFKineticTerms} allows us to construct the Lagrangian in
terms of superfields and the functions 
\texttt{Theta2Component} and \texttt{Theta\-bar2\-Component} to select the
$\theta\!\cdot\!\theta$ and $\thetabar\!\cdot\!\thetabar$ components of the
expansion of the superfield Lagrangian in
terms of the Grassmann variables, respectively. Specifying 
the arguments of the function
\texttt{VSFKineticTerms},
\begin{verbatim}
LV = Theta2Component[ VSFKineticTerms[ VSF ] ] + 
    Thetabar2Component[ VSFKineticTerms[ VSF ] ]
\end{verbatim}
allows us to compute the kinetic and gauge interaction Lagrangian terms associated
with the vector superfield represented by the symbol {\tt VSF}.

To summarize, implementing a (renormalizable) Lagrangian density for a supersymmetric theory in
\feynrules\ only requires the definition of the superpotential \texttt{SuperW},
since the rest of the Lagrangian computation is automatic.
The full (supersymmetric) Lagrangian is thus calculated by issuing 
\begin{verbatim}
LC=Theta2Thetabar2Component[CSFKineticTerms[]]; 
LV=Theta2Component[VSFKineticTerms[]] + 
      Thetabar2Component[VSFKineticTerms[]];
LW=Theta2Component[SuperW]+Thetabar2Component[HC[SuperW]];
Lag = LC + LV + LW;
\end{verbatim}

The Lagrangian density obtained in this way still depends on the
auxiliary $F$ and $D$-fields that can be eliminated by inserting the
solutions of their equations of motion back into the Lagrangian.
This operation is automated in
\feynrules\ by means of the \texttt{SolveEqMotionD} (this eliminates the $D$-fields),
\texttt{SolveEqMotionF} (this eliminates the $F$-fields) and \texttt{SolveEqMotionFD}
(this eliminates all auxiliary fields) functions. More precisely, all the auxiliary fields 
that could appear in the Lagrangian represented by the symbol \texttt{Lag}
are eliminated by typing one of the two commands
\begin{verbatim}
SolveEqMotionFD[Lag]
SolveEqMotionF[SolveEqMotionD[Lag]]
\end{verbatim}

Finally, the component fields of a supermultiplet are 
Weyl fermions and need to rewritten in terms of four-component spinors.
This can be achieved using the {\tt WeylToDirac[]} function described in the 
beginning of this section.

For more information on the superspace module of \feynrules\ we refer to 
refs.~\cite{Duhr:2011se,Fuks:2012im} and to Table \ref{tab:autolag} which
collects all the functions presented in this section.

\begin{table}
\bgfb
\multicolumn{2}{c}{\textbf{Table~\ref{tab:autolag}: Constructing supersymmetric
Lagrangians.}}\\
\texttt{CSFKineticTerms[csf]} &
Derives kinetic and gauge interaction terms
associated with the chiral superfield {\tt csf}. If called without any
argument, a sum over the whole chiral content of the theory is performed.\\
\texttt{VSFKineticTerms[vsf]} &
Derives kinetic and gauge interaction terms
associated with the vector superfield {\tt vsf}. If called without any
argument, a sum over the whole gauge content of the theory is performed.\\
\texttt{SolveEqMotionFD[} $\mathcal{L}$ {\tt ]}
& Computes and solves the equations of motion for all
auxiliary fields. The solutions are then inserted in the Lagrangian $\mathcal{L}$.\\
\texttt{SolveEqMotionD[} $\mathcal{L}$ {\tt ]}
& Computes and solves the equations of motion for the
auxiliary $D$-fields. The solutions are then inserted in the Lagrangian $\mathcal{L}$.\\
\texttt{SolveEqMotionF[} $\mathcal{L}$ {\tt ]}
& Computes and solves the equations of motion for the
auxiliary $F$-fields. The solutions are then inserted in the Lagrangian $\mathcal{L}$.\\
\texttt{WeylToDirac[} $\mathcal{L}$ {\tt ]}
& Reexpresses a Lagrangian $\mathcal{L}$, containing
two-component Weyl fermions, in terms of
four-component fermions. \\
\egfb
\textcolor{white}{\caption{\label{tab:autolag}}}
\end{table}


\section{Running \feynrules}
\label{sec:running}
After the model description is created and the Lagrangian constructed, it can be loaded into \feynrules\ and the Feynman rules obtained.
\subsection{Loading \feynrules}
\label{sec:loadFR}
The first thing that must be done when using \feynrules\ is to load the package
into a \mathematica\ session.  This should be done before any of the model 
description is loaded in the kernel and so should be the first line of the 
\mathematica\ notebook\footnote{In other words, if the model description is done 
in a \mathematica\ notebook, it should come after \feynrules\ is loaded.}.  In 
order to load \feynrules, the user must first specify the directory where it
is stored and then load it by issuing 
\texttt{
$\$$FeynRulesPath = SetDirectory[ $<$the address of the package$>$ ];\\
$<<$ FeynRules\`}

\subsection{Loading the model file}
\label{sec:loadMF}
After the \feynrules\ package has been loaded\footnote{The user may want to 
change 
the current directory of \mathematica\ at this point.  Otherwise, all the files
and directories generated by \feynrules\ may end up in the main \feynrules\ 
directory.}, the model can be loaded using the command \verb+LoadModel+,
\begin{verbatim}
LoadModel[ < file.fr >, < file2.fr >, ... ]
\end{verbatim}
The model may be contained in one model file or split among several files.  For \feynrules\ to run properly, the extension of each model file should be \verb+.fr+.  If the model description is entered directly in the \mathematica\ notebook, the 
list of files is then empty.  In this case, \verb+LoadModel[]+ has to be executed
after all the lines of the model description are loaded into the kernel.

Any time the model description changes, the model must be reloaded.  Currently, this means that the \mathematica\ kernel must be quit and the \feynrules\ package and model must be reloaded from the beginning.  An exception to this is the Lagrangian.  It can be changed and extended without having to reload the model information.

In the rest of this section, we describe the main utilities included in \feynrules\ which are summarized in Table \ref{fig:Loading Model Files}.

\subsection{Extracting the Feynman rules}
\label{sec:Feynman Rules}
After the model is loaded and the Lagrangian is defined, the Feynman rules can be extracted using the command \verb+FeynmanRules+.  For the rest of this section, we use the QCD Lagrangian defined in Eq.~(\ref{eq:QCD Lagrangian}) as an example.  The 
Feynman rules can be generated by means of the command\footnote{Since the vertices list may be very long, it is usually desirable to end this statement with a semicolon.}:
\begin{verbatim}
vertsQCD = FeynmanRules[ LQCD ];
\end{verbatim}
The vertices derived by \feynrules\ are then written out on the screen and stored internally in the variable \verb+vertsQCD+. The function \verb+FeynmanRules+ has several options, that are described below.

The user can instruct \mathematica\ to not write the Feynman rules to the screen with the option \verb+ScreenOutput+ set to {\tt False}, 
\begin{verbatim}
vertsQCD = FeynmanRules[ LQCD, ScreenOutput -> False];
\end{verbatim}
In this case, the Feynman rules are generated, stored in \verb+vertsQCD+, but not displayed on screen.

In the two previous examples, the flavors were not expanded.  For example, the 
preceding commands will only generate a single quark-gluon vertex ({\tt dq dqbar 
G}). It is often desirable to perform a flavor expansion, \ie, to obtain 
separately the vertices {\tt d dbar G}, {\tt s sbar G}, and {\tt b bbar G}. To 
achieve this, the user can employ the \verb+FlavorExpand+ option. This option can be used 
to specify individual flavor indices to be expanded over, as in 
\verb+FlavorExpand->Flavor+ where only the \verb+Flavor+ index is expanded over 
(and not any other flavor indices if defined).  It may also refer to a list of 
flavor indices to be expanded, as in \verb+FlavorExpand->{Flavor,SU2W}+. 
Similarly, it 
may be used to expand over all the flavor indices as in \verb+FlavorExpand->True+. Note that it is always possible to expand over the flavors at a later stage using the {\tt FlavorExpansion[ ]} function, i.e., 
\begin{verbatim}
vertices = FeynmanRules[ L ];
vertices = FlavorExpansion[ vertices ];
\end{verbatim}
which is equivalent to calling {\tt FeynmanRules[ L, FlavorExpand -> True ]}. Note that calling the {\tt FlavorExpansion[ ]} function in general runs faster than the {\tt FlavorExpand} option. The difference is that when using the {\tt FlavorExpand} option with {\tt FeynmanRules}, the flavors are expanded before the vertices are computed, so \feynrules\ computes the vertices for the individual flavors separately.

At this stage we have to make a comment about the {\tt Unfold[ ]} function introduced in Section~\ref{sec:indices}. If an index type is wrapped in the {\tt Unfold[ ]} command, then \feynrules\ will always expand over these indices. Moreover, any index which expands field in terms of non-physical states must be wrapped in {\tt Unfold[ ]}.  As an example, the adjoint index carried by the weak gauge bosons must always be expanded over, because the physical states are the photon and the $W^\pm$ and $Z$ bosons, while for quarks it can be interesting to keep the vertices in a compact form, keeping explicit flavor indices in the vertices. Note that, in order to speed up the computation of the vertices, (some of) the interfaces perform the flavor expansion using the {\tt FlavorExpansion[ ]} command.

The list of Feynman rules can be quite long and it may sometimes be desirable to extract just one or a few vertices.  There are several options available that limit the number of vertices computed by \feynrules:
\begin{itemize}
\item {\tt MaxParticles -> n} instructs {\tt FeynmanRules} to only derive vertices whose number of external legs does not exceed {\tt n}. The option {\tt MinParticles} works in a similar way.
\item {\tt MaxCanonicalDimension -> n} instructs {\tt FeynmanRules} to only derive vertices whose canonical dimension does not exceed {\tt n}. The option {\tt MinCanon- icalDimension} works in a similar way.
\item \verb+SelectParticles ->{{...}, {...},...}+ instructs \verb+FeynmanRules+ to only derive the vertices specified in the list.  For example, the command:
\begin{verbatim}
FeynmanRules[ LQCD, SelectParticles ->{{G,G,G}, {G,G,G,G}}]
\end{verbatim}
only leads to the derivation of the three and four-point gluon vertices.
\item \verb+Contains -> { ... }+ instructs \verb+FeynmanRules+ to only derive the vertices which involve all the particles indicated in the list.
\item \verb+Free -> { ... }+ instructs \verb+FeynmanRules+ to only derive the vertices which do not involve any of the particles indicated in the list.
\end{itemize}

\verb+FeynmanRules+, by default, checks whether the quantum numbers that have been defined in the model file are conserved for each vertex. This check can be turned off by setting the option \verb+ConservedQuantumNumbers+  to \verb+False+. Alternatively, the argument of this option can be a list containing all the quantum numbers \feynrules\ should check for conservation.

The Feynman rules constructed are stored internally as a list of vertices where each vertex is a two-component list.  The first element enumerates all the particles that enter the vertex while the second one gives the analytical expression for the vertex. We illustrate this for the quark-gluon vertex,

\verb+{{{G, 1}, {qbar, 2}, {q, 3}},+ $i g_s\, \delta_{f_2f_3}\, \gamma^{\mu_1}_{s_2s_3}\, T^{a_1}_{i_2i_3}$ \verb+}+

Moreover, each particle is also given by a two-component list, the first element 
being the name of the particle while the second one the label given to the indices referring to this particle in the analytical expression.

As the list of vertices computed by \feynrules\ can be quite long for complicated models, the \verb+SelectVertices+ routine can be employed.
  The options shared with the function {\tt FeynmanRules} are \verb+MaxParticles+, \verb+MinParticles+, \verb+SelectParticles+, \verb+Contains+ and \verb+Free+ and 
can be invoked as
\begin{verbatim}
vertsGluon = SelectVertices[vertsQCD, 
                    SelectParticles->{{G,G,G},{G,G,G,G}}];
\end{verbatim}

It is sometimes convenient to construct the Feynman rules by parts, for instance, when one splits the QCD Lagrangian into three pieces as in
\begin{verbatim}
LQCD = LGauge + LQuarks + LGhosts;
\end{verbatim}
The Feynman rules can be constructed all at once as in the previous examples, or they can be constructed separately as in:
\begin{verbatim}
vertsGluon = FeynmanRules[ LGauge ];
vertsQuark = FeynmanRules[ LQuarks ];
vertsQuark = FeynmanRules[ LGhosts ];
\end{verbatim}
They can later be merged using the function \verb+MergeVertices+ 
\begin{verbatim}
vertsQCD = MergeVertices[ vertsGluon, vertsQuark, vertsGhosts ];
\end{verbatim}
This merges the results obtained for \verb+vertsGluon+, \verb+vertsQuark+ and {\tt verts- Ghosts} into a single list of vertices.  If there are two contributions to the same vertex, they will be combined into one vertex.

\subsection{Manipulating Parameters}
\label{sec:updateparams}
The parameters are also an important part of the model and several functions 
allow to manipulate them.  The numerical values of any parameter or function of
one or several parameters can be obtained with the use of the 
\verb+NumericalValue+ function, as for example in
\begin{verbatim}
NumericalValue[ Sin[ cabi ]]
\end{verbatim}
where \verb+cabi+ is the Cabibbo angle.  This returns the numerical value of the sine of the Cabibbo angle.

In the case the user wants to change the value of a subset of external parameters,
the function \verb+UpdateParameters+ can be used such as in 
\begin{verbatim}
UpdateParameters[ gs -> 0.118 , ee -> 0.33 ]
\end{verbatim}
where \verb+gs+ and \verb+ee+ are the strong and electromagnetic coupling constants, respectively.

In order to write and read the numerical values of the parameters to and from a 
file, \feynrules\ comes with the two functions \verb+WriteParameters+ and 
\verb+ReadParameters+.  By default, they write and read the parameters to and from the file named \verb+M$ModelName+ with a ``.pars'' appended, but this can be changed with the options \verb+Output+ and \verb+Input+ as in:
\begin{verbatim}
WriteParameters[Output -> "parameters.pars"]
ReadParameters[Input -> "parameters.pars"]
\end{verbatim}
The routine \verb+WriteParameters+ writes out the external and internal parameters (including masses and widths) to the specified file, while the function \verb+ReadParameters+ only reads in the external parameters (including masses and widths).  This gives the user another way to change the values of the external parameters.  The user can modify the values of the parameters included in the parameter file created by \verb+WriteParameters+, and then read them back in using \verb+ReadParameters+.  Any changes made to the internal parameters are ignored.

In addition to the native \feynrules\ parameter files ({\tt .pars}), there is a second way to update numerical values of external parameters. Indeed, 
\feynrules\ is equipped with its own reading and writing routine for Les Houches Accord (LHA)-like parameter files~\cite{Skands:2003cj,Allanach:2008qq}. For example,
issuing the command
\begin{verbatim}
ReadLHAFile[ Input -> "LHA-file" ]
\end{verbatim}
reads the LHA-like parameter file \verb+"LHA-file"+ and updates the numerical values of all the external parameters of the model in the current \mathematica\ session. Similarly, the command
\begin{verbatim}
WriteLHAFile[ Output -> "LHA-file" ]
\end{verbatim}
writes all the numerical values to file in a LHA-like format.

\begin{table}
\bgfbalign
\multicolumn{2}{c}{\textbf{Table~\ref{fig:Loading Model Files}: \feynrules\ 
  commands}}\\
\multicolumn{2}{l}{{\tt LoadModel[}f1.fr, f2.fr, \ldots{\tt ]}}\\
  & Loads and initializes the model defined by the model files f1.fr, f2.fr,... .\\
\multicolumn{2}{l}{{\tt FeynmanRules[}$\mathcal{L}$, options{\tt ]}}\\
  & Calculates the Feynman rules associated with the Lagrangian $\mathcal{L}$.  
  The list of available options are given in Table \ref{fig:FR and V Options}.\\
\multicolumn{2}{l}{{\tt SelectVertices[}verts, options{\tt ]}}\\ 
  & Selects a subset of the vertices contained in verts.  The list of 
  available options are given in Table \ref{fig:FR and V Options}.\\
\multicolumn{2}{l}{{\tt MergeVertices[}v1, v2, \ldots{\tt ]}}\\
 & Merges the vertex lists v1, v2,\ldots into one single list.\\
\multicolumn{2}{l}{{\tt NumericalValue[} f[pars] {\tt ]}}\\
  & Gives the numerical value of f[pars] where f is some function and pars is some set of parameters of the model.\\
\multicolumn{2}{l}{{\tt UpdateParameters[}{$p1\rightarrow v1$}, $p2\rightarrow v2$,\ldots{\tt ]}}\\ 
  & Changes the values of the parameters p1,p2,\ldots to v1,v2,\ldots.\\
\multicolumn{2}{l}{{\tt WriteParameters[}options{\tt ]}}\\
  & Writes the numerical values of the internal and external parameters (including
  masses and widths) to a text file whose the filename consists of the value of
  {\tt M\$ModelName} with the extension {\tt .pars} appended, unless otherwise 
  specified by means of the {\tt Output} option.\\
\multicolumn{2}{l}{{\tt ReadParameters[}options{\tt ]}}\\
  & Reads the external parameters from a text file. By default, the text file is 
  named as in {\tt WriteParameters} and must have the same format as the one 
  created by {\tt WriteParameters}. Another input filename can be specified by 
  means of the {\tt Input} option.\\
\multicolumn{2}{l}{{\tt WriteLHAFile[ Output -> } file {\tt]}}\\
& Writes the numerical values of all external parameters to the text file file in an LHA-like format.\\
\multicolumn{2}{l}{{\tt ReadLHAFile[ Input -> } file {\tt]}}\\
& Reads the numerical values of the external parameters from the LHA-like text file file.\\  
\egfbalign
\textcolor{white}{\caption{\label{fig:Loading Model Files}}}
\end{table}

\begin{table}
\bgfb
\multicolumn{2}{c}{\textbf{Table~\ref{fig:FR and V Options}: {\tt FeynmanRules}
   and {\tt SelectVertices} options}}\\
{\tt ScreenOutput} & Option of {\tt FeynmanRules}. If turned to {\tt False}, the 
  derived Feynman rules do not appear on the screen. The default is {\tt True}.\\
{\tt FlavorExpand} & Option of {\tt FeynmanRules}. Expands over the flavor indices
  specified by the list which this option is referring to. If turned to {\tt True},
  all flavor indices are expanded.\\
{\tt ConservedQuantumNumbers} & Option of {\tt FeynmanRules}. Checks whether the 
  quantum numbers specified by the list which this option is referring to are 
  conserved. If {\tt True} ({\tt False}), the conservation of all (no) quantum 
  numbers is checked. The default is {\tt True}.\\
{\tt MinParticles} & Only vertices involving at least the specified number of external fields are derived.\\
{\tt MaxParticles} & Only vertices involving at most the specified number of external fields are derived.\\
{\tt MinCanonicalDimension} &   Option of {\tt FeynmanRules}. Only vertices of at least the specified canonical dimension are derived.\\
{\tt MaxCanonicalDimension} &  Option of {\tt FeynmanRules}. Only vertices of at most the specified canonical dimension are derived.\\
{\tt SelectParticles} & Calculates only the vertices specified in the list which
  this option is referring to.\\
{\tt Contains} & Only the vertices which contain at least the particles contained 
  in the list which this option is referring to are derived.\\
{\tt Free} & Only the vertices which do not contain any of the particles contained
  in the list which this option is referring to are derived.
\egfb
\textcolor{white}{\caption{\label{fig:FR and V Options}}}
\end{table}

\subsection{Manipulating superspace expressions}\label{sec:manipsusy}
The $N=1$ superspace is defined by supplementing to the ordinary spacetime
coordinates $x^\mu$ a Majorana spinor
$(\theta_\alpha, \thetabar^\alphadot)$. These extra coordinates are represented 
in \feynrules\ by the symbols \texttt{theta} and \texttt{thetabar} 
\be
  \texttt{theta[alpha]} \leftrightarrow \theta_\alpha 
  \quad \text{and} \quad
  \texttt{thetabar[alphadot]} \leftrightarrow \bar\theta_{\dot\alpha} 
\nn \ee
By convention, both spin indices are assumed to be lower indices.
The position of the spin indices can be modified by employing the
rank-two antisymmetric tensors represented by the objects \texttt{Deps} and
\texttt{Ueps}, which equivalently take as arguments left-handed or
right-handed spin indices. For instance, one could implement the following
expressions in a \mathematica\ session as
\be \bsp
  \bar\theta^\alphadot = \epsilon^{\alphadot\betadot}\bar\theta_{\betadot} 
    \leftrightarrow  &\ \texttt{Ueps[alphadot,betadot] thetabar[betadot]}  \\ 
  \theta_\alpha = \epsilon_{\alpha\beta} \epsilon^{\beta\gamma} \theta_\gamma 
    \leftrightarrow &\  \texttt{Deps[alpha,beta] Ueps[beta,gamma]
    theta[gamma]}  
\esp \nn \ee

Proper computations in superspace require to
keep track, on the one hand, of the position of the spin indices and, on the
other hand, of the fermion ordering. To this aim, \feynrules\ always assumes
that an explicit spin index is a
lower index. Moreover, fermion ordering information is provided by using the 
syntax \texttt{nc[chain]}, 
where {\tt chain} stands for any ordered sequence of fermions (with lower spin
indices).
As a simple example, we show a possible implementation for the scalar product
$\lambar \!\cdot\! \lambar^\prime$, where $\lambar$ and $\lambar'$ are
right-handed Weyl fermions,
\be
  \lambar \!\cdot\! \lambar^\prime \leftrightarrow 
     \texttt{nc[lambdabar[bd], lambdabarprime[ad]] Ueps[bd,ad]} 
\nn\ee
This way of inputting superspace expressions is however highly unpractical for
longer expressions. Therefore, \feynrules\ contains an environment \texttt{ncc}
similar to the \texttt{nc} environment, the difference lying in the fact that all 
spin indices and $\epsilon$-tensors can be omitted, 
\be
   \lambar \!\cdot\! \lambar^\prime \leftrightarrow 
    \texttt{ncc[lambar, lambarprime]}\vspace*{-.1cm}
\nn\ee
The \texttt{ncc} environment is automatically handled by \feynrules. The result
of the command above consists of the same expression, but given in
canonical form, employing rank-two antisymmetric tensors and two-component
spi\-nors with lowered indices, ordered within a \texttt{nc} environment.

Another consequence of the mandatory usage of this canonical form is that the
\mathematica\ output associated to a superspace expression is in general
difficult to read. To improve the readability, it is possible to force
\feynrules\ to form invariant products of spinors and to simplify products of
Grassmann variables according to the relations
\be
  \theta^\alpha \theta^\beta = -\frac12 \theta \!\cdot\! \theta \epsilon^{\alpha \beta}\ ,
    \quad 
  \thetabar^\alphadot \thetabar^\betadot = \frac12 \thetabar \!\cdot\! \thetabar
    \epsilon^{\alphadot \betadot} \quad\text{and}\quad
  \theta^\alpha \thetabar^\alphadot = \frac12 \theta \sigma^\mu \thetabar
    \sibar_\mu{}^{\alphadot\alpha}\ , 
\ee
deduced from the Grassmann algebra fulfilled by the variables $\theta$ and
$\thetabar$. This operation is achieved by means of the function 
\texttt{ToGrassmannBasis} which takes any expression \texttt{exp}, depending on
the superspace coordinates, as an argument, 
\begin{verbatim}
  ToGrassmannBasis[ exp ] 
\end{verbatim}
First of all, this function rewrites the expression \texttt{exp} in terms of a restricted
set of scalar products involving Grassmann variables and Pauli matrices, and, second,
optimizes the index naming scheme used. This allows to
collect
and simplify \mathematica\ expressions that are equal up to the names of
contracted indices, such as,  \eg,
\begin{verbatim}
  Dot[lambda[al], lambda[al]] -  Dot[lambda[be], lambda[be]] 
\end{verbatim}
The output of the \texttt{ToGrassmannBasis} function consists therefore of
expressions very
close to their original forms. This method also works on tensorial quantities
containing non-contracted spin indices. As a result, the
\texttt{To\-Grass\-mann\-Ba\-sis} method matches the free indices 
either to single fermions or to chains
containing one fermion and a given number of Pauli matrices. For instance, 
applying the \texttt{ToGrassmannBasis} function on
$(\sibar^\nu\lambda)_\alpha$, $\lambda$ being a left-handed Weyl fermion, leads
to 
\beq\bsp
  &\  \texttt{ToGrassmannBasis[ nc[lambda[b]] * sibar[nu,ad,b] ]}\\
  &\ \Rightarrow 
    \texttt{nc[ TensDot2[sibar[nu,ad,b], lambda[b]][up, Right,ad] ]}
\esp\nn\eeq
One observes that a chain containing one Pauli matrix and the fermion $\lambda$
has been formed and stored in a \texttt{TensDot2} environment. The structure
related to this environment is defined by 
\begin{verbatim}
  TensDot2[ chain ][pos, chir, name]
\end{verbatim}
where \texttt{chain} is a
sequence of one Weyl fermion and possibly one or several Pauli matrices,
\texttt{pos} is the \texttt{up} or \texttt{down} position of the free spin
index, \texttt{chir} indicates if it is a left-handed (\texttt{Left})
or right-handed (\texttt{Right}) index and \texttt{name} denotes the symbol
associated to the index.

The optimization of the index naming scheme performed by the
\texttt{To\-Grass\-mann\-Ba\-sis} function can also be applied directly on any
expression, even if not concerned with superspace computations. In this case, it
is enough to type 
\begin{verbatim}
  OptimizeIndex[ expression ] 
\end{verbatim}

\begin{table}
\bgfb
\multicolumn{2}{c}{\textbf{Table~\ref{tab:superspace}: Superspace conventions in
\feynrules}}\\
{\tt theta[al]} & The Grassmann variable $\theta_\alpha$.\\
{\tt thetabar[aldot]} & The Grassmann variable $\bar\theta_\alphadot$. \\
{\tt Ueps[al, be]} & The rank-two antisymmetric tensor with upper indices 
  $\epsilon^{\alpha\beta}$. \\
{\tt Deps[al, be]} & The rank-two antisymmetric tensor with lower indices 
  $\epsilon_{\alpha\beta}$. \\
{\tt nc[seq] } & Ordered sequence of fermionic field(s), labeled by {\tt
  seq}. The indices of the fields are explicitly written.\\
{\tt ToGrassmannBasis[exp]} 
 & This function rewrites any function {\tt exp}
of the superspace coordinates in a human-readable form.\\
{\tt OptimizeIndex[exp]} 
 & This function optimizes the index naming scheme used in
  the expression {\tt exp}. \\
{\tt Tonc[exp]} & This function transforms expressions to their
  canonical form, with lower spin indices and $\epsilon-$tensors. \\
{\tt TensDot2[chain][pos,chir,name]} &\\
& This environment contains a sequence, labeled by {\tt chain}, of one Weyl
fermion and possibly one or several Pauli matrices. The symbols {\tt pos}, {\tt
chir} and {\tt name} are the upper (\texttt{up}) or lower (\texttt{down})
position, the chirality (\texttt{Left} or \texttt{Right}) and
the name of the free index.\\
\egfb
\textcolor{white}{\caption{\label{tab:superspace}}}
\end{table}
\renewcommand{\arraystretch}{1}

The basis associated with the output of
the \texttt{To\-Grassmann\-Basis} function can also be used to input superspace
expressions. There are only two rules to follow. First, 
products of spinors, connected or not by one or several Pauli matrices, are
always written as
\begin{verbatim}
  ferm1[sp1].ferm2[sp2] chain[sp1,sp2] 
\end{verbatim}
In this expression, we have introduced two Weyl fermions \texttt{ferm1} and
\texttt{ferm2} and the symbol \texttt{chain} stands for a series of Pauli
matrices linking the spin indices \texttt{sp1} and \texttt{sp2}. 
Next, the implementation of expressions involving fermions carrying a free spin index must
always employ the \texttt{nc} and \texttt{TensDot2} environments as described
above.  The conversion to the canonical form can subsequently be performed by
means of the \texttt{Tonc} function, 
\begin{verbatim}
  Tonc[ expression ]
\end{verbatim}

The list of functions and environments useful for manipulating and inputting
superspace expressions is collected in Table \ref{tab:superspace}.

\subsection{\label{sec:susy computations}Functions dedicated to superspace computations}
The \feynrules\ package comes with a set of predefined functions dedicated to
superspace computations. First, \feynrules\ offers the possibility to employ the 
generators
$Q_\alpha$ and $\Qbar_\alphadot$ of the supersymmetric algebra
as well as the superderivatives $D_\alpha$ and $\Dbar_\alphadot$ to perform
superspace computations. In our conventions \cite{FuksRausch}, these operators
read 
\be\bsp
     Q_\alpha = -i\Big(\frac{\del}{\del\theta^\alpha} 
        + i \sigma^\mu{}_{\alpha\alphadot}
    \thetabar^\alphadot \del_\mu\Big) \  , 
  &\qquad \Qbar_\alphadot = i \Big(\frac{\del}{\del\thetabar^\alphadot} + i \theta^\alpha
    \sigma^\mu{}_{\alpha \alphadot} \del_\mu\Big) \ , \\
     D_\alpha = \frac{\del}{\del\theta^\alpha}- i \sigma^\mu{}_{\alpha\alphadot} \thetabar^\alphadot 
     \del_\mu \ , 
  &\qquad \Dbar_\alphadot =  \frac{\del}{\del\thetabar^\alphadot} - i\theta^\alpha
     \sigma^\mu{}_{\alpha\alphadot}\del_\mu \ .
\esp\ee%
and can be called in \feynrules\ by typing  
\be \bsp
 &\ Q_\alpha (\text{expression}) \leftrightarrow \texttt{QSUSY[expression,alpha]} 
 \ , \\ 
 &\ \bar Q_\alphadot  (\text{expression}) \leftrightarrow 
    \texttt{QSUSYBar[expression,alphadot]}   \ , \\ 
 &\ D_\alpha (\text{expression}) \leftrightarrow 
      \texttt{DSUSY[expression,alpha]} \ , \\ 
 &\  \bar D_\alphadot
    (\text{expression})  \leftrightarrow \texttt{DSUSYBar[expression,alphadot]}\ .
\esp \nn \ee
The expression on which these operators act must be provided in its canonical
form, employing the \texttt{nc} environment. Next, the computation of the variation of a
quantity $\Phi$ under a supersymmetric transformation of parameters
$(\varepsilon, \bar\varepsilon)$ can be achieved by issuing
\be
  \delta_\varepsilon \Phi = i \big(\varepsilon\!\cdot\! Q + \Qbar \!\cdot\!
\bar\varepsilon\big) \Phi \ .
\ee
However, we recommend the user to employ the shortcut function {\tt DeltaSUSY} 
rather than a more complicated function of {\tt QSUSY} and {\tt QSUSYBar},
\begin{verbatim}
  DeltaSUSY [ expression , epsilon ] 
\end{verbatim}
The symbol \texttt{expression} stands for any function of
superfields and/or component fields while \texttt{epsilon} refers to the
left-handed piece of the supersymmetric transformation parameters, given
without any spin index. There are ten predefined quantities that can play the
role of the transformation parameters and that can be employed by typing
the symbol \texttt{eps\it{x}}, \texttt{\it{x}} being an
integer between zero and nine.

\begin{table}
\bgfb
\multicolumn{2}{c}{\textbf{Table~\ref{tab:spacef}: Superspace functionalities}}\\
{\tt QSUSY[exp,al] } & Computes the action of the supercharge
    $Q_\alpha$ on the expression {\tt exp}. This expression must be given under
    its canonical form and the symbol {\tt al} refers to the spin index of the
    supercharge.\\
{\tt QSUSYBar[exp,ad] } & Same as \texttt{QSUSY}, but for the supercharge $\bar
  Q_\alphadot$. \\
{\tt DSUSY[exp,al] } & Same as {\tt QSUSY}, but for the superderivative
  $D_\alpha$.\\
{\tt DSUSYBar[exp,ad] } & Same as {\tt QSUSYBar}, but for the superderivative
$\bar D_\alphadot$.\\
{\tt SF2Components[exp]} 
  & Expands the expression {\tt
    exp} in terms of the Grassmann variables and simplifies the result to a
    human-readable form. The output consists of a two-component list,
    the complete series and a new list with all the individual
    coefficients.\\ & \\
\multicolumn{2}{l}{\textbf{Shortcuts to the individual component fields}}\\
The series. & {\tt GrassmannExpand[exp]} \\
The scalar term. & {\tt ScalarComponent[exp]}\\
The $\theta_\alpha$ term. & \texttt{ThetaComponent[exp, a]} \\
The $\bar\theta_{\alphadot}$ term. & \texttt{ThetabarComponent[exp, ad]} \\
The $\theta\sigma^\mu\bar\theta$ term.&  \texttt{ThetaThetabarComponent[exp, mu]}\\
The $\theta^2$ term.& \texttt{Theta2Component[exp]}\\
The $\bar\theta^2$ term.& \texttt{Thetabar2Component[exp]} \\
The $\theta^2\bar\theta_\alphadot$ term. & \texttt{Theta2ThetabarComponent[exp, ad]}\\
The $\bar\theta^2\theta_\alpha$ term.&\texttt{Thetabar2ThetaComponent[exp, a]}\\
The $\theta^2\bar\theta^2$ term. & \texttt{Theta2Thetabar2Component[exp]}
\egfb
\textcolor{white}{\caption{\label{tab:spacef}}}
\end{table}

Superfield expressions can always be expanded as a series 
in terms of the Grassmann
variables $\theta$ and $\thetabar$
via the \texttt{SF2\-Components} routine,
\begin{verbatim}
  SF2Components [ expression ] 
\end{verbatim}
This expands in a first step all the chiral and vector superfields appearing in
the quantity \texttt{expression} in terms of their component fields and the
usual spacetime coordinates. Next, the function \texttt{ToGrassmannBasis} is
internally called so that scalar products of Grassmann variables are
formed and the expression is further simplified and rendered
human-readable. 
The output of the \texttt{SF2Components} function consists of a two-component
list of the form 
\begin{verbatim}
  { Full series , List of the nine coefficients }
\end{verbatim}
The first element of this list (\texttt{Full series}) is the complete
series expansion in terms of the Grassmann variables, which could also directly
be obtained by typing in a \mathematica\ session
\begin{verbatim}
  GrassmannExpand [ expression ] 
\end{verbatim}
The second element of the list above contains a list with the nine
coefficients of the series, \ie, the scalar piece independent of the Grassmann
variables, followed by the coefficients of the $\theta_\alpha$,
$\thetabar_\alphadot$, $\theta \sigma^\mu\thetabar$, $\theta\!\cdot\!\theta$,
$\thetabar\!\cdot\!\thetabar$, $\theta\!\cdot\!\theta \thetabar_\alphadot$,
$\thetabar\!\cdot\!\thetabar \theta_\alpha$ and $\theta\!\cdot\!\theta
\thetabar\!\cdot\!\thetabar$ terms. Each of these could also be obtained using
the shortcut functions 
\be\bsp
&\  \texttt{ScalarComponent [ expression ] } \ , \\
&\  \texttt{ThetaComponent [ expression ] } \ , \\
&\  \texttt{ThetabarComponent [ expression ] } \ , \\
&\  \texttt{ThetaThetabarComponent [ expression ] } \ , \\
&\  \texttt{Theta2Component [ expression ] } \ , \\
&\  \texttt{Thetabar2Component [ expression ] } \ , \\
&\  \texttt{Theta2ThetabarComponent [ expression ] } \ , \\
&\  \texttt{Thetabar2ThetaComponent [ expression ] } \ , \\
&\  \texttt{Theta2Thetabar2Component [ expression ] } \ .
\esp\ee
where {\tt expression} stands for any superspace expression written in terms
of superfields and/or component fields.
In order to specify the free spin or Lorentz index related to some of these
coefficient, the user has the option to append to the argument of
the functions related to fermionic (vectorial) coefficients the name of a spin
(Lorentz) index.

All the functions are summarized in Table \ref{tab:spacef}.

\subsection{Mass spectrum generation with \feynrules}\label{sec:asperge}

When mixing relations and vacuum expectation values are declared as presented in Section \ref{sec:mixdecl}, the mass matrices
included in a Lagrangian denoted by {\tt Lag} can be extracted by typing
\begin{verbatim}
 ComputeMassMatrix[ Lag, options ] 
\end{verbatim}
where \texttt{options}
stands for optional arguments. If no option is provided, the function calculates
all the mass matrices included in the Lagrangian for which the numerical value of the mixing
matrix is unknown. By issuing 
\begin{verbatim}
 ComputeMassMatrix[ Lag, Mix ->"l1" ] 
\end{verbatim}
\feynrules\ only extracts the mass matrix associated with the mixing relation denoted by the
label \texttt{"l1"}. Replacing this label by a list of labels leads to the computation of multiple mass matrices
simultaneously. Finally, to avoid information to be printed on the screen during the computation of the mass matrices,
it is enough to include the optional argument \texttt{ScreenOutput -> False} in the commands above.

The results of the method above can be retrieved
through the intuitive functions \texttt{MassMatrix}, \texttt{GaugeBasis}, \texttt{MassBasis},
\texttt{MixingMatrix}, \texttt{BlockName} and \texttt{Ma\-trix\-Symbol} which all
take as argument the label of a mixing relation. A wrapper is also available,
\begin{verbatim}
  MixingSummary [ "l1" ]
\end{verbatim}
which sequentially calls all these functions and organizes the output in a human-readable
form.

The \feynrules\ method to extract analytically a mass matrix
can also be employed to compute any matrix $M$ defined by
\be
  {\cal L}_{\rm mass} = {\cal B}_2^\dag \ M\  {\cal B}_1 \ ,
\ee
where ${\cal B}_1$ and ${\cal B}_2$ are two field bases.
The calculation
of the matrix $M$ is achieved by issuing 
\begin{verbatim}
  ComputeMassMatrix[ Lag, 
      Basis1 -> b1, Basis2 -> b2 ]
\end{verbatim}
where the symbols \texttt{b1} and \texttt{b2} are associated with the bases
${\cal B}_1$ and ${\cal B}_2$ given as two lists of fields. 
In this case, the printing functions introduced above are however not
available.

\subsection{Decay width computation with \feynrules}\label{sec:decays} 
In Section~\ref{sec:modelfile} we saw that it is possible to define the width of every particle appearing in the model file. While \feynrules\ itself does not require the knowledge of the width at any stage during the computation of the Feynman rules, this information is required when outputting a model to a matrix element generator for which a translation interface exists\footnote{An exception to this is that \textsc{CalcHEP} can calculate the widths on the fly.  So, the widths are not required for the \textsc{CalcHEP} interface. 
In addition, one should note that
newer versions of \madgraph\ 5 are also capable of computing widths on the fly. These calculations often include non-negligible
$N$-body decay channels with $N>2$, in contrast to the \feynrules\ module strictly limited to $N=2$.}
 (See Section~\ref{sec:interfaces} for more details).

In this section we shortly review the capability of \feynrules\ to compute all the two-body decays that appear inside a model. For more details we refer to Ref.~\cite{decaypaper}.
At leading-order the two-body partial width of a heavy particle of mass $M$ into two particles of mass $m_1$ and $m_2$ is given by
\begin{equation}\label{eq:gamma}
\Gamma = \frac{1}{2|M|S}\int {\rm d} \Phi_2\,|\mathcal{M}|^2\ ,
\end{equation}
where $S$ denotes the phase space symmetry factor and ${\rm d}\Phi_2$ the usual two-particle phase space measure
\begin{equation}
{\rm d}\Phi_2 = (2\pi)^4\,\delta^{(4)}(p_1+p_2-P)\,\frac{{\rm d}^4p_1}{(2\pi)^3}\,\frac{{\rm d}^4p_2}{(2\pi)^3}\,\delta_+(p_1^2-m_1^2)\,\delta_+(p_2^2-m_2^2)\ .
\end{equation}
In this expression, $P=(M,\vec 0)$ denotes the four-momentum of the heavy particle in its rest frame and $p_1$ and $p_2$ are the momenta of the decay products in the same frame\footnote{
The absolute value in Eq.~\eqref{eq:gamma} comes 
from the fact that in certain BSM models involving Majorana fermions it is possible to choose the phases of the fermion fields such that the mass is made negative.}. 
The matrix element of a two-body decay is a constant, and so the partial width is simply the product of the (constant) matrix element and a phase space factor
\begin{equation}\label{eq:Gamma_1_to_2}
\Gamma = \frac{\sqrt{\lambda(M^2,m_1^2,m_2^2)}\,|\mathcal{M}|^2}{16\,\pi\,S\,|M|^3}\,,
\end{equation}
where $\lambda(M^2,m_1^2,m_2^2)= (M^2-m_1^2-m_2^2)^2-4m_1^2m_2^2$.
The matrix element in turn is easily obtained by squaring a three-point vertex by means of the polarization tensors of the external particles. While the polarization tensors are model independent and only depend on the spin of the  particle, the three-point vertices are computed by \feynrules. We therefore have all the ingredients to evaluate Eq.~\eqref{eq:Gamma_1_to_2}. In the rest of this section we describe the functions that allow to compute two-body partial widths from a list of vertices.

Let us assume that we have computed all the vertices associated with a given Lagrangian {\tt L} in the usual way,
\begin{verbatim}
vertices = FeynmanRules[ L ];
\end{verbatim}
We can then immediately compute all the two-body partial widths arising from {\tt vertices} by issuing the command
\begin{verbatim}
decays = ComputeWidths[ vertices ];
\end{verbatim}
The function {\tt ComputeWidths[]} selects all three-point vertices from the list {\tt vertices} that involve at least one massive particle and no ghost fields and/or Goldstone bosons. Next, the vertices are squared by contracting them with the polarization tensors of the external particles and multiplied by the appropriate phase space factors, this computation relying on the unitarity gauge choice.
The output of {\tt ComputeWidths[]} is a list of lists of the form
\begin{equation*}
\textrm{{\tt\{\{}}\phi_1,\, \phi_2,\,\phi_3 \textrm{{\tt \}, }} \Gamma_{\phi_1\to\phi_2\,\phi_3} \textrm{{\tt \}}}\,,
\end{equation*}
where the first element of the list contains the initial state ($\phi_1$) and the two decay products ($\phi_2$ and $\phi_3$) and the second element is the analytic expression for the corresponding partial width.

Some comments are in order about the function {\tt ComputeWidths[]}. First, the output contains the analytic results for all possible cyclic rotations of the external particles $\{\phi_1,\phi_2,\phi_3\}$ with a massive initial state, independently of whether a given decay channel is kinematically allowed, because certain channels might be closed for a certain choice of the external parameters but not for others. Second, we note that the output of {\tt ComputeWidths[]} is also stored internally in the global variable \verb+FR$PartialWidths+.
The use of this global variable will become clear below. Every time the function {\tt ComputeWidths[]} is executed, the value of the global variable will be overwritten, unless the option {\tt Save} of {\tt ComputeWidths[]} is set to {\tt False} (the default is {\tt True}). In most of the cases, sums over
internal gauge indices are left explicit and non-simplified,
exceptions being related to those involving fundamental and adjoint representations of $SU(3)$ in the case
the user adopts the conventions of Section~\ref{sec:conventionsSMgauge}.

Besides the main function {\tt ComputeWidth} that allows to compute the two-body decays, \feynrules\ is equipped with a set of functions that allow to access the output list produced by the main function.
For example, the following intuitive commands are available:
\begin{quote}
{\tt PartialWidth[ \{}$\phi_1, \phi_2, \phi_3$ {\tt\}, decays ]};\\
{\tt TotWidth[ }$\phi_1${\tt, decays ]};\\
{\tt BranchingRatio[ \{}$\phi_1, \phi_2, \phi_3${\tt\}, decays ]};
\end{quote}
In the following we only discuss {\tt PartialWidth[]}, the use of the other two functions is similar. \feynrules\ first checks, based on the numerical values of the particles defined in the model file, whether the decay $\phi_1\to\phi_2\,\phi_3$ is kinematically allowed, and if so, it will calculate the corresponding partial width $\Gamma_{\phi_1\to\phi_2\,\phi_3}$ from the list {\tt decay}. The second argument of {\tt PartialWidth[]} is optional and could be omitted. In that case the partial widths stored in the global variable \verb+FR$PartialWidth+ will be used by default.

Finally, it can be useful to update the information coming from the original particle declarations by replacing the numerical value of the widths of all particles by the numerical values obtained by the function {\tt TotWidth}, which can be achieved by issuing the command
\begin{verbatim}
UpdateWidths[ decays ];
\end{verbatim}
where, as usual, the argument {\tt decays} is optional. After this command has been issued, the updated numerical results for the widths will be written out by the translation interfaces to matrix element generators.


\section{A Simple Example}
\label{sec:simpleexample}

In this section we present an example of how to implement a model into \feynrules\ and how to use the code to obtain the Feynman rules. While the model does not have immediate phenomenological relevance, the implementation is complete in the sense that we discuss in detail all the 
necessary steps. We emphasize that the model does not exploit all the features of \feynrules. For more advanced examples, we recommend to consult the models already implemented~\cite{FRweb} or the details given in Refs.~\cite{Christensen:2009jx,%
Duhr:2011se,Fuks:2012im}. 

The model under consideration is a variant of the $\phi^4$ theory. It displays all the most relevant features the user might encounter when implementing a new model. In particular, it shows how to
\begin{itemize} 
\item[-] define indices of various types,
\item[-] to perform expansions over `flavor' indices,
\item[-] define mixing matrices,
\item[-] perform the rotation from the gauge eigenstates to the mass eigenstates (without employing the mass diagonalization package for which we refer to Ref.~\cite{Alloul:2013fw}),
\item[-] add gauge interactions to the model.
\end{itemize}
In the following, we perform the implementation step by step, adding one feature at a time, in order to clearly show how to deal with the various concepts. A user who follows these steps all the way down to the end will achieve a fully functional \feynrules\ implementation of this model, which may serve as a starting point for \hisher{} own model implementation.

\subsection{The model}
The model we consider is a variant of the $\phi^4$ theory. More precisely we consider two complex scalar fields $\phi_i(x)$, $i=1,2$, interacting through the Lagrangian
\beq\label{eq:Lscal}
\mathcal{L}_{scal} = \partial_\mu\phi^\dagger_i\partial^\mu\phi_i - \phi^\dagger_i\mathcal{M}_{ij}\phi_j + (\phi^\dagger_i\lambda_{ij}\phi_j)^2\,,
\eeq
where the mass matrix $\mathcal{M}$ and the coupling matrix $\lambda$ are assumed to be real and symmetric,
\beq
\mathcal{M}=\left(\begin{array}{cc}
m_1^2 & m_{12}^2/2\\
m_{12}^2/2 & m_2^2
\end{array}\right) {\rm~~and~~}
\lambda=\left(\begin{array}{cc}
\lambda_{11} &\lambda_{12} \\
\lambda_{12} & \lambda_{22}
\end{array}\right)\,.
\eeq
As the mass matrix is not diagonal, the fields $\phi_i$ are not mass eigenstates.
They are related to the mass eigenstates $\Phi_i$ via an orthogonal transformation,
\beq\label{mixing}
\phi_i = U_{ij}\,\Phi_j\ ,
\eeq
where $U$ denotes the orthogonal matrix that diagonalizes the mass matrix 
$\mathcal{M}$,
\beq
  U^T\mathcal{M}\,U = \left(\begin{array}{cc}M_1^2&0\\0&M_2^2\end{array}\right)\ .
\eeq
In general we cannot diagonalize $\mathcal{M}$ and $\lambda$ simultaneously, so 
that after diagonalization, the couplings are explicitly dependent on the mixing.
In the expression above, the eigenvalues of the mass matrix are denoted by $M_1^2$ and $M_2^2$, and without loss of generality we may assume that $M_1^2<M_2^2$. The matrix U is then determined by the (normalized) eigenvectors of $\mathcal{M}$. The eigenvalues and eigenvectors can easily be computed in this case using the {\tt Eigenvalues[]} and {\tt Eigenvectors[]} functions of \mathematica\footnote{We stress that, in general, the diagonalization procedure can be very complicated and one needs to resort to external numerical codes to compute the mass eigenvalues and the rotation matrices or use the \asperge\ package (see Section \ref{sec:aspergeinter}).}. 
For example, the eigenvalues of $\mathcal{M}$ are given by
\beq\label{eq:eigenvalues}
M_{1,2}^2 = \frac{1}{2}\left(m_1^2 + m_2^2\mp\sqrt{(m_1^2-m_2^2)^2+m_{12}^4}\right)\,.
\eeq
The rotation matrix $U$ can be parametrized by a single angle $\theta$,
\beq
U = \left(\begin{array}{cc}
-\sin\theta & \cos\theta\\ \cos\theta& \sin\theta\end{array}\right)\,,
\eeq
where the angle reads
\beq\label{eq:angle}
\sin\theta = \frac{m_{12}^2}{\sqrt{m_{12}^4 + \left(m_1^2-m_2^2+\sqrt{(m_1^2-m_2^2)^2+m_{12}^4}\right)^2}}\,.
\eeq
We now describe how this model can be implemented into \feynrules.

\subsection{Preparation of the model file -- model information}
Each \feynrules\ model file begins by summarizing the model information, which 
acts as its electronic signature. The user can give his/her model a name 
(a string) and include additional information on the model, such as contact 
details of the author(s) of the model file, references to the literature and the 
version number of the implementation of the model. We emphasize that, although 
this information is optional, it can be useful to keep track of modifications, and 
we strongly encourage users to include it for every implementation. 
In our case, the model information could be entered as follows:
\begin{verbatim}
M$ModelName = "Example_Model";

M$Information = {
  Authors      -> {"A. Alloul", "N.D. Christensen", "C. Degrande", 
                   "C. Duhr", "B. Fuks"},
  Institutions -> {"IPHC / U. Strasbourg", "U. Pittsburgh", 
                   "U. Illinois", "ETH Zurich", "CERN"},
  Emails       -> {"adam.alloul@iphc.cnrs.fr", "neilc@pitt.edu", 
                   "cdegrand@illinois.edu", 
                   "duhrc@itp.phys.ethz.ch", "fuks@cern.ch"},
  Date         -> "April 1st, 2013",
  References   -> {"The FeynRules Manual"}
};
\end{verbatim}

\subsection{Preparation of the model file -- index declarations}
Although the model information is optional and may or may not be included in the front matter of a model file, the declaration of all types of indices that appear in the model is mandatory. In our model, there is only one type of index, the index labeling the scalar field $\phi_i$, ranging from 1 to 2. It is therefore sufficient to include, at the beginning of the model file,
\begin{verbatim}
IndexRange[ Index[Scalar] ] = Range[2];
IndexStyle[ Scalar, i];
\end{verbatim}
The first line defines an index of type {\tt Scalar} that takes values in the range \verb+{1,2}+. Note that the name {\tt Scalar} for the index can be chosen freely as long as it does not conflict with existing symbols used in the same \mathematica\ session. The second line instructs \feynrules\ that indices of type scalar should be printed as symbols starting by the letter $i$.

\subsection{Declaration of the objects -- parameters}
We now describe the declaration of all the parameters that appear inside the model. The Lagrangian of Eq.~\eqref{eq:Lscal} depends on six free real  parameters -- three for each real symmetric matrix. In addition, in Eqs.~\eqref{eq:eigenvalues} and \eqref{eq:angle}, we have defined the eigenvalues and the mixing angle as functions of the three mass parameters appearing in the Lagrangian. We therefore need to define nine different parameters in the \feynrules\ model file, out of which only six are independent. Note that there is a freedom in how one chooses the independent parameters. Here we follow the convention that the independent parameters are those that appear in the Lagrangian of Eq.~\eqref{eq:Lscal}.

We start by implementing the independent, or external, parameters. This can be done as follows,
\begin{verbatim}
M$Parameters = {
   
   lam == {
        ParameterType    -> External,
        ComplexParameter -> False,
        Indices          -> {Index[Scalar], Index[Scalar]},
        Value            -> {lam[1,1] -> 0.9,
                             lam[1,2] -> 0.1,
                             lam[2,1] -> 0.1,
                             lam[2,2] -> 0.9},
        Description  -> "Scalar quartic coupling matrix"
        },
        
    MM == {
        ParameterType    -> External,
        ComplexParameter -> False,
        Indices          -> {Index[Scalar], Index[Scalar]},
        Value            -> {MM[1,1] -> 100^2,
                             MM[1,2] ->  10^2/2,
                             MM[2,1] ->  10^2/2,
                             MM[2,2] ->  200^2},
        Description -> "Mass matrix"
        }
  };
  \end{verbatim}
Let us make some comments about the declaration of the external parameters. First, the matrices $\mathcal{M}$ and $\lambda$ are implemented as parameters {\tt MM} and {\tt lam}, each of them carrying two indices of type {\tt Scalar}. The values of the components of the matrix must be given one by one. Finally, the option {\tt Description}, whose value is a string describing the parameter, is purely optional and could have been omitted.

The implementation of the internal parameters is similar to the case of the external ones, and we append to \verb+M$Parameters+ the following elements,
\begin{verbatim}
M1 == {
   ParameterType    -> Internal,
   ComplexParameter -> False,
   Value            -> Sqrt[1/2 (MM[1, 1] + MM[2, 2] - 
                       Sqrt[(MM[1,1]-MM[2,2])^2+4 MM[1,2]^2])],
   Description      -> "Small mass eigenvalue"
   },
   
M2 == {
   ParameterType    -> Internal,
   ComplexParameter -> False,
   Value            -> Sqrt[1/2 (MM[1, 1] + MM[2, 2] +
                       Sqrt[(MM[1,1]-MM[2,2])^2+4 MM[1,2]^2])],
   Description      -> "Large mass eigenvalue"
   },
   
sinth == {
   ParameterType    -> Internal,
   ComplexParameter -> False,
   Value            -> 2 MM[1,2]/Sqrt[4 MM[1,2]^2 + (MM[1,1]-
        MM[2,2] + Sqrt[(MM[1,1]-MM[2,2])^2+ 4 MM[1,2]^2])^2],
   Description      -> "Sine of the mixing angle"
   }
\end{verbatim}   
These declarations are exactly the same as for the external parameters, except that the {\tt Value} options now refer to the algebraic relations of 
Eqs.~\eqref{eq:eigenvalues} and~\eqref{eq:angle}. Finally, we also need to 
implement the rotation matrix $U$ by appending to \verb+M$Parameters+,
\begin{verbatim}
UU == {
   ParameterType    -> Internal,
   ComplexParameter -> False,
   Indices          -> {Index[Scalar], Index[Scalar]},
   Value            -> {UU[1,1] -> -sinth,
                        UU[1,2] -> Sqrt[1-sinth^2],
                        UU[2,1] -> Sqrt[1-sinth^2],
                        UU[2,2] -> sinth},
    Description     -> "Mixing matrix"
    }
\end{verbatim}

\subsection{Declaration of the objects -- fields}
Next we turn to the declaration of the fields that appear in the Lagrangian. We need to declare separately the gauge eigenstates $\phi_i$ and the mass eigenstates $\Phi_i$. Then, we define a replacement rule that instructs \feynrules\ how to perform the rotation from the gauge to the mass basis.

We start by defining the mass eigenstates by including in 
\verb+M$ClassesDescription+,
\begin{verbatim}
S[1] == {
    ClassName     -> Phi,
    ClassMembers  -> {Phi1, Phi2},
    SelfConjugate -> False,
    Indices       -> {Index[Scalar]},
    FlavorIndex   -> Scalar,
    Mass          -> {{M1, Internal}, {M2, Internal}}
    }
\end{verbatim}
This defines a scalar field {\tt Phi} carrying an index of type {\tt Scalar}. 
The symbol {\tt Phibar} for the corresponding conjugate field is automatically 
defined. The {\tt FlavorIndex} option specifies that indices of the type 
{\tt Scalar} label the elements of the list {\tt ClassMembers}\footnote{This 
information seems redundant at this stage. It is however mandatory because a field
may carry more than a single index (see Section \ref{sec:example-gauge}). 
In that case, the {\tt FlavorIndex} option 
singles out the one type of index in the {\tt Indices} list which labels the 
{\tt ClassMembers}.}. Since the masses of the different class member have
already been defined previously as internal parameters, the values of the masses 
of the class members have been set to {\tt Internal}.

Next, the declaration of the gauge eigenstates is performed as follows
\begin{verbatim}
S[2] == {
    ClassName     -> phi,
    ClassMembers  -> {phi1, phi2},
    SelfConjugate -> False,
    Indices       -> {Index[Scalar]},
    FlavorIndex   -> Scalar,
    Unphysical    -> True,
    Definitions   -> {phi[i_] :> Module[{j}, UU[i,j] Phi[j]]}
    }
\end{verbatim}
Compared to the mass eigenstates, the {\tt Mass} option has been replaced by a 
pair of options. First, the option \verb+Unphysical -> True+ identifies gauge 
eigenstates. It has no other effect than to instruct \feynrules\ not to output this field to any Feynman diagram calculators (which in general work at the level of mass eigenstates). Secondly, we use the {\tt Definitions} option to define a replacement rule that rotates the gauge basis to the mass basis. This teaches \feynrules\ to replace every occurrence of \verb+phi[i_]+, where \verb+i_+ is a \mathematica\ pattern representing some arbitrary index, by the expression \verb+UU[i,j] Phi[j]+. Note the appearance of the environment {\tt Module} in the right-hand side of the replacement rule. A {\tt Module} is an internal \mathematica\ command that takes two arguments
\begin{enumerate}
\item a list of symbols \verb+{a,b,c,...}+,
\item an expression {\tt X}.
\end{enumerate}
The {\tt Module} evaluates {\tt X} and replaces every occurrence of the symbols in the list \verb+{a,b,c,...}+ by new symbols \verb+a$+$n$, \verb+b$+$n$, \verb+c$+$n$, \ldots, where $n$ is an integer such that the resulting new symbols are unique. In this way, it is ensured that the contracted index {\tt j} introduced each
time the module is called is unique and does not conflict with any other symbol of the same name.

\subsection{The Lagrangian and the Feynman rules}
After we have defined the parameters and the fields of the model, we can write 
down its Lagrangian.  It is sufficient to translate Eq.~\eqref{eq:Lscal} in terms 
of the symbols introduced in the previous subsections,
\begin{verbatim}
Lscal = del[phibar[i], mu] del[phi[i], mu] - 
        phibar[i] MM[i,j] phi[j] + 
        (phibar[i1] lam[i1, i2] phi[i2]) * 
        (phibar[j1] lam[j1, j2] phi[j2])
\end{verbatim}
Let us note that we cannot write \verb+(phibar[i1] lam[i1, i2] phi[i2])^2+ because this expression is ambiguous,
\beq
(\phi_i\lambda_{ij}\phi_j)^2 \neq \phi_i^2\,\lambda_{ij}^2\phi_j^2\,.
\eeq

The first thing we should do is check that our mass matrix diagonalization worked properly.  After loading
the \feynrules\ package and the model, as presented in Sections \ref{sec:loadFR}
and \ref{sec:loadMF}, we issue the command
\begin{verbatim}
CheckMassSpectrum[ Lscal ]
\end{verbatim}
If everything was done properly, \feynrules\ will state that all mass terms are diagonal and will check their numerical values against the numerical values for \verb+M1+ and \verb+M2+, which were set as the masses of \verb+Phi1+ and \verb+Phi2+, respectively.

We have now all the ingredients to compute the Feynman rules of the model. Loading
the \feynrules\ package and the model, as presented in Sections \ref{sec:loadFR}
and \ref{sec:loadMF}, issuing then the command
\begin{verbatim}
FeynmanRules[ Lscal ]
\end{verbatim}
returns the vertex (all particles are considered in-going)
\begin{equation*}
\Phi_{i_1}, \Phi_{i_2}, \Phi_{i_3}^\dagger, \Phi_{i_4}^\dagger:\quad
i\lambda_{jk}\lambda_{lm}\left(U_{ki_1}\,U_{mi_2}+U_{mi_1}\,U_{ki_2}\right)\left(U_{ji_3}\,U_{li_4}+U_{ji_4}\,U_{li_3}\right)\,.
\end{equation*}
By default the vertex is outputted for particle classes, \ie, all indices labeling the members are symbolic. It is possible to obtain the Feynman rules for the individual class members by applying the function {\tt FlavorExpansion[]} to the output of {\tt FeynmanRules[]}, or by typing 
\begin{verbatim}
FeynmanRules[ Lscal, FlavorExpand -> True ]
\end{verbatim}

\subsection{Extending the model  --  Gauge interactions}\label{sec:example-gauge}
So far our model does not contain any gauge interactions. In the following, we 
describe how we can easily extend the model to include such interactions. To be more precise, we assume that the scalar fields $\phi_i$ transform in the fundamental representation of some $SU(3)$ gauge group. The Lagrangian of this model is then extended from Eq.~\eqref{eq:Lscal} by including the kinetic term for the gauge boson, denoted by $G_\mu^a$, and by replacing the ordinary space-time derivative by the covariant derivative
\beq
D_\mu = \partial_\mu -i\,g_s\,T^a\,G_\mu^a\,,
\eeq
where $g_s$ is the gauge coupling and $T^a$ denotes the generators of the fundamental representation of $SU(3)$. The BRST-invariant Lagrangian then reads, in Feynman gauge,
\beq\bsp
\mathcal{L} = &\,-\frac{1}{4}F_{\mu\nu}^aF^{\mu\nu}_a +\partial_\mu\bar{c}^aD^\mu c^a - \frac{1}{2}(\partial^\mu G_\mu^a)^2 \\
&\,+D_\mu\phi^\dagger_{ik}D^\mu\phi_{ik} - \phi^\dagger_{ik}\mathcal{M}_{ij}\phi_{jk} + (\phi^\dagger_{ik}\lambda_{ij}\phi_{jk})^2 
\,,
\esp\eeq
where $k$ denotes the $SU(3)$ fundamental index carried by the field, and where 
$c$ is the ghost field associated with the gauge boson. We now illustrate that 
gauging a model in \feynrules\ is not much more complicated than the procedure 
sketched above.

We start by defining new indices which label the fundamental and adjoint representations of the gauge group, at the beginning of the model file, 
\begin{verbatim}
IndexRange[ Index[Colour] ] = Range[3];
IndexStyle[ Colour, k ];

IndexRange[ Index[Gluon] ] = Range[8];
IndexStyle[ Gluon, a ];
\end{verbatim}
where {\tt Gluon} and {\tt Colour} represent adjoint and fundamental $SU(3)$ indices, respectively.
We define a new external parameter {\tt gs} in \verb+M$Parameters+ corresponding to the gauge coupling,
\begin{verbatim}
gs == {
   ParameterType    -> External,
   ComplexParameter -> False,
   Value            -> 1.22
   }
\end{verbatim}
Next we turn to the declaration of the fields. First we need to extend the definition of the scalar field to include a gauge index of type {\tt Colour} in the list
of indices carried by the field. In other words, we replace the right-hand side of the option {\tt Indices} by {\tt \{Index[Scalar], Index[Colour]\}}. We also need to extend the {\tt Definitions} option to include the second field index,
\begin{verbatim}
Definition -> {phi[i_,n_] :> Module[{j}, U[i,j] Phi[j,n]]}
\end{verbatim}
Finally, we also need to include the declarations of the gauge boson field $G$ and the ghost field $c$ in \verb+M$ClassesDescription+,
\begin{verbatim}
V[1] == {
    ClassName     -> G,
    Indices       -> {Index[Gluon]},
    SelfConjugate -> True,
    Mass          -> 0
    },
    
 U[1] == { 
    ClassName       -> ghG, 
    SelfConjugate   -> False,
    Indices         -> {Index[Gluon]},
    Ghost           -> G,
    QuantumNumbers  -> {GhostNumber -> 1}, 
    Mass            -> 0
  }
\end{verbatim}

At this stage we have defined the indices, the parameters and the particles relating to the gauge group. 
In addition, we also need to introduce group theory objects such as structure constants and representation matrices. This is achieved by declaring a new instance of the gauge group class in \verb+M$GaugeGroups+,
\begin{verbatim}
M$GaugeGroups = {

SU3C == {
   Abelian           -> False,
   GaugeBoson        -> G,
   StructureConstant -> f,
   Representations   -> {T, Colour},
   CouplingConstant  -> gs
  }  
  
};
\end{verbatim}
Once the gauge group is declared, we can simply write down the Lagrangian using the functions {\tt FS} and {\tt DC} for the gauge field strength
tensors and the covariant derivatives,
\begin{verbatim}
L = -1/4 FS[G, mu, nu, a] FS[G, mu, nu, a] + 
    del[ghGbar[a], mu] DC[ghG[a], mu] -
    1/2 (del[G[mu, a], mu]) (del[G[nu, a], nu])+
    DC[phibar[i,k], mu] DC[phi[i,k], mu] - 
    phibar[i,k] MM[i,j] phi[j,k] + 
    (phibar[i1,k] lam[i1, i2] phi[i2,k]) *
    (phibar[j1,l] lam[j1, j2] phi[j2,l])
\end{verbatim}
Issuing the command {\tt FeynmanRules[ L ]} now returns all the interaction vertices of the model, including the gauge interactions of the gluon field and the scalar fields.

\subsection{Implementing the mixing declaration}
Another possible way to implement this model into {\feynrules} is to let it handle automatically the mixings among the gauge eigenstates. In order to do so, one should remove the option \texttt{Definitions} from the declaration of the scalar field \texttt{S[2]} and, instead, provide the list \texttt{M\$MixingsDescription} with the right options. This alternative allows {\feynrules} to extract automatically the mass matrix $\mathcal{M}$ and further export it to the {\asperge} package for a numerical diagonalization. In practice the \texttt{M\$ClassesDescription} should only comprise the following two declarations
\begin{verbatim}
	S[1] == {
	    ClassName     -> Phi,
	    ClassMembers  -> {Phi1, Phi2},
	    SelfConjugate -> False,
	    Indices       -> {Index[Scalar]},
	    FlavorIndex   -> Scalar,
	    Mass          -> {{M1, External}, {M2, External}},
	    PDG           -> {9000001, 9000002}
	    };
	S[2] == {
	    ClassName     -> phi,
	    ClassMembers  -> {phi1, phi2},
	    SelfConjugate -> False,
	    Indices       -> {Index[Scalar]},
	    FlavorIndex   -> Scalar,
	    Unphysical    -> True
	    }
\end{verbatim}
while the list \texttt{M\$MixingsDescription} should be created with the following lines
\begin{verbatim}
	Mix["sca"] == { 
	  GaugeBasis -> {phi[1], phi[2]},
	  MassBasis -> {Phi[1], Phi[2]},
	  MixingMatrix -> UU,
	  BlockName -> UMIX,
	  Inverse -> True}
\end{verbatim}
In the above lines, we have set the masses \texttt{M1} and \texttt{M2} as external as they are to be calculated by the {\asperge} code and added a PDG code for each of the mass eigenstates. We gave this mixing the label \texttt{"sca"}, the mixing matrix \texttt{UU} and the SLHA-block \texttt{UMIX}. To be compliant with both {\feynrules} conventions and equation \eqref{mixing}, that is 
\beq\label{mixing}
\phi_i = U_{ij}\,\Phi_j\ ,
\eeq
we have added the option \texttt{Inverse} with the attribute \texttt{True}. 

In the parameters declaration, one should remove the entries \texttt{M1, M2, sinth} and \texttt{UU} as they correspond to quantities that will be calculated by the {\asperge} code. Finally, as none of the scalar fields introduced here acquires a vacuum expectation value, one may skip the declaration of the corresponding variable \texttt{M\$vevs}. \\

Now that the declaration of the model fits the requirements of the mass diagonalization package, one can issue the commands 
\begin{verbatim}
	WriteASperGe[L];
	RunASperGe[];
\end{verbatim}
in order to generate the source code for {\asperge}, diagonalize the mass matrix \texttt{UU} and update the values for \texttt{M1}, \texttt{M2} and all the components of the mixing matrix \texttt{UU}.


\section{Interfaces}
\label{sec:interfaces}
So far we have only discussed how to implement a model into \feynrules\ and how to
compute the associated interaction vertices. After the Feynman rules have been obtained, the user is typically interested in the phenomenology of the model, which requires the evaluation of Feynman diagrams. There are many tools that allow to evaluate Feynman diagrams automatically. 
While these tools are in general very flexible and allow the user the generate, at least in principle, Feynman diagrams for any model satisfying basic quantum field theory requirements, most of the tools only have a limited number of new physics models implemented. Implementing a new model into any of these tools generally requires the user to implement the interaction vertices one at the time. In addition, each Feynman diagram generator has its own format, making the task of implementing a new model into a Feynman diagram generator tedious and error-prone.

\feynrules\ is equipped with interfaces to various Feynman diagram generators which allow to export the Feynman rules to any of these tools (for which an interface exists) in the form of a set of text files specific to each code. Currently the following interfaces exist,
\begin{itemize}
\item[-] \calchep/\comphep,
\item[-] \feynarts/\formcalc,
\item[-] \sherpa,
\item[-] UFO (Universal \feynrules\ output),
\item[-] \whizard/{\sc Omega}.
\end{itemize}
The UFO format is a generic model format that stores model information in an 
abstract way in the form of {\sc Python}\ objects~\cite{Degrande:2011ua}. More information is provided
in Section~\ref{sec:ufo}.

All interfaces are invoked with commands like

{\tt Write}X{\tt Output[}$\mathcal{L}_1$,$\mathcal{L}_2$,\ldots, options{\tt]}

where X is replaced with a particular label referring to the interface name. 
The main limitations of the interfaces rely in the fact that they can only be used to implement particles and vertices which are natively supported by the corresponding Feynman diagram generator. As an example, each generator makes implicit assumptions on the spins or color representations of the particles present in the model, and/or on the form or the dimension of the interaction vertices. As a consequence, only those models that are compliant with all those constraints are supported by a given tool. The \feynrules\ interfaces are tailored to the different Feynman diagram generators, and they therefore allow to check on a case by case basis whether or not a given model is compliant with a given tool. For example, if an interface detects that a given interaction vertex is not compliant, the vertex is discarded and a warning is printed on the screen. The user then has the possibility to switch to a different Feynman diagram generator for which the restriction is not present by using a different interface without having to change his/her \feynrules\ implementation.

In this section we give a brief account of how to run the \feynrules\ interfaces and we discuss their features and limitations. While some of the interfaces have been the subject of separate publications \cite{Degrande:2011ua,Christensen:2010wz}, we include a summary on how to run all the interfaces for the sake of completeness.

\subsection{Conventions}
While \feynrules\ itself is completely agnostic of the underlying model and does not make any \textit{a priori} assumptions on the form of the particles and parameters that are defined in the model file, Feynman diagrams generators usually have some information, in particular about the Standard Model, hard-coded. For example, most Feynman diagram generators have the running of the strong and/or weak coupling constants implemented, which allows one to use the value of $\alpha_s$ specific to a given process. In order for this to work, the value of $\alpha_s$ must be stored in a variable whose name is hard-coded into the code.
Similarly, many programs have color implemented in an implicit way and/or perform the color algebra internally. 
As a consequence, the \feynrules\ interfaces must communicate information about the Standard Model parameters and gauge groups to the Feynman diagram generators in a specific format. For this to work properly, the user has to follow a set of conventions when writing a model file which are detailed in the rest of this section.

\subsubsection{Name restrictions}\label{sec:namerestr}
While \mathematica, and hence \feynrules, are case sensitive, many Feynman diagram calculators are not. Consequently, the model builders should avoid to use names that only differ by case. 
In addition, \feynrules\ allows the use of Greek letters to define the names of parameters and fields, which cannot be exported to the Feynman diagram calculators.
However, the user can change the name of a parameter or field to be outputted by 
the interfaces by using the parameter option \verb+ParameterName+. This allows to 
output a name matching the requirements of the Feynman diagram calculators. 
Similar options, called {\tt ParticleName} and {\tt AntiParticleName}, are available
for the particle classes. They allow the user to specify the string (or list of 
strings) that should be used in the external programs. Finally, the \TeX-form 
associated with these names can also be specified via the options 
{\tt TeXParticleName} and {\tt TeXAntiParticleName} of the particle class and 
{\tt TeX} of the parameter class.
An example follows:
\begin{verbatim}
ParticleName     -> {"ne", "nm", "nt"},
AntiParticleName -> {"ne~", "nm~", "nt~"}
\end{verbatim}
A more detailed name can also be attached, as a string or list of strings, to any 
particle, by means of the attribute {\tt FullName} of the particle class, such as 
in
\begin{verbatim}
FullName -> "Photon"
\end{verbatim}

\subsubsection{Particle Data Group numbering scheme (PDG) codes}\label{sec:PDG}
Many programs use the Particle Data Group numbering scheme~\cite{Beringer:2012zz} to refer to the particles inside the code. In addition, many codes have information on the particles and their quantum numbers hard-coded. It is therefore crucial that new models respect the PDG numbering scheme whenever possible and we strongly encourage the users to use the existing PDG codes.

 \feynrules\ allows the user to assign PDG codes to all the particles in the model file. For example, the PDG codes for the up-type quarks can be defined by adding the following option to the up-type particle class (which has the $u$, $c$ and $t$ quarks as members),
\begin{verbatim}
PDG -> {2,4,6}
\end{verbatim}
If a particle class has only one member, the curly brackets on the right-hand side may be omitted.
If a user does not define a PDG code for a particle, then \feynrules\ automatically assigns to it a PDG code starting from 9000001.

\subsubsection{Decay widths}
When computing Feynman diagrams it is in general important, already at tree-level, to include the total width into the propagator of a massive particle in order to avoid singularities in the phase space integration. For this reason most Monte Carlo tools take the total width of a particle as a numerical input for tree-level computations. \feynrules\ allows the user to include the numerical value for the width into the model file via the option {\tt Width} (For more details see Section~\ref{sec:parts}). If done in this way, the numerical values for the widths of all the particles are transmitted to the Monte Carlo codes.

The widths of all the particles are, however, in general unknown at the moment of the implementation in \feynrules, because the evaluation of the (tree-level) width of a particle requires the evaluation of Feynman diagrams. There are four different ways the user can include the widths for the Feynman diagram calculators:
\begin{enumerate}
\item He/she may use the function {\tt TotWidth[}particle{\tt ]} introduced in Section~\ref{sec:decays} to compute the total two-body decay width of particle.
\item He/she may export the model to a Feynman diagram calculator via one of the interfaces without including the widths of the particles\footnote{If no numerical value is given in the model, the widths are automatically set to a default value of 1~GeV.} and then use the tool itself to compute all particle partial widths in all possible decay channels. He/she can then update the model file with the corresponding numerical values.
\item Some Feynman diagram calculators offer the possibility to evaluate the widths of the particles on the fly without the need of specifying their value in the model file. For more information we refer to the documentation of the various tools.
\item There are dedicated tools that compute the widths and the branching fractions for specific new physics models (\eg, the Minimal Supersymmetric Standard Model).
If such a tool exists for the model under consideration, the user may simply want to update his model by reading in the output files of any of these codes.
\end{enumerate}
In all cases we want to stress that the branching fractions and the widths are highly benchmark-dependent and need to be reevaluated every time the numerical value of an external parameter is changed.

\subsubsection{Parameter input}\label{sec:paramin}
In Section~\ref{sec:parameters} we have shown how to assign a numerical value 
to an external parameter using the {\tt Value} option. These values are then 
transmitted to the Feynman diagram calculators when running the \feynrules\ 
interfaces. In practice, one often wants to scan over the parameter space 
of a new model. It is however highly inefficient to rerun \mathematica\ every time the user wants to change the numerical value of some external parameter. Rather, it is more convenient to update the numerical parameters at runtime in the Feynman diagram calculator. 

Many Monte Carlo tools rely on a Les Houches (LH)-like format to input the numerical values of the external parameters. This format is inspired by the Supersymmetry-Les Houches-Accord (SLHA)~\cite{Skands:2003cj,Allanach:2008qq} which defines a standard format for the numerical parameters in the Minimal Supersymmetric Standard Model and certain extensions thereof. In this format numerical values are grouped into certain blocks, and each parameter is identified inside its own block by one or more integer numbers, called
counters.

\feynrules\ internally stores the numerical values of all external parameters in a form which closely follows the SLHA format. In particular, each external parameter is assigned to a block and counter, which can be specified using the {\tt BlockName} and {\tt OrderBlock} options of the parameter class. For example, the definition of the strong coupling constant can be assigned to the third entry of the block named {\tt SMINPUTS} by adding the following options to the definition of the parameter
\begin{verbatim}
BlockName        -> SMINPUTS,
OrderBlock       -> 3,
\end{verbatim}
The value of {\tt BlockName} can be an arbitrary \mathematica\ symbol, while the value of {\tt OrderBlock} can be either an integer or a list of integers. The {\tt BlockName} and {\tt OrderBlock} attributes are optional. If omitted, the parameter 
are assigned automatically to the block {\tt FRBlock}. Moreover, masses and widths are automatically assigned to the blocks {\tt MASS} and {\tt DECAY}, the identifier inside the block being the PDG code of the particle.
By convention, the entries of a matrix defines a block on its own, the identifiers inside the block being the position inside the matrix. 

By convention, all the parameters defined inside a LH-like format are real. Complex external parameters have then to be implemented by splitting them into their real and imaginary parts. As an illustration, we consider a complex external parameter $a = a_R+ia_I$, where $a_R$ and $a_I$ are real. This parameter is thus implemented into the LH-like format by defining $a$ as internal and $a_R$ and $a_I$ as external.

\subsubsection{Definition of Standard Model parameters and gauge groups}\label{sec:conventionsSMgauge}
The parameters and the gauge groups of the Standard Model have a special significance in most Monte Carlo codes. In particular they must be implemented using certain conventions and giving specific names to certain variables, in order to ensure that, \eg, the strong and/or electroweak couplings are run correctly and the color algebra is performed correctly. As a consequence, despite the fact that \feynrules\ is generic and does not distinguish between the Standard Model and non-Standard Model parameters when computing the Feynman rules, the interfaces have an explicit dependence on them. The Standard Model input parameters and gauge groups must therefore be implemented into a \feynrules\ model file following certain conventions.

First, the strong coupling constant $g_s$ and its square over $4\pi$ should be declared as in the following example,
\begin{verbatim}
aS    == { 
   ParameterType    -> External,
   BlockName        -> SMINPUTS, 
   OrderBlock       -> 3,
   Value            -> 0.1184, 
   InteractionOrder -> {QCD,2},
   Description      -> "Strong coupling constant at the Z pole"
  },

gs == { 
   ParameterType    -> Internal, 
   Value            -> Sqrt[4 Pi aS],
   InteractionOrder -> {QCD,1},  
   ParameterName    -> G,
   Description      -> "Strong coupling constant at the Z pole"
  },
\end{verbatim}
We note that $\alpha_s$ is implemented as an external parameter while $g_s$ is defined as an internal parameter derived from $\alpha_s$. Similarly, the electroweak coupling is defined as follows,
\begin{verbatim}
aEWM1 == { 
   ParameterType    -> External, 
   BlockName        -> SMINPUTS, 
   OrderBlock       -> 1, 
   Value            -> 127.9,
   InteractionOrder -> {QED,-2},
   Description      -> "Inverse of the EW coupling constant at the Z pole"
  },

Gf == {
   ParameterType    -> External,
   BlockName        -> SMINPUTS,
   OrderBlock       -> 2,
   Value            -> 1.16637*^-5, 
   InteractionOrder -> {QED,2},
   Description      -> "Fermi constant"
  },
  
aEW == {
   ParameterType    -> Internal,
   Value            -> 1/aEWM1,
   InteractionOrder -> {QED,2},
   Description      -> "Electroweak coupling constant"
  },
  
ee == { 
   ParameterType    -> Internal, 
   Value            -> Sqrt[4 Pi aEW], 
   InteractionOrder -> {QED,1}, 
   Description      -> "Electric coupling constant"
  }
\end{verbatim}
The external parameters in the electroweak sector are the inverse of the electromagnetic coupling at the $Z$ scale, $\alpha_{EW}(M_Z)^{-1}$ and the Fermi constant $G_F$. The reason for choosing $\alpha_{EW}(M_Z)^{-1}$ as the external input parameter, and not $\alpha_{EW}(M_Z)$ itself is only to be compliant with the SLHA. The electromagnetic coupling at the $Z$ scale and the electric charge $e$ are declared as internal parameters. Finally, it is also encouraged to choose the mass of the $Z$ boson as an external parameter, still following the SLHA conventions.

The QCD gauge group has a special status in \feynrules\ as it defines quantities that have to be dealt with in a special way by the interfaces. This ensures, \eg, that the color algebra is correctly performed by the Feynman diagram calculators. The correct definition of the QCD gauge groups is 
\begin{verbatim}
SU3C == { 
    Abelian           -> False, 
    CouplingConstant  -> gs, 
    GaugeBoson        -> G,
    StructureConstant -> f, 
    Representations   -> {{T, Colour}, 
                         {T6, Sextet}},
    SymmetricTensor   -> dSUN
  } 
  \end{verbatim}
This definition requires the gluon field, defined in \verb+M$ClassesDescription+, to be called {\tt G}. The gluon field carries an index of type {\tt Gluon}, which represents the adjoint representation of QCD. So far, there is, at least to our knowledge, no Monte Carlo code for which an interface exists that can deal with representation other than ${\bf 1}$, ${\bf 3}$, ${\bf 6}$ and ${\bf 8}$ (and their complex conjugate representations, if applicable). In the example above, we have defined two representations of the QCD gauge group, the fundamental and sextet representations, which act via the generators {\tt T} and {\tt T6} on indices of type {\tt Colour} and {\tt Sextet}, respectively. A full list of the group theoretical object of QCD that can be used to construct a Lagrangian can be found in Table~\ref{tab:QCD_objects}. Furthermore, the indices labeling the different representations should be defined at the beginning of the model file as in 
\begin{verbatim}
IndexRange[ Index[ Colour ] ] = Range[3];
IndexRange[ Index[ Sextet ] ] = Range[6];
IndexRange[ Index[ Gluon ] ] = Range[8];
\end{verbatim}
The function {\tt NoUnfold[]} might be added in the right-hand side if desired.  This is only used by the 
\feynarts\ interface.

In addition to the QCD charges of the particles, some Monte Carlo programs also use the information on the electric charge of a particle. The electric charge should be defined in the {\tt QuantumNumbers} option of the particle class as {\tt Q} to ensure that the information is correctly transmitted to the matrix element generators by the \feynrules\ interfaces.

\begin{table}
\bgfb
\multicolumn{2}{c}{\textbf{Table~\ref{tab:QCD_objects}: Group-theoretical objects}}\\
 {\tt T[a,i,j]} & Generators of the fundamental representation, $T^a_{ij}$.\\
 {\tt T6[a,m,n]} & Generators of the sextet representation, $T^a_{6,ij}$. \\
 {\tt f[a,b,c]} & $SU(3)$ structure constants, $f^{abc}$.\\
 {\tt dSUN[a,b,c]} & Totally symmetric tensor $d^{abc}$.\\
 {\tt Eps[i,j,k]} & Totally antisymmetric tensor $\epsilon_{ijk}$ connecting three (anti)triplet indices.\\
 {\tt K6[m,i,j]} & Clebsch-Gordan coefficient connecting a sextet and two anti-triplets, considered incoming.\\
  {\tt K6bar[m,i,j]} & Clebsch-Gordan coefficient connecting a anti-sextet and two triplets, considered incoming.
\egfb
\textcolor{white}{\caption{\label{tab:QCD_objects}}}
\end{table}

\subsubsection{Switching between unitary and Feynman gauge}
Some Monte Carlo programs allow the user to generate QCD processes both in unitary and in Feynman gauge. 
Having a model that can run in both gauges can be a powerful way to check that the implementation is gauge invariant by generating matrix elements in both gauges. In addition, some codes run faster in one gauge than in the other.
It is therefore desirable to implement a model both in unitary and Feynman gauge whenever possible. 
At this stage, the user has to implement by hand all the terms in the Lagrangian related to the gauge fixing procedure.
In practice, every model implemented in Feynman gauge can be transformed to unitary gauge by removing all the vertices that involve ghost fields and/or Goldstone bosons. It can therefore be useful to add a Boolean variable \verb+$FeynmanGauge+ into the model, which, if set to {\tt False}, removes all the terms from the Lagrangian which depend on ghost fields and Goldstone bosons. Furthermore, some interfaces use this variable to communicate to the matrix element generator whether or not the model can be run in Feynman gauge. For these reasons, we strongly recommend the use of the variable \verb+$FeynmanGauge+ in each model implementation. For an illustration of how to use it, we recommend to look at the implementation of the Standard Model shipped with this package.

\subsubsection{Interaction orders}\label{sec:intorder}

Perturbative expansions allow one to compute an amplitude at a given order in each of the coupling constants. Some programs use this expansion to order contributions to a process according to their relative importance and eventually only keep the largest one\footnote{Currently, only the UFO interface uses this feature.}. For example, $u\bar u\to t \bar t$ has three tree-level diagrams, corresponding to the gluon, the $Z$ boson and the photon in the $s$-channel. However, the contribution of the electroweak bosons is only a small correction to the leading strong contribution due to the hierarchy between the strong and the electroweak coupling constants. \madgraph\ will only compute the strong contributions by default, \ie, the amplitude proportional to $g_s^2 e^0$  and not the one proportional to $g_s^0 e^2$. The largest allowed power of each coupling can also be specified, which allows to keep the electroweak diagrams as well.
As a consequence, the program needs to know how many powers of $g_s$ and $e$ are present in each vertex to compute the dependence of the diagrams. These powers are computed in \feynrules\ from the analytic expressions of the vertices and the dependence in the coupling constants of each parameter. The latter is given in the parameter description through the \verb+InteractionOrder+ option. For example, \verb+QCD+ is defined as the power of $g_s$. Therefore $\alpha_s$ has 
\begin{verbatim}
InteractionOrder -> {QCD,2}
\end{verbatim}
because it is proportional to $g_s^2$. The second order defined for the Standard Model is \verb+QED+ which counts the powers of $e$. Both electroweak coupling constants $g$ and $g'$ have 
\begin{verbatim}
InteractionOrder -> {QED,1}
\end{verbatim}
while $\alpha_{EM}$ has \verb+InteractionOrder -> {QED,2}+. In addition, model builders can define  new orders for other couplings similarly.

The default behavior of the Monte Carlo generator comes from the knowledge of the hierarchy of couplings. This is defined in the \feynrules\ model as an additional list :
\begin{verbatim}
M$InteractionOrderHierarchy = {
  {QCD,1},
  {QED,2}
}
\end{verbatim}
The numbers associated with the couplings give their relative importance. In the above example, one power of $e$ is considered equivalent to two powers of $g_s$.  If a new vector connects the up and top quarks with a new physics coupling $g_{NP}$, one can possibly set $g_{NP}$ order to \verb+InteractionOrder -> {NP,1}+ and the hierarchy can be defined as
\begin{verbatim}
M$InteractionOrderHierarchy = {
  {QCD,1},
  {NP,1},
  {QED,2}
}
\end{verbatim}
as for the strong interactions to avoid that new physics is removed by default. Finally, the user can also define a maximum value for the power of a coupling to be allowed in each diagram,
\begin{verbatim}
M$InteractionOrderLimit = {
  {NP,2}
}
\end{verbatim}
This is used to indicate to the generator that the largest power $g_{NP}$ appearing in an  amplitude is two. For example, the interaction of the gluon with the Higgs boson through a top loop can be added as an effective operator in the large top mass limit. The production by gluon fusion of a single Higgs boson can be simulated in this way by tree-level event generators. However, this vertex cannot be used twice for double Higgs boson production by gluon fusion because it does not include all the contribution at this order in $\alpha_s$ (box-type diagrams with a top loop are missing).

\subsubsection{Drawing Feynman diagrams}\label{sec:propagators}
Some of the programs interfaced to \feynrules\ allow to draw Feynman diagrams
associated with a given process. The way the propagators are drawn can be
defined at the \feynrules\ level by means of the options 
{\tt Propagator\-La\-bel},
{\tt PropagatorType} and {\tt PropagatorArrow} of the particle class. The first of 
these attributes allows to modify the label attached to an internal line of a 
Feynman diagram and takes a string (or a list of strings if one has several class
members) as an argument. The second attribute refers to the type of line to employ
when drawing the propagator. By default, it is inferred from the spin of the
particle but this can be modified by the user, who can set the argument of 
{\tt PropagatorType} to {\tt ScalarDash} (dashed line), {\tt Sine} (wavy line),
{\tt Straight} (straight line), {\tt GhostDash} (dotted line), or {\tt Curly}
(gluonic line).  Finally, setting the option {\tt PropagatorArrow} to {\tt True} or
{\tt False} allows to put an arrow or not on the propagator.

\subsection{The \asperge\ interface}\label{sec:aspergeinter}

From the information concerning the mixing relations among the different
fields of the model (see Section~\ref{sec:mixdecl}), \feynrules\ is capable of generating a package dubbed
\asperge\ which allows to diagonalize, at tree-level, the
associated mass matrices. The interface between \feynrules\ and \asperge\ can be called
by typing in the \mathematica\ session
\begin{verbatim}
  WriteASperGe[ Lag, Output -> dirname ]
\end{verbatim}
In this expression, the first argument (\ie, the symbol \texttt{Lag}) is mandatory
and refers to the model Lagrangian whereas the second argument (\ie, the replacement
rule \texttt{Output -> dirname}) is optional and indicates the name of the
directory where the \asperge\ files should be created. If this option is not specified,
the directory \texttt{ModelName\textunderscore MD} is used by default, 
\texttt{Mo\-del\-Na\-me} being the name of the \feynrules\ model. 

This interface works in several steps. First, all the mass matrices are
extracted from the Lagrangian. This is achieved through the function 
\texttt{Com\-pu\-te\-Mass\-Ma\-trix} introduced in Section
\ref{sec:asperge}. Second, the interface writes a set of model-independent
files:
\begin{itemize}
 \item the three \cpp\ source and header files
   \texttt{MassMatrix.cpp}, \texttt{Mass\-Ma\-trix.hpp} and \texttt{Matrix.hpp}
   dedicated to matrices, their properties and their diagonalization;
 \item the three \cpp\ source files,  \texttt{Par.cpp}, \texttt{CPar.cpp},
   \texttt{RPar.cpp}, together with the associated header files,
    that contain the definition of the internal format used by \asperge\ with respect
    to the model parameters;
 \item the two source files \texttt{ParSLHA.cpp}, \texttt{SLHA\-Block.cpp}, together
    with the associated header files, which contain the mapping of the internal format
    used by \asperge\ and the SLHA structure used by \feynrules;
  \item the two files \texttt{tools.cpp} and \texttt{tools.hpp} which are dedicated to
    printing and string manipulation routines;
  \item a makefile allowing to compile \asperge.
\end{itemize}
Finally, the interface creates four model-dependent files:
\begin{itemize}
  \item the main \cpp\ program, \texttt{main.cpp}, which starts with the declaration of the
    different mass matrices of the model, then proceeds with their diagonalization and 
    eventually maps the eigenvalues to the PDG codes of the physical fields;
  \item The two parameter files \texttt{Parameters.cpp} and \texttt{Parameters.hpp}, which
    contain the SLHA structure with all the model external parameters
    as implemented in the \feynrules\
    model, followed by the relations linking the external and internal parameters\footnote{For a correct
    running of \asperge, the internal
    parameters cannot depend on the particle masses and mixing matrices.}. The definitions of the mass matrices are also provided in these two files;
  \item the data file \texttt{Ex\-ter\-nals.dat} (stored in the
    subdirectory \texttt{input}) that contains the numerical values of
    the external parameters of the model and that the user can modify according to his/her needs.
\end{itemize}

Technically speaking, matrix diagonalization as performed by \asperge\
is based on a symmetric bi-diagonalization of the mass matrices, followed by their QR reduction.
This strategy follows one of the matrix diagonalization algorithms
implemented in the \gsl\ library, which is used by \asperge.
There is a single condition that must be satisfied in order to employ such an
algorithm: only Hermitian matrices can be diagonalized.
Consequently, \asperge\ will take care of diagonalizing
the matrices $M^{\dagger} M$ and $M M^{\dagger}$ 
when charged fermions are involved, since their mass matrix $M$ is by construction
non-her\-mi\-ti\-an. This allows to obtain left-handed and
right-handed fermion mixing matrices separately, as well as the squared mass eigenvalues.
Consequently, before running the \asperge\ package, the user has to  verify that
the \gsl\ libraries are installed on \hisher\ system,
together with the \texttt{g++} compiler which is employed to compile \asperge.
Otherwise, the makefile generated by the interface must be edited accordingly.

Once compiled, \asperge\ can be executed by typing in a shell
\begin{verbatim}
  ./ASperGe <inputfile> <outputfile>
\end{verbatim}
The two arguments of the function refer to the file containing
the numerical value of the external
parameters (\texttt{<input\-file>}, which could be the file
\texttt{Externals.dat} mentioned above)
and the file which will contain the program
results (\texttt{<outputfile>}). These results consist of the input parameters,
followed by the numerical values of the 
mixing matrices, split in terms of their real and imaginary parts (as imposed by
the SLHA conventions). Concerning Majorana particles, the \asperge\ package
imposes the mixing matrices
to be real, so that some mass eigenvalues can be negative.
Finally, the output file also contains the masses of the physical
eigenstates stored in the SLHA block \texttt{MASS}.

It is also possible to indicate to the interface to create a code allowing for the
diagonalization of a given subset of the mass matrices of the model, instead of
all of them. This is achieved via the {\tt Mix} option, already introduced
in the context of the {\tt ComputeMassMatrix} (see Section~\ref{sec:asperge}) function,
and which works in the same way,
\begin{verbatim} 
  WriteASperGe[ Lag, Mix -> {"l1", "l2"} ]
\end{verbatim}
In order to diagonalize specific mass matrices,
the user can type in a shell
\begin{verbatim}
  ./ASperGe <infile> <outfile> m1 m2 ...
\end{verbatim}
where \texttt{m1}, \texttt{m2}, \etc, are the names of the mixing matrices under consideration.
All the information related to the undiagonalized mass matrices is here ignored by the code.

Finally, for practical reasons, both the program compilation and execution can be directly
performed from the \mathematica\ session. The command to use is
\begin{verbatim}
  RunASperGe[ ]
\end{verbatim}
which stores the program output in a file named \texttt{out.dat}. The content of this
file is then directly loaded into the \mathematica\ kernel in order
to update the numerical value of all the model parameters.

\subsection{The \calchep/\comphep\ interface}\label{sec:calchep}
The \calchep/\comphep\ interface can be invoked via the command
 
 {\tt WriteCHOutput[} $\mathcal{L}_1$, $\mathcal{L}_2$, \ldots, options {\tt]} 
 
 where $\mathcal{L}_1$, $\mathcal{L}_2$, \ldots are the various pieces of the model
Lagrangian and options are the options to be passed to the interface. They
are addressed throughout this section and can be found summarized in Table \ref{fig:CH Interface}.
\begin{table}
\bgfb
\multicolumn{2}{c}{\textbf{Table~\ref{fig:CH Interface}: CalcHEP/CompHEP Interface Options}}\\
{\tt CHAutoWidths} & Whether \calchep\ should calculate the widths on the fly, if 
  set to {\tt True} (the default choice). Otherwise ({\tt False}), the values 
  given in the model implementation are used. When the {\tt CompHEP} option is set 
  to {\tt True}, the default behavior of {\tt CHAutoWidths} is however 
  {\tt False}. \\
{\tt CompHEP} & Allows to write a \calchep\ model file ({\tt False}, by default) or
  a \comphep\ model file ({\tt True}).\\
{\tt ModelNumber} & The number to name the model with, set to 1 by default. For 
  example, the particle file is in this case denoted by {\tt prtcls1.mdl}.\\
{\tt Exclude4Scalars} & In some cases, models can have an abundance of vertices 
  with four scalar fields which are phenomenologically irrelevant. These vertices 
  can be discarded by setting this option to {\tt True}, the default choice being 
{\tt False}.\\
{\tt LHASupport} & If set to {\tt True}, the \calchep{} model reads the external variables from a LHA file.  If {\tt False} (the default), the external parameters are stored in {\tt varsN.mdl}.\\
{\tt Output} & This option is only available for the function {\tt WriteCHExtVars}
  and specifies an alternate file to write the external variables to.  The default 
  choice is {\tt varsN.mdl} in the current directory where {\tt N} refers to the 
  model number (see {\tt ModelNumber}).\\
{\tt Input} & This option is only available for the function {\tt ReadCHExtVars} 
  and specifies an alternate file to read the external variables from. The default 
  choice is {\tt varsN.mdl} in the current directory, where {\tt N} refers to the 
  model number (see {\tt ModelNumber}).
\egfb
\textcolor{white}{\caption{\label{fig:CH Interface}}}
\end{table}

When invoked, this interface first creates a directory named {\tt M\$ModelName} 
with {\tt-CH} appended, if it does not already exist. Then, it creates the files 
{\tt prtclsN.mdl}, {\tt varsN.mdl}, {\tt funcN.mdl} and {\tt lgrngN.mdl} where 
{\tt N} is the number of the model.  It is set to 1 by default but this can be 
modified through the option {\tt ModelNumber}. The particle definitions are written
to the file {\tt prtclsN.mdl}, the external parameters (including external masses 
and widths) to the file {\tt varsN.mdl} and the internal parameters (including 
internal masses and widths) to the file {\tt funcN.mdl}.  
At this 
point, the interface derives the Feynman rules associated with three-point and 
four-point interactions and writes them to {\tt lgrngN.mdl}.  The vertex list is
 simplified by renaming the vertex couplings as {\tt x1}, {\tt x2}, {\tt x3}, \etc,
and the definitions of these new couplings are appended to {\tt funcN.mdl}, along 
with the other internal parameters.  Since \calchep\ only computes internal variables once per session, this improves the speed of the phase space integration.

Although \calchep\ and \comphep\ can calculate diagrams in both Feynman and unitary
gauge, they are faster when Feynman gauge is adopted. It is therefore highly 
recommended to implement a new model in Feynman gauge.  However, if a user decides 
to implement the model in unitary gauge, he/she should remember that according to 
the way \calchep\ and \comphep\ have been implemented, the ghosts associated 
with massless non-abelian gauge bosons must be implemented. In particular, 
gluonic ghost fields must always be implemented in either gauge.

One major constraint of the \calchep/\comphep\ system is that the color structure is implicit.  For many vertices (\eg, quark-quark-gluon), this is not a problem.  However, for more complicated vertices, there may be an ambiguity.  For this reason, the developers of \calchep/\comphep\ have chosen to split them up using auxiliary fields.  Although this can be done for very general vertices, it is not yet fully supported in \feynrules.  Currently, only the gluon four-point vertex and squark-squark-gluon-gluon vertices are automatically split up in this way.  Support for more general vertices is expected in the future.

The model files are ready to be used and can be directly copied to the \calchep/\comphep\ {\tt models} directories.  Care should be taken not to overwrite model files that are already present.  Furthermore, \calchep/\comphep\ only load the models whose model numbers are in a consecutive set beginning with 1.  Therefore, for example, if the {\tt models} directory contains models 1, 2 and 3, the user should number his/her new model 4.  Alternatively, in \calchep, the user can import the model by specifying the directory where it is stored by using the command {\tt IMPORT OF MODELS} in the \calchep\ graphical user interface.

The default format for this interface is the \calchep\ format.  A user can direct this interface to write the files in the \comphep\ format by use of the {\tt CompHEP} option. One subtlety should be mentioned here.  If the model is written to the \comphep\ directory and if the user edits the model inside \comphep\ and tries to save it, \comphep\ will complain about {\tt C} math library functions appearing in the model.  Nevertheless, it does understand them. If a model correctly works in \calchep\, it also works in \comphep\ and leads to the same physics results.

\calchep\ has the ability to calculate the widths of the particles on the fly.  By default, the \calchep\ interface writes model files configured for automatic width
 computations, regardless of the numerical values provided in the \feynrules\  model 
file.  This can be turned off by setting the option {\tt CHAutoWidths} to False.  This option is set to False if {\tt CompHEP} is set to True.  The user can also fine-tune it afterwards by setting some widths to be automatic and others to be fixed in the \calchep{} model files.

In some cases, it is preferable to use the LHA format for the external parameters rather than \calchep's {\tt varsN.mdl}.  For this reason, this interface has the option {\tt LHASupport} which, when set to {\tt True}, causes the \calchep{} model files to read the values of the external parameters from an LHA file.  If set to {\tt False} (the default), the external variables are read from {\tt varsN.mdl} as usual for \calchep.  More information about the \calchep{} support for LHA parameter files can be found in \cite{Belanger:2010st}.

Some models, such as the most general minimal supersymmetric extension of the Standard Model have so many four-scalar vertices that it takes a very long time for \feynrules\ to calculate all of the Feynman rules. In many cases, these vertices are not
relevant for the physics being studied and can be discarded by setting the option {\tt Exclude4Scalars} of the interface to {\tt True}, the default behavior being {\tt False}. 

The \calchep\ interface also contains a set of functions that read and write the external parameters from and to the \calchep\ variable file  {\tt varsN.mdl}.  After loading the model into \feynrules, the external parameters can be updated from
a \calchep\ model by executing the function {\tt ReadCHExtVars[} options {\tt]}.  This function accepts all the options of the \calchep\ interface, plus the 
option {\tt Input} which instructs \feynrules\ where to find the \calchep\ 
variable file.  The default is a file named {\tt varsN.mdl}, located in the current
working directory.  In the case \comphep\ variable file is read, the option 
{\tt CompHEP} should be set to {\tt True}.

The current values of the external parameters in \feynrules\ can also be written to a \calchep\ external variable file {\tt varsN.mdl} using {\tt WriteCHExtVars[} options {\tt]}.  This allows to bypass writing out the entire model if only the model parameters are changed.

Neither \calchep{} nor this interface support {\tt InteractionOrder} defined in Sec.~\ref{sec:intorder}.  However, the strong coupling constant $g_s$ must be used explicitly in the Lagrangian.  This interface will factor it out of each vertex and put it in the {\tt Factor} column of {\tt lgrngN.mdl}.  The electric charge {\tt ee} should also be used explicitly.  If its power is the same for each term of a vertex, it will also be factored out and put in the {\tt Factor} column.

Further details of how \calchep{} uses these model files can be found in Ref.~\cite{Belyaev:2012qa}.

\subsection{The \feynarts\ interface} \label{sec:feynarts}

The \feynarts\ model files are generated from a \feynrules\ model implementation
by issuing the command:

{\tt WriteFeynArtsOutput[} $\mathcal{L}_1$, $\mathcal{L}_2$, \ldots, options {\tt]}

where $\mathcal{L}_1$, $\mathcal{L}_2$, \ldots\ are the pieces of the model 
Lagrangian and options are options among those listed in Table 
\ref{fig:FA Interface}.  This function creates a directory denoted by 
\verb+M$ModelName_FA+ with three files. Each of these files is named as 
\verb+M$ModelName+, followed
with a different extension among {\tt .gen},  {\tt .mod} and {\tt .pars} appended.
Another filename root may be specified with the \verb+Output+ option of the interface. 
This directory can then be moved directly to the \feynarts\ model directory and used in \feynarts\ as any other built-in model.

\begin{table}
\bgfb
\multicolumn{2}{c}{\textbf{Table~\ref{fig:FA Interface}: Options of the \feynarts\ 
interface}}\\
\multicolumn{2}{l}{All the options of {\tt FeynmanRules} can be employed, 
in addition to:}\\
{\tt Output} & The name of the \feynarts\ directory and files generated by the interface.  The default is {\tt M\$ModelName} with \texttt{\textunderscore FA} appended.\\
{\tt DiracIndices} & Whether to write spin and Dirac indices in the generic file. 
  If {\tt Automatic}, they are added only if the model Lagrangian contains 
  interactions with more than two fermions. The default value is {\tt Automatic}.\\
{\tt CouplingRename} & Whether to rename vertices with a new name and store the definitions in {\tt M\$FACouplings}.  The default value is {\tt True}.
\egfb
\textcolor{white}{\caption{\label{fig:FA Interface}}}
\end{table}

The \feynarts\ generic file ({\tt .gen}) created in this way contains the following information:
\begin{itemize}
\item[-] The kinematical indices as in the standard \feynarts\ file 
  {\tt lorentz.gen}, except if Dirac\footnote{The spin indices of \feynrules\ 
  ({\tt Spin}) are called Dirac indices in \feynarts.} indices are necessary. In 
  this case, they are consistently added for the fermionic particles.
\item[-] {\tt\$FermionLines = True} if there are no spin indices or {\tt\$FermionLines = False} otherwise, such that fermion flows are correctly used in \feynarts. 
\item[-] A copy of the simplification, truncation and last generic rules of {\tt lorentz.gen}.
\item[-] Generic propagators as in {\tt lorentz.gen} or with the Dirac indices for the fermion propagators (if necessary).
\item[-] The basis of the vertex structures, given as kinematic vectors. The 
  entries of these vectors are all the possible Lorentz structures of an 
interaction vertex 
among specific generic particles (scalar fields, spinors, \etc). In other words,
it includes the parts of the vertices containing Lorentz indices, spin indices, 
momenta, \etc 
\item[-] The flipping rules, \ie, the rules that are applied to reverse the flow in a vertex.
\end{itemize}

The \feynarts\ model file ({\tt .mod}) contains:
\begin{itemize}
\item[-] A copy of the index declarations implemented in the \feynrules\ model file
with the \texttt{Unfold} removed. In contrast, the \texttt{NoUnfold} tag is kept 
and used in \feynarts. It avoids the creation of one diagram and one amplitude for each possible value of the associated indices.
\item[-] A copy of the {\tt M\$ClassDeclarations} list, as implemented in the 
\feynrules\ model file, but with the attributes not relevant for \feynarts\ 
removed.
\item[-] Some additional functions used by \feynarts.
\item[-] The list {\tt M\$CouplingMatrices}. Each element of this list consists of 
a set of coefficients to be mapped to the generic kinematic vectors given in the {\tt .gen} file.
\item[-] The list {\tt M\$FACouplings}, containing replacement rules allowing
to express the vertex coefficients in terms of the parameters defined in the \feynrules\ model file.
\end{itemize}

Finally, the parameter file ({\tt .pars}) contains:
\begin{itemize}
\item[-] The list {\tt M\$ExtParams} containing replacement rules of the external parameters by their values.
\item[-] The list {\tt M\$IntParams} containing replacement rules of the internal parameters by their values (in terms of other model parameters).
\item[-] The lists {\tt M\$Masses} and {\tt M\$Widths} containing replacement rules
for the masses and widths of the particles.
\end{itemize}

The generic file and the model file must be both specified when calling the 
\feynarts\ {\tt InsertFields} function via the options {\tt GenericModel} and 
{\tt Model}, respectively. The parameter file can be loaded in \mathematica\ 
if necessary according to the needs of the user.

The \feynarts\ interface fully expands the vertices before extracting the Lorentz structures. Consequently, each Lorentz structure appearing in the kinematic vectors is never a sum or a difference. 
Moreover, for fermion vertices, a Lorentz structure always ends with a chirality 
projector. The interface completes the kinematic vectors such that they close eventually up to a sign under any permutation of the indices of identical particles (at the generic level), as required by \feynarts. 
The quantity {\tt M\$FlippingRules} contains all the flipping rules for zero to two gamma matrices and one chirality projector. Products of more than two gamma matrices are hence always reduced using 
\begin{equation}
  \gamma^\mu\gamma^\nu\gamma^\rho = \eta^{\mu\nu}\gamma^{\rho} - \eta^{\mu\rho}\gamma^{\nu} + \eta^{\rho\nu}\gamma^{\mu} - i \epsilon^{\mu \nu \rho\sigma} \gamma_{\sigma}\gamma_{5} \ .
\end{equation}
In the case all spin indices are explicit, {\tt M\$FlippingRules} is irrelevant and
consists thus of an empty list.

If the option of the interface {\tt DiracIndices} is set to {\tt True}, it can be used to force the presence of Dirac indices even if there are at most two fermions in each interaction vertex. 
An error message is printed to the screen if the {\tt DiracIndices} option is set to {\tt False} and if the model contains vertices with more than two fermions. The default {\tt Automatic} value implies the explicit presence of Dirac indices only if vertices with more than two fermions are found.

Eventually, \formcalc\ will algebraically simplify and manipulate the amplitudes created by \feynarts.  Replacing long analytic expressions by symbols at the level of the
vertices makes
these manipulations  more efficient.  Therefore, when the Feynman rules are exported
to a \feynarts\ model file by \feynrules, vertex expressions 
are replaced by symbols, denoted by {\tt gc}$i$ with $i$ being an integer number,
multiplying the various Lorentz structures.
The definitions of these {\tt gc}$i$ quantities are stored in the list of 
replacement rules 
{\tt M\$FACouplings}, an example element of this list being \verb+gc3 -> ee/sw+.
The user may then apply (or not) this list to the results computed by 
\feynarts\ and \formcalc\ at any time.  To turn this renaming off when calling
the \feynarts\ interface, the option \verb+CouplingRename+ has to be set 
to {\tt False}. The parameters can be further replaced by their numerical values
by employing the lists of replacement rules {\tt M\$ExtParams}, 
{\tt M\$IntParams}, {\tt M\$Masses} and {\tt M\$Widths} that contain the values of 
the different parameters of the model. 

As shown in Section \ref{sec:parts}, it is possible to gather particles with similar 
properties into classes. This allows \feynarts\ to generate fewer diagrams and 
\formcalc\ to receive fewer terms to simplify when a computation is started. The 
specific couplings, masses and widths of each particle can then be inserted only 
after performing the calculation, made in this way more efficient and faster.

The \feynrules\ model format is inspired from the \feynarts\ one and also allows to 
group particles into classes. Obviously, the resulting advantage at the level
of \formcalc\ is only present if the user writes the model and its Lagrangian in a 
`classy way'. However, the model can still 
be exported to \feynarts\ if the Lagrangian breaks this rule. The interface then
defines one new class for each particle as soon as a vertex depending on a specific 
class member is found. Only the new classes are subsequently transmitted to 
\feynarts. Their list  can be obtained with the function {\tt NewFeynArtsClasses[]}.

\subsection{The \sherpa\ interface}

The \sherpa\ interface can be called with the following command,

{\tt WriteSHOutput[}$\mathcal{L}_1$, $\mathcal{L}_2$,\ldots, options{\tt]}

where ${\cal L}_1$, ${\cal L}_2$, \ldots, are the different pieces of the model
Lagrangian.
The interface does not accept any special options besides the usual options available to {\tt FeynmanRules[]}, the option {\tt Exclude4Scalars} described in Section 
\ref{sec:calchep}, and the option {\tt Output} that allows to set the output directory. Issuing the above command produces a directory that contains the following text files:
\begin{itemize}
\item[-] {\tt feynrules.dat}: A static file, setting up the model in \sherpa.
\item[-] {\tt Particle.dat}: The list of all particles together with their properties.
\item[-] {\verb+param_card.dat+}: LH-like file defining the numerical values of the external parameters.
\item[-] {\verb+ident_card.dat+}: File linking the entries in {\verb+param_card.dat+} to the variables used in the \sherpa\ code.
\item[-] {\verb+param_definition.dat+}: File containing analytical expressions for all the internal parameters.
\item[-] {\tt Interactions.dat}: File defining all the interaction vertices with their couplings.
\end{itemize}
The master switch to use a \feynrules\ generated model within \sherpa\ is
\begin{verbatim}
MODEL = FeynRules
\end{verbatim}
to be set either in the (model) section of the \sherpa\ run card or on the command line once the \sherpa\ executable is called. For further information on the structure of the \sherpa\ interface, we refer to Ref.~\cite{Christensen:2009jx}.

Although the \sherpa\ interface has been developed such as to be able to handle interactions that are as general as possible, there are several limitation on the kind of models that can be handled by the interface. The main limitation comes from the kind of structures the matrix element generator of \sherpa\ can handle. Currently, only Lorentz and color structures which are already included in the Standard Model or in the Minimal Supersymmetric Standard Model are supported by the interface\footnote{Currently, the use of fermion number violating interactions with the \sherpa\ interface is discouraged.}. In addition, the \sherpa\ interface can only handle fields with spin 0,1/2 or 1 transforming in the trivial, fundamental or adjoint representations of the QCD gauge group. Finally, one should mention that QCD showers are only invoked for colored particles present in the SM or the Minimal Supersymmetric Standard Model (MSSM). New colored states are not hadronized, and they should be decayed before entering the hadronization stage.

\subsection{\label{TeX Interface}The \TeX-Interface}

\feynrules\ comes with an extended \TeX-interface that can be invoked with the 
command 

{\tt WriteLaTeXOutput[} $\mathcal{L}_1$, $\mathcal{L}_2$, \ldots, $V_1$, $V_2$, \ldots, options {\tt]} 

where $\mathcal{L}_1$, $\mathcal{L}_2$, \ldots are the different pieces of the model
Lagrangian and  $V_1$, $V_2$, \ldots are vertex lists (as output by 
{\tt FeynmanRules}). The options options supported by the interface can be 
found summarized in Table \ref{fig:TeX Interface} and are addressed throughout this
section.
The \TeX-interface can be invoked in several ways according to the arguments of the
function {\tt WriteLaTeXOutput[]}. It can be either called with no arguments, in 
which case the interface writes the \TeX-files describing the model but does not 
include any part of the model Lagrangian or vertex list, with only Lagrangians, with
only vertex lists, with only options or with any combination of Lagrangians, vertex 
lists and options.

\begin{table}
\bgfb
\multicolumn{2}{c}{\textbf{Table~\ref{fig:TeX Interface}: TeX Interface Options}}\\
  {\tt Overwrite} & Determines whether the interface must overwrite ({\tt True}) the 
  generated \TeX\ files, if existing, or not ({\tt False}). If set to {\tt Automatic}
  (the default value), the user is asked, for each file, whether it has to be 
  overwritten.\\
{\tt Paper} & This option determines what kind of paper format to use.  Choices are 
  {\tt letter} (the default value), {\tt legal} and {\tt a4}.\\
\egfb
\textcolor{white}{\caption{\label{fig:TeX Interface}}}
\end{table}

The Lagrangian pieces and vertex lists can be entered in two ways.  Taking the 
example of the Lagrangian {\tt L1}, the interface can be simply called as in 
{\tt WriteLaTeXOutput[ L1}, \ldots {\tt]}. A \TeX-name can also be associated with
the Lagrangian. In this case, the interface is called as in
{\tt WriteLaTeXOutput[} {\tt\{Subscript[L,Gauge],L1\}}, \ldots {\tt]}
where {\tt Subscript[L,Gauge]} denotes the Lagrangian {\tt L1}. Vertex lists behave similarly and can be entered directly or with a name as in {\tt WriteLaTeXOutput[} \ldots, {\tt \{Subscript[V,Gauge],V1\}}, \ldots {\tt]}.  If entered, these names will be used for the section names.  If they are not used, names will be generated as {\tt L1}, {\tt L2}, \etc\ for the Lagrangians and {\tt V1}, {\tt V2}, \etc\ for the vertex lists.

The files that this interface generates are split among the various sections of the \TeX\ document.  The main file is named {\tt M\symbol{36}ModelName} with the extension {\tt .tex} appended.  This file instructs \LaTeX \ to include the other files, 
\begin{itemize}
\item title.tex:  Contains the title.
\item abstract.tex:  Contains the abstract.
\item introduction.tex:  Contains the introduction.
\item symmetries.tex:  Contains information about the gauge symmetries and indices.
\item fields.tex:  Contains information about the fields.
\item lagrangians.tex:  Contains the Lagrangians.
\item parameters.tex:  Contains information about the parameters.
\item vertices.tex:  Contains the vertices.
\item bibliography.tex:  Contains the bibliography.
\end{itemize}

After being written by the interface, the user can modify these files in any way he/she wishes.  The next time the interface runs, it will first check whether the files exist.  If they do not, it will generate them in the usual way.  If they do exist, the interface will ask the user whether they should be overwritten or not.  In this way, the user can make changes to files and not have them overwritten.  On the other hand, if the user wishes to have all of the files overwritten, he/she can set the option {\tt Overwrite} to {\tt True}.  However, if the user would like none of the files to be overwritten, he/she can set the option {\tt Overwrite} to {\tt False}.  In this case, the interface skips any files that already exist.  The default is for the interface to ask the user what to do for each file which already exists.

The paper in common use depends on the country a person lives in.  For this reason, the option {\tt Paper} was created.  Its defaults value is {\tt letter}, 
but this option can also be set to {\tt a4} or {\tt legal} as in 
{\tt Paper -> "a4"}.


\subsection{The UFO interface}
\label{sec:ufo}
The UFO interface can be called with the following command,

{\tt WriteUFO[}$\mathcal{L}_1$, $\mathcal{L}_2$,\ldots, options{\tt]}

where ${\cal L}_1$, ${\cal L}_2$, \ldots, are the different pieces of the model
Lagrangian.
The interface accepts, in addition to all the options of {\tt FeynmanRules[]} and 
the option {\tt Exclude4Scalars} described in Section 
\ref{sec:calchep},  various options which are summarized in Table~\ref{tab:ufo_options} (see also Ref.~\cite{Degrande:2011ua}).
When run, the interface computes all the vertices and stores all the information about the model in the form of a \python\ module which can be linked to existing Feynman diagram calculators. Currently, the UFO is used by \aloha~\cite{deAquino:2011ub}, 
{\sc MadAnalysis} 5~\cite{Conte:2012fm} and \madgraph\ 5 \cite{Alwall:2011uj}
and will be used in the future by \gosam~\cite{Cullen:2011ac,Cullen:2011xs} and 
\herwig++~\cite{Bahr:2008pv}. For details about the structure of the \python\ module and the syntax used to store the analytical expressions for the Lorentz and color structures of the vertices, we refer to Ref.~\cite{Degrande:2011ua}. Here it suffices to say that any vertex coupling fields of spins no greater than two and transforming in the representations {\bf 1}, {\bf 3}, {\bf 6} or {\bf 8} of the QCD gauge group can be implemented into the UFO format. 
This is done by decomposing each vertex into a color $\otimes$ spin basis using the master formula
\beq\label{eq:generic_vertex}
  \begin{cal}V\end{cal}^{a_1\ldots a_n, \ell_1\ldots\ell_n}(p_1,\ldots,p_n) =
    \sum_{i,j}C_i^{a_1\ldots a_n}\,G_{ij}\,L_j^{\ell_1\ldots\ell_n}(p_1,\ldots,p_n)
     \ , 
\eeq
where $C_i^{a_1\ldots a_n}$ and $L_j^{\ell_1\ldots\ell_n}(p_1,\ldots,p_n)$ denote tensors in color and spin space respectively, and $G_{ij}$ is a matrix of coupling constants. The analytic expression for the color and spin tensor are stored inside the UFO, and can be transformed into {\sc Fortran} and/or \cpp routines using the \aloha\ package~\cite{deAquino:2011ub}.

Note that for the UFO files to work properly, all coupling constant should be assigned an interaction order (see Section \ref{sec:intorder}).

\begin{table}
\bgfbalign
\multicolumn{2}{c}{\textbf{Table~\ref{tab:ufo_options}: Options accepted by the UFO interface}}\\
\\
{\tt Output} & A string specifying the output directory. The default is {\tt M\$ModelName} with {\tt \textunderscore UFO} appended.\\
{\tt Input} & A list of vertices that will be merged with the vertices computed from the Lagrangian before transforming them into the UFO format. The default is an empty list. \\
{\tt AddDecays} & If {\tt True}, the analytic expressions for all two-body decays are included into the file {\tt decays.py}. The default is {\tt True}.\\
\egfbalign
\textcolor{white}{\caption{\label{tab:ufo_options}}}
\end{table}

\subsection{\label{sec:whizard}The \whizard\ interface}
The \whizard{} interface can be called with the following command,

{\tt WriteWOOutput[}$\mathcal{L}_1$, $\mathcal{L}_2$,\ldots, options{\tt]}

where ${\cal L}_1$, ${\cal L}_2$, \ldots, are the different pieces of the model
Lagrangian.
The interface accepts, in addition to all the options of {\tt FeynmanRules[]} and 
the option {\tt Exclude4Scalars} described in Section 
\ref{sec:calchep},  various options which are summarized in Table~\ref{fig:Whizard Interface}.  The full details of this interface can be found in Ref.~\cite{Christensen:2010wz}.  We will summarize some of the most important points.

\begin{table}
\bgfb
\multicolumn{2}{c}{\textbf{Table~\ref{fig:Whizard Interface}: Options of the \whizard\ 
interface}}\\
\multicolumn{2}{l}{All the options of {\tt FeynmanRules} can be employed, 
in addition to:}\\
{\tt Input} & A list of vertices from {\tt FeynmanRules} to use instead of a Lagrangian.\\
{\tt Output} & The name of the directory where the \whizard\ files should be written by the interface.  The default is determined from {\tt M\$ModelName}.\\
{\tt WOModelName} & The name by which the model will be known to \whizard.  The default is determined from {\tt M\$ModelName}.\\
{\tt WOGauge} & The gauge of the \whizard\ files.  Choices are {\tt WOFeynman}, {\tt WORxi} and {\tt WOUnitarity} (the default).\\
{\tt WOGaugeParameter} & The parameter used to determine the gauge.  The default is {\tt Rxi}.\\
{\tt WOAutoGauge} & Whether to automatically assign Goldstone boson masses in Feynman and $R_\xi$ gauges and automatically append the parameter $\xi$ to the parameter list in $R_\xi$ gauge.\\
{\tt WORunParameters} & Which parameters are required to be computed for each phase space point.  The default is {\tt aS} and {\tt G}.\\
{\tt WOFast} & Whether to increase the number time consuming checks to determine if a vertex is supported ({\tt False}).  The default is {\tt True}.\\
{\tt WOMaxCouplingsPerFile} & The maximum number of couplings written to a single Fortran file.  The default is $500$.\\
{\tt WOVerbose} & Whether to enable verbose output including more extensive information for the vertices that are not supported.  The default is {\tt False}.
\egfb
\textcolor{white}{\caption{\label{fig:Whizard Interface}}}
\end{table}

The \whizard\ interface can currently handle spins $0$, $1/2$, $1$ and $2$.  The color representations supported include the singlet, triplet, antitriplet and octet.  The full list of vertices supported by this interface can be found in Table 1 of Ref.~\cite{Christensen:2010wz} and comprises a large number of the possible vertices for these spins and color representations.  When an unsupported vertex is identified, a warning message is printed, the vertex is skipped and the interface continues to process the other vertices.  The option {\tt WOFast} can be set to {\tt False} to include additional checks to attempt to match model vertices to supported operators.   A new version of \whizard\ and this interface is planned that will support all the vertices supported by \feynrules.

Since the strong coupling and all parameters depending on it must be evaluated for each scale, the \whizard\ interface flags these parameters to be recalculated for each phase space point while all other parameters are only calculated once for a given simulation.  The list of parameters recalculated for each parameter point can be extended by use of the option {\tt WORunParameters} which is given a list of the parameters to be recalculated.

This interface supports the unitary, Feynman and $R_\xi$ gauges.  The user can choose the gauge by setting the option {\tt WOGauge}.  The choices are {\tt WOFeynman}, {\tt WORxi} and {\tt WOUnitarity} (which is the default).  If the $R_\xi$ gauge is chosen, the gauge parameter must be set using the {\tt WOGaugeSymbol} option.  If either Feynman or $R_\xi$ gauges are chosen, the option {\tt WOAutoGauge} can be set to {\tt True}.  In this case, the gauge symbol will automatically be created and the masses of the Goldstone bosons will automatically be determined.  For this to work, the model must be implemented in Feynman gauge.  Ghosts are ignored by this interface.

By default this interface writes the \whizard\ model files to the directory based on {\tt M\$ModelName} but can be changed by use of the option {\tt Output}.  The name that \whizard\ uses to load the model is also based on {\tt M\$ModelName} but can be changed with the option {\tt WOModelName}.  Once the interface is done, assuming \whizard\ was installed in the standard way, the user can make the model available to \whizard\ by change into the directory where the model files were written and issuing the commands {\tt ./configure} followed by {\tt make install}.  (If this directory already had an old set of model files that were overwritten, it may be necessary to call {\tt make clean} before compiling.)  This will install the model files in the {\tt .whizard} subdirectory of the user's home directory.  At this point, this model can be used exactly like any built-in \whizard\ model.  

If the user only wants to use a new set of values for the external parameters but the model has not changed in any other way, the command {\tt Write\-WO\-Ext\-Pa\-rams[filename]} can be used.  It will write the current values to a sindarin file called {\tt filename} which can be used in a \whizard\ session.


\section{Running time} \label{sec:bench}
Compared to the previous versions of the code, the core module of \feynrules\ \FRversion\
has been greatly improved with respect to speed. In particular,
the internal treatment of lists and/or long expressions has been deeply
modified to benefit from the various strengths of the \mathematica\
platform. The validation of the new routines have been performed by comparing
results as outputted by version 1.6 of the code to those retrieved by the new version.

In addition, the execution speed of the program has also been improved thanks to
parallelization so that
\feynrules\ can now use more than one CPU core, if available on the system.
By default, the maximum number of available cores is employed and
\feynrules\ is loaded on each of them, together with one \mathematica\ slave
kernel per core. This behavior
can be modified according to the user's needs by setting
one of two dedicated variables accordingly.
First, in the case the user wants to run \feynrules\ on a single
CPU core, he/she has to issue, in a \mathematica\ session, the command
\begin{verbatim}
  FR$Parallel = False
\end{verbatim}
before loading the \feynrules\ package.
Second, in the case the user wants to employ only a subset of the available CPU power
for \feynrules, he/she can type the command
\begin{verbatim}
  FR$KernelNumber = <n>
\end{verbatim}
where \texttt{<n>} is an integer number smaller than or equal to the maximum number of
available CPU cores.
In the case this command is run before \feynrules\ is started, \texttt{<n>}
\mathematica\ slave kernels are started and each of these are employed by \feynrules.
Otherwise, if typed after \feynrules\ has been loaded, this indicates to \feynrules\
to reduce the number of slave \mathematica\
kernels to \texttt{<n>}. Note that in order to increase the number of kernels, \feynrules\ must be restarted.

At this stage, it is still possible to force \feynrules\ to effectively employ one CPU
core for the calculations that it will perform by typing
\begin{verbatim}
  FR$Parallelize = False
\end{verbatim}
This commands keeps all the slave \mathematica\ kernels running but only employs one of them
for the computations.

\renewcommand{\arraystretch}{1.25}
\begin{table}
\begin{center}
\begin{tabular}{l || l || l| l| l| l}
\hline\hline
  Command & FR 1.6 & FR \FRversion\ - 1 & FR \FRversion\ - 2 & FR \FRversion\ - 4 & FR \FRversion\ - 8\\
\hline\hline
  \texttt{FeynmanRules}  & 5.84 s & 4.98 s & 3.09 s & 2.32 s & 1.93 s\\
  \texttt{WriteCHOutput} & 9.33 s & 9.51 s & 8.05 s & 6.26 s & 5.53 s\\
  \texttt{WriteUFO}      & 9.05 s & 8.82 s & 7.89 s & 6.51 s & 6.05 s\\
\end{tabular}
\textcolor{white}{\caption{\label{tab:runSM}}}
\end{center}
{Table \ref{tab:runSM}: Running time associated with the three \feynrules\ commands
  \texttt{FeynmanRules} (second line), \texttt{WriteCHOutput} (third line) and \texttt{WriteUFO} (fourth line),
where the decay widths are not calculated.
For this comparison, we use the Standard Model whose Lagrangian has been
  calculated before calling these commands.
  We compare version 1.6 of \feynrules\ (second column) to version \FRversion\ when using one (third
  column), two (fourth column), four (fifth column) and eight (last column) CPU cores.}
\end{table}
\renewcommand{\arraystretch}{1.}

In Table \ref{tab:runSM}, Table~\ref{tab:runMSSM} and Table~\ref{tab:runsps1a},
we compare the running times of the code
in three different contexts. Table~\ref{tab:runSM} is dedicated to the Standard Model.
We first compute the model Lagrangian and then execute the \texttt{FeynmanRules} command
without performing the full expansion over flavor indices,
\begin{verbatim}
lag = LSM
FeynmanRules [ lag ]
\end{verbatim}
The results are presented in the second line of the table and correspond to
only the second of the two commands above. Then, we turn to the efficiencies
of the interfaces and take the example of the \calchep\ (third line of the table) and UFO (last line
of the table) interfaces. In the last case, we omit the computation of the decay
widths for the sake of the comparison, as this feature is absent in the version 1.6 of \feynrules.
The running time are those obtained after typing, in a \mathematica\ session,
\begin{verbatim}
WriteCHOutput [ lag ] 
WriteUFO [ lag, AddDecays->False ]
\end{verbatim}
respectively. They correspond to the sum of the times needed to first extract the
Feynman rules, then perform the flavor expansion and eventually translate the rules in terms of the
\calchep\ and UFO formats, respectively. Whereas this mimics the behavior of most users, an exclusive
test of the interfaces can be achieved by computing the Feynman rules separately and providing the vertices
as input to the interfaces. This however goes beyond the tests performed in this section where we
have chosen to stick to the commands mostly employed by the users.
For all the three commands above, we confront results as obtained by means of \feynrules\ 1.6,
the previous stable version, to those obtained with \feynrules\ \FRversion\ when using one, two, four
and eight cores. For the tests, we employ a machine with a 2.3 GHz Intel
Core i7 processor and 16 GB of memory (1600 MHz DDR3). Moreover, the operating system is
{\sc MacOS X} 10.8.4. As can be seen from the table, the speed improvement
between the older and new version of \feynrules\ is significant, especially when more than one core is employed.

\renewcommand{\arraystretch}{1.25}
\begin{table}
\begin{center}
\begin{tabular}{l || l || l| l| l| l}
\hline\hline
  Command & FR 1.6 & FR \FRversion\ - 1 & FR \FRversion\ - 2 & FR \FRversion\ - 4 & FR \FRversion\ - 8\\
\hline\hline
  \texttt{FeynmanRules}  & 325.5 s & 213.7 s &  79.7 s &  62.6 s &  41.0 s\\
  \texttt{WriteCHOutput} & 853.4 s & 618.9 s & 350.8 s & 283.9 s & 204.4 s\\
  \texttt{WriteUFO}      & 436.0 s & 518.5 s & 316.1 s & 273.8 s & 239.7 s\\
\end{tabular}
\textcolor{white}{\caption{\label{tab:runMSSM}}}
\end{center}
{Table \ref{tab:runMSSM}: Same as Table~\ref{tab:runSM}, but for the MSSM. In contrast to the
  Standard Model case, four scalar interactions have been removed when running the commands
  \texttt{WriteCHOutput} and \texttt{WriteUFO}.}
\end{table}
\renewcommand{\arraystretch}{1.}

Table~\ref{tab:runMSSM} addresses the MSSM. As for the Standard Model, the Lagrangian is first computed
separately and then employed together with the \texttt{FeynmanRules} command (second line
of the table),
\begin{verbatim}
lag = Lag
FeynmanRules [ lag ]
\end{verbatim}
Concerning the interfaces, we optimize the output by removing the
hundreds of vertices describing four-scalar interactions which
are, for tree-level computations, in general phenomenologically less relevant.
Moreover, the parameters are taken as those of the
minimal supergravity scenario SPS 1a \cite{Allanach:2002nj}\footnote{
Although this benchmark point is now experimentally excluded, it is still employed as
a standard choice for the default values of the MSSM model files of most of
matrix element generators. This motivates our choice.}.
We refer to the Appendix of Ref.~\cite{Christensen:2009jx} for the adopted numerical values
of the model's free parameters. For the comparison of the running time, we hence issue, in \mathematica,
\begin{verbatim}
WriteCHOutput [ lag , Exclude4Scalars -> True] 
WriteUFO [ lag, AddDecays->False, Exclude4Scalars -> True ]
\end{verbatim}
We find similar conclusions as for the Standard Model case except when comparing
\feynrules\ 1.6 with the newer version run with one single CPU core.
It may indeed seem that the UFO interface in \feynrules\ \FRversion\
is less efficient than in the previous version 1.6. However, this is explained by
the fact that the resulting UFO library
is more efficient concerning the handling of the model parameters. The gain in
time consequently only appears
when performing matrix element computations with, \eg, \madgraph~5.

\renewcommand{\arraystretch}{1.25}
\begin{table}
\begin{center}
\begin{tabular}{l || l || l| l| l| l}
\hline\hline
  Command & FR 1.6 & FR \FRversion\ - 1 & FR \FRversion\ - 2 & FR \FRversion\ - 4 & FR \FRversion\ - 8\\
\hline\hline
  \texttt{WriteCHOutput} & 960.1 s & 1134.0 s & 356.6 s & 305.6 s & 254.2 s\\
  \texttt{WriteUFO}      & 871.9 s &  984.1 s & 316.6 s & 283.3 s & 250.7 s\\
\end{tabular}
\textcolor{white}{\caption{\label{tab:runsps1a}}}
\end{center}
{Table \ref{tab:runsps1a}: Same as Table~\ref{tab:runMSSM}, but for after applying
  restrictions as in the SPS 1a benchmark point (no flavor violation in the fermion
  and sfermion sectors).}
\end{table}
\renewcommand{\arraystretch}{1.}

Finally, in Table~\ref{tab:runsps1a}, we apply, before calling the interface, a set of restrictions
(see Section~\ref{sec:restrictions}) as motivated by the SPS 1a point. In more detail, flavor violation
is the fermion and sfermion sectors is not allowed. This is achieved by issuing, after the computation
of the Lagrangian,
\begin{verbatim}
  WriteRestrictionFile[ ]
  LoadRestriction["ZeroValues.rst"]
\end{verbatim}


\section{\label{sec: web validation}Model Validation and Debugging}

By using \feynrules, the risk of
creating faulty model implementations is reduced. However, possible issues can still arise from either
faulty \feynrules\ input or software bugs. 
We have taken great pains to validate the
\feynrules\ package \cite{Christensen:2009jx}, and we, further, plan to continually extend and
improve these tests in order to make the \feynrules\ package ever more dependable.
On the other hand, when a new model is implemented, it has been up to the author
of that new model to validate it.  In order to ensure a high level of
quality for the models implemented using \feynrules, we have proposed
the following set of validation guidelines \cite{Butterworth:2010ym}:
\begin{itemize}
\item \textbf{Documentation:} Full documentation should be included to ensure traceability and reproducibility.   Relevant information should be included in the model file in the variable {\tt M\$ModelInformation} and include references to the appropriate papers for the model and the model implementation as well as any URLs where further information can be obtained.  Furthermore, versions for the operating system, \mathematica\ and \feynrules\ where any tests were performed should be included.
\item \textbf{Basic sanity checks:} The model implementation should satisfy basic sanity tests such as hermiticity and gauge invariance (see the item below).  The Feynman rules obtained from \feynrules\ should be compared with those in the literature and simple cross sections and/or decay rates should be computed and compared with the literature as well.  The results of these tests should be included in the documentation.
\item \textbf{Testing one generator:} The model should be exported to the format of one of the matrix-element generators and detailed calculations should be performed.  Comparisons should be made with the literature, with the built-in SM calculations for processes where the new physics does not have an effect and with any other implementations of the same model.  If multiple gauges are available in the matrix-element generator, they should be compared.  High energy unitarity cancellation should be tested if applicable (the widths should be set to zero for this test).  The results should be included in the documentation.
\item \textbf{Testing several generators:} A full comparison of a very large set of processes should be done between multiple matrix-element generators supported by \feynrules\ as well as between supported gauges.  Additionally, independently implemented versions of the same model should be compared.  For this step it is important that the widths be set to zero since each matrix-element generator has slightly different methods of dealing with them.  The results should be included in the documentation.
\end{itemize}

Although it is virtually impossible to prove the absolute correctness
of a model implementation, following these guidelines should give
greater confidence.  Unfortunately, in practice the level of model validation has
been irregular.  One of the challenges is that although it is
straightforward to test a handful of Feynman rules by hand,
validating the full set of vertices, of which there can be
thousands, has been difficult.  One approach to this task has been to
automate the calculation of the cross section by the different matrix
element generators (MEG), in different gauges and in different, independent
implementations and compare the results \cite{Christensen:2009jx}.  
We believe that this procedure nicely
complements the sanity tests and the validation of a small set of
Feynman rules by hand.  However, it can be difficult for a user to set
up such an automated test for several reasons.  It requires expertise
in each code that is to be compared as well as a careful matching of
the setups for each code (in particular, the cuts and model input).

It is our goal to attain a very high level of validation for every
model implemented in \feynrules.  To this end, we have begun developing
a new automated web validation which abstracts the details of the
matrix element generator setup \cite{Brooijmans:2012yi}.  
Experts in each matrix element
generator takes care of the installation and running of the MEG on the
server.
The model implementation author accesses the MEG through a web
interface where \heshe\ uploads \hisher\ \feynrules\ model files to the server.  The server uses the uploaded files to generate model files for the supported MEGs and allows the user to run validations on the model implementation.  The server displays the results on the screen and marks processes where inconsistencies are suspected, giving the user an idea of where to look for problems.  If no inconsistencies are found, the user gains greater confidence in \hisher\ model.

Although this software is still in a beta stage and we
feel there is much more that should be done to improve it, 
it is already useful for validation.  For this reason, it has been made public.

\subsection{\label{sec:webv account}URL and Account}

The web interface can be accessed from any web enabled device by
pointing a web browser at the url:
\begin{center}
\begin{verbatim}
http://feynrules.phys.ucl.ac.be/validation
\end{verbatim}
\end{center}

The browser will be immediately redirected to the author's personal
model page.  If not already logged in, the user will be
redirected to the login page.  To obtain a user account, the author
must currently email one of the \feynrules\ authors to request
one.  The models page contains a list with all the models uploaded so far.

\subsection{\label{sec:webv upload model}Uploading a model}

To validate a model, the user must first upload \hisher\ model files.  To do this, the user should click the `New Model' button on his or her model page, which leads 
to a form where the user specifies the properties of the new model to be uploaded.

The first thing the user must specify is the location of \hisher\ model files.  More than one model file can be uploaded as long as all model files can be loaded together using the \verb|LoadModel| function as in
\begin{verbatim}
LoadModel[file1, file2,...]
\end{verbatim}
where \verb|file1, file2,...| are the names of the uploaded files. Although a model file that directly calls code from other files is allowed in \feynrules, it is not supported in the web validation.

The user may also upload restriction files and parameter files if there are any.  Multiple restriction and parameter files upload is possible.  At a later stage, when the MEG files are generated, the user is hence able to choose which restriction files and which parameter files to employ (see Section \ref{sec:webv generate model files}).  Multiple sets of MEG files can be generated, each with a unique set of restriction files and parameter files.  The restriction files must be loadable by the \verb|LoadRestriction| function (see Section \ref{sec:restrictions}) as in 
\begin{verbatim}
LoadRestriction[file1, file2, ...]
\end{verbatim}
where \verb|file1, file2,| \etc, are the restriction files.  The parameter files must be in a LH form and be loadable by the \verb|ReadLHAFile| function (see Section \ref{sec:paramin}) as in
\begin{verbatim}
ReadLHAFile[Input->file]
\end{verbatim}
where \verb|file| is the parameter file.

There is also a text box where the user can specify the Lagrangian of his or her model, in any form allowed by the \verb|FeynmanRules| function (see Section \ref{sec:Feynman Rules}).
For example, taking the Standard Model implementation, the user could equivalently enter any of the three choices
\begin{verbatim}
LSM
{LGauge,LFermion,LHiggs,LYukawa,LGhost}
LGauge+LFermion+LHiggs+LYukawa+LGhost
\end{verbatim}

There is also the option to turn off four-scalar interactions, which is useful for some models where the number of such interactions is impractical (see Section~\ref{sec:calchep}).  Finally, the user can choose between the publicly available version (the default) or the development version of \feynrules.

Once the user finishes filling in the form, he or she can click on the `Submit' button.
The browser uploads the model files and information.  The server organizes the files and runs \feynrules\ on them for the first time, without generating MEG files.  It is just testing whether \feynrules\ can load the model files, the restriction files and the parameter files without any major error.  If there is no error, the user can move on to the validations themselves.  If there are errors, the user is blocked from proceeding. 

The web validation is not designed to help solving syntax errors.  It is the responsibility of the user to test that his or her model files, restriction files and parameter files can all be loaded by \feynrules\ before uploading them to the web validation.  The web validation does not currently support the backup or the versioning of models or validations and so should not be used for long term storage.  It is hoped that backup and versioning will be added in a future version.

After uploading the model files, the browser is sent to the model page for the new model.  This page is where all the information for the model and access to the validations for the model is presented.  At the top of the model page is the user's name.  Clicking on his or her name will take the user back to \hisher\ models page where the new model is listed along with any other models the user has uploaded.

\subsection{\label{sec:webv generate model files}Generating matrix
  element generator files}

Assuming the uploaded model files pass the basic tests described above, the next step is to generate MEG files.  To do this, the user must click on the `Create MEG Files' button.  A small dialog opens and allows to choose a name for the MEG files as well as one or more restriction files, if desired, and a parameter file, if desired.  The user then clicks the `Create' button upon which the web server creates jobs for each MEG and submits them to the cluster queue. 

Each MEG job loads the model and any restriction and parameter files requested.  It then attempts to run the MEG interface on the model.  If it is unsuccessful, that MEG will not go any further for this model, restriction file(s) and parameter file combination.  If it was successful, the server submits another job to the queue where the MEG attempts to generate a cross section for the process $e^+e^-\to\mu^+\mu^-$.  Any non-zero cross section at this stage is taken to indicate that the model, restriction files and parameter files together with the MEG interface are capable of producing compilable code.  If the code does not compile or the result returned is zero, the MEG is again blocked from further calculations with this restriction parameter combination.  If both of these tests are passed, the MEG is made available for further computations.

The status of each job is communicated to the user by two symbols below the MEG name.  The first communicates whether the \feynrules\ interface succeeded.  The second communicates whether the MEG calculation of $e^+e^-\to\mu^+\mu^-$ succeeded.  Possible symbols for these are \verb|Q| which means that the job is queued but not yet completed, a green \checkmark which means the job was successful, a red \text{x} which means the job failed, and a \verb|?| which means the result is unknown (and should not normally happen).  Any MEG that passes both tests can continue even if others failed.

More than one restriction-parameter file combinations can be created according to the needs of the user.  Each can have multiple validations run on them.  It is not necessary to generate the same restriction-parameter file combination multiple times since they will be identical.

All restriction-parameter file combinations are presented on the model page in a
dedicated section.

\subsection{\label{sec:webv run 2-2 validation}Running $2\to2$ validations}

The next step is to create a new validation.  On the model page, a list of restriction parameter MEG combinations are presented.  For each of them, there is a list of validations that the user has already created as well as a `Create New Validation' button.  Clicking on this button takes the user to the new validation page.

On the new validation page, the user specifies the details of the validation beginning with giving it a name.  Next, the user chooses whether he or she would like to compare $2\to2$ processes or another topology.  The user can also choose whether to compare integrated cross sections or phase space points.  At the time of this writing, only $2\to2$ integrated cross section comparisons are supported, other validations being planned for the future.

The default is for the server to compare all $2\to2$ processes which satisfy all the symmetries specified in the model and cannot be rotated into one another.  There is the possibility to reduce the number of processes by use of a restriction menu.  The user can specify at least how many particles of a certain type must appear in the external states of each process.  The user can also specify at least how many particles of a certain type must not appear in the external states of each process.

The restriction categories include the spin of the particle, the indices of the particle and the charges of the particle.  These can be combined to tune the list of processes that results.  For example, the user could specify that they want two or more spin 1/2 fermions, two or more particles with triplet color indices, two or more particles that are not spin 1/2 fermions, and four or more particles that are not scalar fields.  This would result in processes with two quarks and two vector bosons in the Standard Model.

Once the form has been filled out, the user clicks the `Create Validation' button.  The server submits a job to the cluster to generate the processes.  The user is taken to the validation page.  When the job is finished, the user is presented with a list of the processes and allowed to start a validation.

At this point the user can choose the MEG's and gauges he or she would like to run.  Any MEG that passed the tests during the restriction-parameter combination MEG file generation are available.  The user chooses between them and then clicks on either `Start Fresh Validation' or `Finish Validation'.  If `Start Fresh Validation' is chosen, the server first removes any previous result for this validation and begins the validation.  If `Finish Validation' is chosen, the server keeps the old results and runs the validation for any missing results.

The `Finish Validation' option can be useful in many scenarios.  If the user would like to remove one or more of the MEG's or a gauge, he or she can simply turn those off and click `Finish Validation'.  The server will remove the results for the MEG's and gauges that the user has removed.  It will keep all the others and return the results.  In another scenario, the user may wish to add a MEG or a gauge to the results without redoing the results that are already present.  In this case, the user simply checks the boxes for the new MEG's and/or gauges and clicks `Finish Validations'.  The server will keep all the old results and start jobs for the new MEG's and/or gauges.

When a job is started, jobs for all the cross section calculations are submitted to the cluster.  They are mixed with any other jobs on the cluster so that everyone who uses the cluster will see progress at roughly the same rate.  A user's jobs do not have to wait for someone else's validation to finish.  Likewise, the user can run multiple validations at the same time.

The web page for the validation is updated periodically.  The user can refresh his or her browser to see the completed results.  Any jobs that are finished will appear with their cross section.  If all the MEG's and gauges are finished for a particular process, the $\chi^2$ is also presented for that process, calculated as
\begin{equation}
\chi_i^2 = \sum_i \left(\frac{\sigma_i-\sigma_b}{\Delta\sigma_i}\right)^2
\end{equation}
where $\sigma_i$ is the cross section for the $i^{\rm th}$ MEG and gauge choice, $\Delta\sigma_i$ is the Monte Carlo uncertainty in that calculation and $\sigma_b$ is the best value for the cross section which is calculated to minimize the $\chi^2$ associated with the process under consideration
\begin{equation}
\frac{\partial\chi_i^2}{\partial\sigma_b} = 0
\end{equation}
The best value is shown on the web page along with the other results and the $\chi^2$ for the analysis of the user.

The user can hover his or her mouse over a result to see the cross section for that MEG and gauge as well as the Monte Carlo error and the number of standard deviations it is away from the best value.  
Additionally, the user can click on the cross section from a MEG to see details from the calculation.

The results are ordered from highest $\chi^2$ to lowest to make it easier for the user to spot potentially problematic processes which may guide the user to problematic areas of his or her model files.  Very high values are almost certainly a sign of problems.  Smaller values may be the result of statistical fluctuations.

The results of the $\chi^2$ for each process are binned and plotted as a histogram at the top of the validation page.  In addition, the theoretical $\chi^2$ distribution (with $n-1$ degrees of freedom where $n$ is the number of MEG's and gauges compared) is plotted alongside the results of the calculations.  We find that when a model is correctly implemented, the results are better than or as good as the theoretical $\chi^{2}$ curve. If there are results from the calculation far outside the theoretical $\chi^{2}$ curve, it is usually an indication of problems with the model.  The user can click on this figure to download an \texttt{.eps} version.

We find that the greatest success occurs if multiple MEG's are compared in just one gauge or when gauge invariance is checked separately with just one MEG at a time.  As already mentioned, multiple validations can be created for each restriction parameter combination.

A \LaTeX/PDF output is planned but not yet implemented.

\subsection{Independent Model Implementations}

If a model has been implemented independently, it is important to check that the new implementation agrees with the old.  For this purpose, the web validation allows to compare to ``stock" implementations.   The user starts this process on \hisher\ main model page under the heading ``Stock Models" by clicking ``Add New Stock Model" which brings up a dialog box for the new stock model.  The user enters the name that this stock model should be known by, chooses the matrix element generator, the model, the gauge, uploads model files as necessary, uploads a particle translation file, and uploads zero or more parameter files.  We will now make some comments about each of these.

The user has the option to choose one of the built-in models from the matrix element generators or to use a version located on their hard drive.  To use their own version, they should choose ``Upload My Own Model Files".  At this point, they can upload the model files as a single gzipped tar-ball.  For \calchep, this should be no directory in this tar-ball.  For \madgraph, this should be a tar-ball of the UFO files including the main containing directory.  For \whizard, this should be a tar-ball of the directory containing all the code including the makefile.  

Where supported, the user can choose the gauge for this stock model.  If both gauges are desired, the stock model should be uploaded twice, once for each gauge.  

Next, the user must supply a particle translation file which tells the web validation platform what the names of each particle are, if different from the \feynrules\ names.  This is a pure text file with each particle on a separate line.  For each particle, the first string should be the \feynrules\ name followed by a comma followed by the stock name.  Lines beginning with hashes are comments and ignored.  Here is an example:
\begin{verbatim}
#Standard Model particles
#FR , CH
a    , A
g    , G
ve   , ne
ve~ , Ne
\end{verbatim}
This file should not be zipped when uploaded.  

Finally, sometimes the same stock model is desired to be run with several different parameter files.  This can be done by uploading one or more parameter files in the native format of the matrix element generator.  For \calchep, this is a {\tt varsN.mdl} file.  For \madgraph, this is a {\tt param\_card.dat} file.  For \whizard, this is a {\tt sindarin} file.  If the model files uploaded already have the parameter point required for the validations, no parameter files need be uploaded.

Once all these files are uploaded, the server tries to run the process $e^+,e^-\to\mu^+,\mu^-$ with the stock model.  If it gets a non-zero result, this model becomes available for validations.  To use it in a validation, check the box next to it before running the validation.

We should mention that for this to work, the user has to painstakingly make sure all the parameters and definitions are set exactly the same.  This means, for example, that all the widths have to be set to zero and all the masses and parameters have to be fixed to the same value as in the \feynrules\ model files.

\subsection{\label{Debugging}Model Debugging}
When bugs in the model files occur, it can be difficult to track them down and fix them.  In this subsection, we would like to describe some of the methods the user can follow to accomplish this.  The first problem that users often face is syntax errors.  After writing a model file, the user attempts to load it in a \feynrules\ \mathematica\ session but gets a list of errors.  This is usually caused by simple syntax errors, although sometimes the \mathematica\ syntax is ok but something has been implemented incorrectly.  When this happens, the user should comment out everything in the \feynrules\ model files that is new since the last time the model files worked.  Then, the user should iteratively uncomment small pieces of the model file and retest loading it into \feynrules.  This process should be continued until the smallest possible chunk that breaks it is found.  Once this is done, the user can find the typo and fix it.  There may be several problems like this that need to be solved one-at-a-time.  

Once the model loads without error messages, the user should next test the hermiticity of their Lagrangian, the proper diagonalization and normalization of the kinetic and mass parts of the Lagrangian as well as the Feynman rules in the literature.  If problems are found, they should be fixed before moving on.  After this, the user should use the export interface of their choice.  If problems are found at this stage or at the stage of using the resulting model files in the Feynman diagram calculator, the user should review the appropriate subsection of Section~\ref{sec:interfaces} to make sure they have followed all the requirements for that package.  Inconsistencies should be corrected before moving on.  Once the model appears to be running, the user should do a thorough check between gauges and matrix-element generators on the web validation.  

We have found that following this list of steps enables most bugs to be removed.  If the user still has issues after following these guidelines, \heshe\ may request help from the \feynrules\ authors.  However, the user should reduce the problem to the minimal case where it is present and show that \heshe\ has put forth a proper effort to uncover the problem themselves.  In addition, the version of their operating system, \mathematica, \feynrules, and any matrix-element generator used should be included in any request for help.

\section{Conclusions}
\label{sec:conclusion}

The \feynrules\ package allows a model builder to implement \hisher\ new theoretical model in a unified format, independent of the matrix element generator it will be used in.  The user inputs a list of fields, parameters, symmetry groups and Lagrangian.  \feynrules\ then computes the Feynman rules and outputs them to the format appropriate for the Feynman diagram calculator of choice.  Interfaces for \calchep/\comphep, \feynarts/\formcalc, \madgraph\, \sherpa\ and \whizard/\ohmega\ are currently available which allow \feynrules\ models to be used natively with these Feynman diagram calculational packages without any modification to the packages.  This has made these model implementations more robust, dependable and transferable.  Since the first version of \feynrules, many new features have been added to \feynrules\ requiring the release of a new major version with this accompanying manual.  

With these new features, \feynrules\ has created a platform where the tree-level phenomenology of large varieties of new physics models can easily be studied. Recently, however, new automated tools allow the user to generate events at next-to-leading order (NLO) in perturbation theory~\cite{Cullen:2011ac,Berger:2008sj,Hirschi:2011pa,Cascioli:2011va,Bevilacqua:2011xh,Badger:2012pg}. These tools require not only the input of the tree-level vertices of the model, but in addition the user has to provide all the one-loop ultra-violet counterterms required to renormalize the one-loop amplitude. Furthermore, depending on the underlying algorithm to construct the one-loop matrix element, additional tree-level-like vertices need to be included to reproduce the correct rational terms in the one-loop amplitude~\cite{Draggiotis:2009yb,Garzelli:2009is,Garzelli:2010qm,Shao:2012ja}. While all of these new vertices formally go beyond a simple tree-level computation from a Lagrangian, they are nevertheless universal (for a given model) and can be computed once and for all. Future versions of \feynrules\ will allow the user to compute these ingredients required for NLO computations automatically from the tree-level Lagrangian of the model, thus allowing for an automated event generation at NLO accuracy for large classes of new physics models.

\section*{Acknowledgments}

We are grateful to Johan Alwall, Priscila de Aquino, Karen de Causmaecker, Nicolas Deutschmann, Jorgen D'Hondt, Camilo Garcia-Cely, David Grellscheid, Thomas Hahn, Valentin Hirschi, Fabio Maltoni, Olivier Mattelaer, Kentarou Mawatari, Bettina Oexl, Gizem \"Ozt\"urk, Michel Rausch de Traubenberg, Tho\-mas Reiter, J\"urgen Reuter, Christian Speckner, Tim Stelzer and Yoshitaro Takaesu  for discussion and collaborations on the different interfaces. We acknowledge the support of the IT team at UCLouvain (Vincent Boucher, Jer\^ome de Favereau, Pavel Demin).
The {\sc FeynRules} team is grateful to the IPHC laboratory for support in organizing the 2010 and 2012 {\sc FeynRules} meetings at Mont Sainte-Odile.
This work was supported in part by the Research Executive Agency (REA) of the European Union under the Grant Agreement number PITN-GA-2010-264564 (LHCPhenoNet) and
MCnetITN FP7 Marie Curie Initial Training Network PITN-GA-2012-315877.
A.A. acknowledges partial support of a PhD fellowship of the French ministry for education and research.
A.A. and B.F were supported in part by the French ANR 12 JS05 002 01 BATS@LHC and by the Theory-LHC-France initiative of the CNRS/IN2P3.
Ce. D. was supported in part by the U. S. Department of Energy under Contract No. DE-FG02-13ER42001 and by the NSF grant  PHY-0757889.
 N.D.C. was supported in part by the LHC-TI under U.S. National Science Foundation, grant NSF-PHY-0705682, by PITT PACC, and by the U.S. Department of Energy under grant No. DE-FG02-95ER40896 and C.D. was supported by the ERC grant ``IterQCD''.

\newpage

\bibliography{FRBib}

\end{document}